\documentclass[11pt,a4paper]{article}
\pdfoutput=1
\usepackage{jheppub}
\usepackage{latexsym,amsfonts,amsmath,amssymb}
\usepackage{bbm}
\usepackage{graphicx}
\usepackage{color}
\usepackage{caption}
\usepackage{subcaption}
\usepackage[normalem]{ulem}
\usepackage{comment}
\usepackage{url}
\usepackage{slashed}
\usepackage{tabu}
\usepackage{multirow}

\newcommand{\CK}{\mathcal{K}}
\newcommand{\CL}{\mathcal{L}}

\newcommand{\CN}{\mathcal{N}}

\newcommand{\CV}{\mathcal{V}}

\newcommand\cD{{\cal D}}

\newcommand\cF{{\cal F}}

\newcommand\cL{{\cal L}}

\newcommand\cN{{\cal N}}

\newcommand\cV{{\cal V}}

\renewcommand{\Im}{{\rm Im}}

\newcommand{\Tr}{\mbox{Tr}}

\newcommand{\IR}{\mathbb{R}}

\newcommand{\IZ}{\mathbb{Z}}

\newcommand{\half}{\frac{1}{2}}
\newcommand{\ndt}{\noindent}

\newcommand{\nn}{\nonumber}

\def\e{\epsilon}

\def\i{\mathrm{i}}

\def\p{\partial}

\def\bea{\begin{eqnarray}}
\def\eea{\end{eqnarray}}
\def\be{\begin{equation}}
\def\ee{\end{equation}}
\def\bse{\begin{subequations}}
\def\ese{\end{subequations}}

\newcommand{\D}{\mathcal{D}}

\newcommand{\bem}{\begin{pmatrix}}
\newcommand{\eem}{\end{pmatrix}}

\renewcommand{\=}{\;  = \;}
\def\+{\, + \,}

\def\wt{\widetilde}
\def\wh{\widehat}
\def\bar{\overline}

\def\rt2{\sqrt{2}}

\renewcommand{\Im}{\mbox{Im}}

\newcommand{\SBH}{S^\text{qu}_\text{BH}}

\def\s{\sigma}
\def\g{\gamma}

\def\a{\alpha}
\def\b{\beta}
\def\d{\delta}

\def\l{{\lambda}}
\def\o{{\omega}}
\def\w{{\omega}}

\def\ve{\varepsilon}

\def\nv{n_\text{v}}
\def\nh{n_\text{h}}

\newcommand{\qb}{Q_\text{brst}}
\newcommand{\qeq}{Q_\text{eq}}

\newcommand{\db}{\delta_\text{brst}}
\newcommand{\deq}{\delta_\text{eq}}

\newcommand{\ba}{\begin{array}}
\newcommand{\ea}{\end{array}}

\newcommand{\nbd}{n_\text{bdry}}

\title{Twisting and localization in supergravity: \\ equivariant cohomology of BPS black holes}

\author{Imtak Jeon$^{a}$} 
\author{and Sameer Murthy$^b$}

\affiliation{$^a$ Harish-Chandra Research Institute Chhatnag Road,\\ 
   Jhusi, Allahabad 211019, India}
\affiliation{$^b$ Department of Mathematics, King's College London,\\
  The Strand, London WC2R 2LS, U.K}

\emailAdd{imtakjeon@gmail.com}
\emailAdd{sameer.murthy@kcl.ac.uk}

\abstract{We develop the formalism of supersymmetric localization in supergravity using  
the deformed BRST algebra defined in the presence of a supersymmetric background as 
recently formulated in~\cite{dWMR}.
The gravitational functional integral localizes onto the cohomology of a global 
supercharge~$\qeq$, obeying~$\qeq^{2}=H$, where~$H$ is a global symmetry of the background.
Our construction naturally produces a twisted version of supergravity whenever supersymmetry can be realized off-shell. 
We present the details of the twisted 
graviton multiplet and ghost fields for the superconformal formulation of four-dimensional~$\CN=2$ supergravity. 
As an application of our formalism, we systematize the computation of the exact quantum 
entropy of supersymmetric black holes. In particular, we compute the one-loop determinant 
of the~$\qeq \CV$ deformation operator for the off-shell fluctuations of the Weyl multiplet
around the~$AdS_{2} \times S^{2}$ saddle. This result, which is consistent with the corresponding 
large-charge on-shell analysis, is needed to complete the first-principles computation of the quantum entropy. 
}

\begin{document}

\maketitle

\section{Introduction}
\label{sec:introduction}

The program of computing the exact macroscopic quantum entropy of supersymmetric black holes in string theory has 
made good strides since its inception~\cite{Dabholkar:2010uh, Dabholkar:2011ec}. 
In the simplest cases one can go much beyond the leading asymptotic analysis of~\cite{Sen:1995in, Strominger:1996sh}, 
and compute exact integer degeneracies from a continuum calculation in macroscopic string theory, thus 
providing valuable lessons about the quantization of the gravitational degrees of freedom of a black hole. This has been 
made possible by the application of the technique of supersymmetric localization to supergravity in the near-horizon~$AdS_{2}$ 
region of BPS black holes. The remarkable success of this idea indicates that it may be very useful in a larger class of situations 
beyond that of BPS black holes in asymptotically flat space considered in~\cite{Dabholkar:2010uh, Dabholkar:2011ec}. 
One could think of applying similar ideas to other BPS black holes or, more broadly, to calculate exact bulk 
functional integrals in a generic~$AdS_{d+1}/CFT_{d}$ setting, thus giving rise to an exact understanding of a sector of holography.

The main idea of localization, as is well-known by now, is to consider a fermionic operator~$Q$ that is a symmetry
of the theory~\cite{Duistermaat:1982vw, Berline:1982, Atiyah:1984px, Witten:1988ze, Witten:1988xj}. 
One deforms the theory by a~$Q$-exact operator and the functional integral reduces to an integral 
over the set of critical points of this deformation~$Q\CV$. With an appropriate choice of~$\CV$, one obtains the critical points 
to be the set of all off-shell field configurations annihilated by~$Q$. Equivalently, one twists all the fields by the spinorial 
generator of~$Q$, and then the functional integral can be written as an integral over the space of twisted or cohomological 
variables that are in manifest representations of the supersymmetry algebra. The twisting procedure also greatly 
simplifies the calculation of the one-loop determinants of the deformation operator involved in localization. 
In its most powerful equivariant version, we have a supersymmetric theory defined on a background space 
that admits a fermionic charge obeying the off-shell algebra~$Q^{2}= H$, 
with~$H$ being a compact bosonic generator 
acting on the background space as well as the field space~\cite{Nekrasov:2002qd}. 

Despite its successes mentioned above, localization in supergravity has always suffered from some formal issues as well
as practical problems. In this paper we address and resolve two of the foundational issues:~(1) What is the meaning 
of~$Q$ in supergravity?~(2) What are the correct twisted variables of supergravity? The heart of the difficulties in both these 
problems lies in the non-linear nature of supergravity. As we explain, the answers to both questions depends on the existence 
of a supersymmetric background, which we assume to be a non-compact space with an asymptotic boundary, that is 
used to define the global symmetries. We focus on asymptotically Anti de Sitter space here, but our construction should 
also apply to other spaces like asymptotically flat space. In the rest of the introduction, we explain these two questions, their resolution,
and their consequences in some detail.

\subsection{A global supercharge~$\qeq$ in supergravity}

The main formal issue underlying localization in supergravity is 
how to define a rigid supercharge in the quantum theory of supergravity 
in which the metric and gravitini are fluctuating in the functional integral. 
This is sometimes expressed as the slogan that all symmetries in 
supergravity are gauge symmetries, or that there is no global (super)symmetry in (super)gravity. 
One can of course overcome this by considering a space with a boundary, interpreting the 
boundary conditions on the fields as a fixed background, and integrating over the fluctuations. 
The (super)symmetries of the background are now our global symmetries. 

The question really is how to implement these background symmetries 
on all the fluctuating quantum fields of the gravitational theory. In ordinary gauge theories, 
there is a well-understood method to split the gauge transformations between the background and quantum fields. 
In contrast, the background field method in supergravity has not been well-developed until now, 
the technical hurdle being the field-dependence of the structure functions of the gauge algebra of supergravity. 
The general formalism to perform covariant quantization in the presence of a background/boundary for generic gauge 
algebras was put forward in the paper~\cite{dWMR}. Here we flesh out this idea in the context of supergravity. 
In particular, by choosing a background field configuration that admits a global supersymmetry algebra, we 
construct a rigid fermionic generator~$\qeq$ which is a deformation of the BRST algebra and 
obeys~$\qeq^{2}=H$ 
with~$H$ being an isometry of the background. We then use this fermionic symmetry to localize the functional integral.

\subsection{The variables of twisted supergravity and~$\qeq$-cohomology}

The core problem with the construction of the twisted variables is that  
fixing a metric background partially fixes the gauge in the gravitational multiplet, and 
the gauge-fixing condition is generically not compatible with supersymmetry. 
This incompatibility shows up, for example, as a mismatch between the off-shell 
bosonic and fermionic degrees of freedom. 
There are three ways in which this problem is fixed in rigid supersymmetric gauge theories: 
(i) use the full superfield, (ii) fully fix the gauge symmetry (explicitly breaking covariance), 
or (iii) do a covariant quantization by introducing ghosts, and choosing a combination of the global~$Q$ 
and the BRST charge~$\qb$ to perform the twist~\cite{Baulieu:1988xs}.  
In our current situation, (i) we do not 
generally have a superfield formalism for supergravity\footnote{In special cases like~$\CN=1$ supergravity in 
four dimensions, one can exploit the superfield formulation of supergravity as in~\cite{Grisaru:1981xm, Grisaru:1983rg}.}, 
and (ii) fully gauge-fixing the gauge symmetries of supergravity is not technically easy, nor is it particularly elegant.  
The third route is general and systematic, but unlike for Yang-Mills theory, generically we only have the 
nilpotent BRST charge $\qb$ in supergravity. Once again, the introduction of a 
background with its global symmetry~$\qeq$ solves this problem.

In this paper we construct the cohomological classification of the fields of the Weyl-multiplet of~$\CN=2$ superconformal 
gravity~\cite{deRoo:1980mm,deWit:1980tn,deWit:1984px} combined with the ghosts for the gauge symmetries of the theory. 
We find that this is a multiplet consisting of 94~bosonic 
+ 94~fermionic degrees of freedom, paired up under the fermionic generator~$\qeq$.\footnote{Representations of superalgebras in 
Yang-Mills theories preserving various fractions of 
supersymmetry~\cite{Berkovits:1993hx, Baulieu:2007ew} have been used in a powerful manner for the 
localization of these theories~\cite{Pestun:2007rz}. 
As far as we are aware, the analogous construction in supergravity around non-trivial supersymmetric backgrounds is new.} 
This is our construction of twisted supergravity. 
There have been previous discussions involving supergravity 
fields and ghosts, and their twisting, from different points of view, mainly involving relations to topological 
theories~\cite{Baulieu:2012jj,Bae:2015eoa,Costello:2016mgj,Imbimbo:2018duh}. 
As far as we are aware, the explicit construction of the transformation rules of the rigid supercharge 
in the supergravity-ghost system around non-trivial supersymmetric backgrounds is new. 
The papers~\cite{Bae:2015eoa,Imbimbo:2018duh} also consider applications to localization relating to the 
problem of finding backgrounds obeying the localization equations, on which rigid gauge theories can be defined. 
In contrast, our construction here allows us to go beyond the supersymmetric solutions and 
actually perform the integral over all fluctuations of supergravity.

\vspace{0.3cm}

As a consequence of our formalism, we can exhibit the equivariant cohomology in the $AdS_{2} \times S^{2}$ near-horizon 
background of BPS black holes. In this case the algebra is~$\qeq^{2}= L_{0}-J_{0} \equiv  H$ 
where~$L_{0}$ is the rotation of the Euclidean~$AdS_2$ Poincare disk and~$J_{0}$ is a rotation of~$S^2$. 
Around the fixed points of~$H$ the symmetry generators can be embedded in the~$SO(4)$ rotation group,
and the equivariant cohomology becomes that of the rigid theory. 
We use our cohomological classification of fields to compute, using index theory, 
the off-shell one-loop determinant of the deformation operator~$Q\CV$ in the localization procedure for the Weyl multiplet, 
following the formalism of~\cite{Pestun:2007rz, Hama:2012bg}. 
In the black hole context, there is an additional subtlety due to the non-compactness of~$AdS_2$ and the 
related ``boundary modes"~\cite{Banerjee:2009af}. A careful treatment of these modes leads to a modification of the usual index analysis. 
Combined the corresponding 
one-loop computation for vector multiplets and hypermultiplets~\cite{Gupta:2015gga, Murthy:2015yfa}, this 
gives a complete answer for~$\CN=2$ supergravity coupled to vector multiplets.

\vspace{0.3cm}

The plan of the paper is as follows. In Section~\ref{sec:BRST} we review the construction of the deformed 
BRST algebra presented in~\cite{dWMR}, and apply it to generic off-shell supergravity theories. In Section~\ref{sec:twisted} 
we focus on~$d=4$, $\CN=2$ supergravity in the superconformal formalism and present the twisted variables 
and the complete set of transformations under~$\qeq$. In Section~\ref{sec:BH} we apply our formalism to 
the~$AdS_{2} \times S^{2}$ near-horizon background of supersymmetric black holes and compute the 
one-loop contribution of the Weyl multiplet to the localization formula. This includes an analysis of the 
boundary modes of~$AdS_2$ in our supersymmetric formalism.
In Section~\ref{sec:Outlook} we 
conclude with an outlook and some speculations about how our ideas can be useful in various directions. 
In the appendices we record our conventions and the details of the superconformal supergravity that we use,
in an attempt to make this paper reasonably self-contained.

\section{Deformed BRST cohomology in supergravity \label{sec:BRST}}

In this section we summarize our ideas of equivariant cohomology in supergravity. 
We begin this section by a review of~\cite{dWMR} in which the formalism for an equivariant BRST cohomology for an arbitrary 
gauge algebra in the presence of a background is constructed. We briefly review the set up and the key equations of~\cite{dWMR},  
and make some comments on the physical interpretation of the BRST variation equations. 
In the second subsection we review the formulation of supergravity as a gauge theory and write down the general 
form of the variations of all the fields and ghosts, ending with the twisted algebra. In the third subsection we 
consider supergravity coupled to vector multiplet matter fields and show how we can recover known results 
as a limit of our formalism. 

\subsection{Review of the general formalism}

We work in the context of a gauge theory whose fields are generically denoted by $\phi^i$.
We follow the notations and conventions of~\cite{Gomis:1994he} unless explicitly mentioned otherwise. 
The infinitesimal gauge transformations are of the form:
\begin{equation}
  \label{eq:gauge-transf}
  \delta\phi^i \=\xi^{\,\a} R_\a{}^{i}(\phi) \,, 
\end{equation}
where $R_\a{}^i$ may include derivatives acting on the (bosonic or fermionic\footnote{Our conventions for placement of 
the Grassman variables is different from~\cite{dWMR}. It is chosen to allow us to take away the Grassman
parameter easily when we define the charge from the variation rules.}) parameters $\xi^\a(x^{\mu})$.
We consider theories where the gauge transformations obey off-shell closure, which is expressed by
\begin{equation}
  \label{eq:closure}
\xi^{\,\g}_{[2} \,\xi^{\,\b}_{1]} R_{\b}{}^{j}\, \partial_j R_{\g}{}^{i} \= \frac12  \xi^{\,\g}_{2} \,\xi^{\,\b}_{1}  f_{\b \g}{}^\a \, R_\a{}^i  \,.
\end{equation}
The gauge transformations also obey the Jacobi identity:
\begin{equation}
  \label{eq:Jacobi}
 \xi_1^{\,\a}\xi_2^{\,\b}\xi_3^{\,\g} R_\g{}^j \partial_j f_{\b\a}{}^{\d}- \xi_1^{\,\a}\xi_2^{\,\b}\xi_3^{\,\g} f_{\g\b}{}^{\s}f_{\s\a}{}^{\d}+ \mbox{cyclic in $(1,2,3)$}  \= 0 \,.
\end{equation}
The equation~\eqref{eq:closure} defines the structure functions~$f_{\a\b}{}^\g(\phi^i)$. In Yang-Mills theories, these reduce to the structure constants 
of the gauge group. In supergravity, these functions depend in a non-trivial manner on the fields, and many of the complications 
of supergravity arises from this dependence.

We are interested in a set up where the fields are decomposed into background and quantum fields as
\begin{equation}
  \label{eq:bg-split}
  \phi^i \= \mathring\phi{}^{\,i} + \widetilde\phi{}^{\,i}\,.
\end{equation}
Correspondingly we can restrict some of the gauge transformations to a subgroup parameterized 
by~$ \mathring{\xi}^{\,\a}$,
and get background transformation of the form
\begin{equation}
  \label{eq:split-transfo} 
  \mathring{\delta}\mathring{\phi}^i \= \mathring{\xi}^{\,\a} R(\mathring{\phi})_\a{}^i\,. 
\end{equation}
The action of the background transformations on the quantum fields~$\wt \phi{}^{\,i}$ is then a difference of the 
transformation~\eqref{eq:gauge-transf} on the full field and the background transformations~\eqref{eq:split-transfo}.

The BRST transformation rules for the background and quantum fields are derived in~\cite{dWMR} by promoting 
the gauge transformations~\eqref{eq:gauge-transf}, \eqref{eq:split-transfo} to BRST variations, and then 
showing that these transformations form a closed algebra, thus leading to a nilpotent operator. Below 
we include a slightly different presentation using the idea of background-freezing\footnote{This idea was 
inspired by its use, with great effect, in various contexts in field theory and string theory~\cite{Seiberg:1993vc}. 
The new point here is to apply it to the ghost system.}. 
The usual BRST transformation rules on the full (background + quantum fields)\footnote{We use the 
notation~$c^{\,\a}$ rather than~$\wt c^{\,\a}$ to denote the quantum ghost as there is only one such field in any theory. 
The field~$\mathring{c}^{\, \a}$, although formally playing the role of the background ghost at the moment, 
will become a fixed parameter rather than a ghost field in our treatment below.} are
\bea \label{eq:brst2again}
\db\,\phi^i & \=&  \Lambda\,(\mathring{c}+c)^\a\, R(\phi)_\a{}^{i}    \, ,  \\ 
\label{eq:brst4again}
\db \, (\mathring{c} + c)^{\a}  & \=&-  \frac12 \, 
(\mathring{c}+c)^\g \Lambda \,(\mathring{c}+c)^{\b}\,f(\phi)_{\b\g}{\!}^\a\,   \, .
\eea
We now insert a factor of~$m_\text{P}$ to separate the classical and quantum 
parts as\footnote{Here we have assumed that the dimension of~$\phi^i$ is one, but it is a general fact that the quantum
fluctuations are suppressed by a positive power of~$m_\text{P}$ and the heuristic argument below goes through. 
The algebra of BRST transformations can be verified independent of these arguments.} 
$\phi = \mathring{\phi} + \frac{1}{m_\text{P}}\widetilde{\phi}$ and~$c^\text{full} := \mathring{c} + \frac{1}{m_\text{P}} c$.
The limit~$m_\text{P} \to \infty$ isolates the BRST transformations acting only on the background fields.
Upon subtracting these background transformations from the full field, we obtain the 
transformation laws of the quantum fields: 
\bea
\label{eq:brst2}
\db\,\widetilde{\phi}^i &\= &\Lambda\,(c + \mathring{c})^\a\, R(\phi)_{\a}{}^i 
  -  \Lambda\,\mathring{c}^{\,\a}\,R(\mathring{\phi})_{\a}{}^i   \, ,  \\
\label{eq:brst4}
\db \, c^{\,\a} &\=& -\frac12 \,(c+\mathring{c})^\g \Lambda \,(c+\mathring{c})^{\b}\, f(\phi)_{\b \g}{\!}^\a\,
  +\frac12 \,  \mathring{c}^{\,\g}\Lambda \,\mathring{c}^{\,\b}\,
  f(\mathring{\phi})_{\b \g}{\!}^\a \, .
\eea
It is clear that the charge~$\db$ is nilpotent, as the transformation rules on the full fields 
as well as on the background is exactly the usual BRST variations, and the variation on the quantum part is 
simply the difference of the two.

The next step is to deform this BRST charge to a new charge~$\deq$ as follows. First we freeze the background 
to some fixed values, which we will take in our application to be the boundary value of the fields, typically a 
solution of the equations of motion of the theory. This can be thought of as a partial gauge-fixing procedure, and 
the corresponding background ghosts
should be set to zero in order for the BRST variations 
to be consistent. 
The only background ghosts that can still have non-zero values are the ones corresponding to isometries of the 
background, which obey
\be \label{isometry}
\mathring{c}^{\,\a}\, R(\mathring{\phi})_\a{}^i \=  0 \, . 
\ee
In the~$AdS/CFT$ type situation mentioned in the introduction, the background fields are fixed by the 
boundary conditions and are not allowed to fluctuate in the functional integral. The isometries above are 
parameterized by background ghosts that are non-normalizable in spacetime, and are 
therefore also fixed in the functional integral. In situations 
where the isometries are normalizable, e.g.~when the spacetimes are compact, we need to introduce
ghosts for ghosts and so on, we will not  consider such situations here in this paper.

The required deformation is obtained by combining this isometry condition with the BRST rules~\eqref{eq:brst2}, \eqref{eq:brst4}:
\bea
\label{eq:new-brst2again}
\deq\, \widetilde{\phi}^i    & \= & \Lambda\,(c + \mathring{c})^\a\, R(\phi)_{\a}{}^i 
 \, ,   \\
 \label{eq:new-brst4again}
\deq \, c^{\, \a} &\=& -\frac12 \,(c+\mathring{c})^\g \Lambda \,(c+\mathring{c})^{\b}\, f(\phi)_{\b\g}{\!}^\a\,  + \frac{1}{2\kappa}  \, 
\mathring{c}^{\,\g}\Lambda \,\mathring{c}^{\,\b}\,
  f(\mathring{\phi})_{\b\g}{\!}^\a  \, .
\eea
Here we have inserted a parameter~$\kappa$ multiplying the deformation term.
The equivariant rules are obtained at~$\kappa=1$, while the ususal BRST 
rules on the full field~\eqref{eq:brst2again},~\eqref{eq:brst4again} are recovered in the limit~$\kappa \to \infty$. 
The equivariant charge obeys the algebra
\be \label{eqalgebra}
\deq^{2} \= \mathring{\delta}_{\mathring{\xi}} \, ,
\ee
where~$\mathring{\delta}_{\mathring{\xi}}$ is the background isometry transformation parameterized by the bilinear\footnote{The order of the grassmann parameter $\Lambda_1$ and $\Lambda_2$ is for $\deq^2 = {\deq}_{1}{\deq}_{2}$.}
\begin{align}
\mathring{\xi}^\a & \= \frac12  \,\Lambda_{2}\,
  \mathring{c}^{\,\g}  \Lambda_{1} \,\mathring{c}^{\,\b}  \, f(\mathring{\phi})_{\b\g}{\!}^\a \, ,
\end{align}
acting on the quantum fields as
\bea 
\mathring{\delta}_{\mathring{\xi}} \,\widetilde{\phi}^i  &\= & \mathring{\xi}^{\,\a}\,R(\phi)_{\a}{}^i \, ,  \\
\mathring{\delta}_{\mathring{\xi}}\, c^{\,\a} &\= &-  c^\g  \, \mathring{\xi}^{\,\b} \, f(\phi)_{\b\g}{\!}^\a  \, .
\eea
These transformations are exactly what we expect according to the representations of the fields and ghosts under the  
isometry transformation around an invariant background: the quantum matter fields~$\wt \phi^i$ transform\footnote{The 
quantum fields generically transform according to the difference of the full transformation~$R$ \eqref{eq:gauge-transf} and the 
background transformation~\eqref{eq:split-transfo}, but in our situation the background transformations are isometries
and therefore have a vanishing action~\eqref{isometry}.}
in the representation~$R_\a{}^{i}$ and the~$c^\a$-ghosts transform in the adjoint representation. 

Now we turn to the anti-ghost~$b_\a$ and the Lagrange-multipliers~$B_\a$. Since we have frozen the background
fields, the background values for these fields can be set to zero. At this point we specialize to our situation of 
interest, namely supergravity backgrounds in which the only background ghosts~$\mathring{c}^\a$ are those 
corresponding to fermionic transformations. In this case we can write the 
transformations on the quantum anti-ghost fields:
\bea \label{btrans}
\deq \, b_{\a} &\=& \Lambda \, B_{\a} \,,\\
\deq \, B_{\a} &\=&  \half \,   \mathring{c}^{\,\s}  \Lambda \,\mathring{c}^{\,\d}  \, f(\mathring{\phi})_{\d \s}{\!}^\b \, f(\phi)_{\b\a}{\!}^\g \, b_{\g}\,.
\label{bBtrans}
\eea
One can check that the commutator of two transformations on these fields also obeys the algebra~\eqref{eqalgebra}
where the background transformations~$\mathring{\delta}_{\xi}$ act as
\be
\deq^2 \, b_{\a} \=  \mathring{\xi}^{\b} f(\phi)_{\b\a}{}^{\g} \, b_{\g} \,,
\ee
as consistent with the fact that~$b_\a$ transforms in the adjoint representation of the full gauge algebra, in 
parallel with the situation for the quantum fields~$\wt \phi^{i}$ and ghosts~$c^{\a}$. 
The algebra also closes in the same way for the Lagrange-multiplier~$B_\a$, i.e.
\be \label{xicircB}
\mathring{\delta}_{\mathring{\xi}}\, B_{\,\a} \=  \mathring{\xi}^{\b} f(\phi)_{\b\a}{}^{\g} \, B_{\g} \, ,
\ee
but this deserves a comment. In a generic theory, if we assume that~$B^\a$ transforms as in~\eqref{bBtrans}, then the 
square of two transformations does not close on~$\mathring{\delta}_{\xi}$ (and contains extra terms with derivatives of 
the structure functions). 
In the construction of~\cite{dWMR}, the closure of the algebra is guaranteed by choosing~$\deq B_{\a}$ to only
involve the \emph{background} structure constant~$f(\mathring\phi)_{\b\a}{\!}^\g$,
instead of the full structure function~$f(\phi)_{\b\a}{\!}^\g$ as in~\eqref{bBtrans}, and as a consequence,
the background transformation~\eqref{xicircB}
also only involves the background structure functions. 
In our supergravity situation, this tension between the closure of the algebra and the``natural" transformation 
of the~$B^\a$ field (as a representation of the full gauge algebra)
does not arise because the relevant function~$f(\phi)_{\b\a}{\!}^\g$ in~\eqref{bBtrans} 
is actually constant. This is because the only functional dependence appears in the commutator of two 
supersymmetries, while the other structure functions are constants. Since we only allow non-zero $\mathring{c}^{\,\a}$ for 
fermionic transformations, the index~$\b$ in the transformation of~$B$ is necessarily bosonic, which therefore implies 
the constancy of~$f(\phi)_{\b\a}{\!}^\g$. 
Thus the quantum and background values are equal, and so~\eqref{bBtrans} is consistent 
with the general construction of~\cite{dWMR}.

The final algebra can be written simply as 
\be \label{eqalgebraagain}
\deq^{2} \= \mathring{\delta}_{\mathring{\xi}} \, ,
\ee
where the background transformation~$\mathring{\delta}_{\mathring{\xi}}$ acts on any quantum field of the theory according
to its representation under the full gauge algebra.


\subsection{Application to supergravity\label{sec:sugra}}

As is well-known, supergravity can be formulated as a gauge theory. The gauge algebra is slightly 
more complicated compared to rigid supersymmetric theories, but falls within the general formalism of the previous section. 
The main technical complication for our purpose, as mentioned in the introduction, is the fact that the commutators of 
the algebra involve \emph{structure functions} rather than structure constants. 
The precise details of the structure functions depends on the theory under consideration, but there is a general 
structure which we now review. (See e.g.~the textbook~\cite{Freedman:2012zz} for a nice introduction.)
Our interest is in off-shell supergravities, and we follow the construction of the conformal supergravity. 

The key symmetries present in any supergravity theory are general coordinate transformations 
(diffeomorphisms) and local supersymmetry transformations. In addition there are other local (bosonic and fermionic) 
symmetries required by consistency. The general coordinate transformations play a special role in the algebra and we 
denote them by~$\delta_{\text{gct}}(\xi)$ where the parameter~$\xi^{\mu}(x^{\nu})$ is the vector field generating the diffeomorphism. 
We collectively denote the rest of the (bosonic and fermionic) gauge transformations by~$\d_{A}(\varepsilon^{A})$. 
Of these, the local supersymmetry transformations are special, they are 
denoted by~$\delta_{Q}(\varepsilon)$ and parameterized by the spinor field~$\varepsilon_{\alpha}(x^{\nu})$. 
The general form of the algebra is as follows:
\bea \label{Nalgebra}
\left[ \delta_{\text{gct}}(\xi_1)\,, \delta_{\text{gct}}(\xi_2)\right] & \= & \delta_{\text{gct}}([\xi_2,\xi_1]) \,,\nn \\[0.5mm]
\left[  \delta_{A}(\varepsilon^A)\,,\delta_{\text{gct}}(\xi)\right] &\= & \delta_A(\xi^\mu \partial_\mu \varepsilon^A) \,, \\[0.5mm]
\left[  \delta_{A}(\varepsilon_1^A)\,,\delta_{B}(\varepsilon_2^B)\right] & \= & \delta_{\text{gct}}( v) +\delta_A (\varepsilon_3^{A}) \,, \nn
\eea
where the parameters on the right-hand side are given by
\be \label{epsbil}
v^\mu \= \varepsilon_2^B \varepsilon_1^A f_{AB}{}{}^{\mu}(\phi) \,,
\; \qquad \varepsilon_3^A \= \varepsilon_2^C \varepsilon_1^B f_{BC}{}^{A}(\phi) \,.
\ee

It is clear from the relations~\eqref{Nalgebra} that the softness of the algebra only appears in the third line, i.e.~in 
the (anti-)commutator of the gauge transformations~$\d_{A}(\varepsilon^{A})$. We now briefly review the origin 
of this softness, as this will be important in the following. 
The starting point to construct the off-shell supergravity gauge algebra is a regular rigid 
super Lie algebra which always includes local translations~$P^{a}$ and local Lorentz 
transformations~$M^{ab}$ (here~$a$ is the local tangent space index), which is then gauged. 
One then has to impose ``conventional'' constraints on the various curvatures. This is a supersymmetric 
generalization of the bosonic constraint which identifies the gauge fields~$e_{\mu}^{a}$ for the local translations 
and~$\o^{ab}_{\mu}$ for the local Lorentz transformations with the vielbein and the spin 
connection, respectively~\cite{Kibble:1961ba}. 
As a consequence of imposing these constraints, the local translations~$P^{a}$ turn into general coordinate 
transformations, and the algebra is modified at a non-linear level. 
The (anti)commutators in the third line of~\eqref{Nalgebra}
that are modified are precisely those that involve a translation on the right-hand side, i.e.~the 
anticommutator of two supersymmetries. 
This anticommutator now contains the general coordinate transformation 
involving the vielbein, as well as the various other gauge transformations of the theory involving 
the corresponding gauge fields. The rest of the~$\d_{A}(\varepsilon^{A})$ transformations are  
homogeneous transformations which rotate the fields and do not produce any translations in their commutators. 
A further field-dependence appears because of the auxiliary fields that are needed to close the supersymmetry algebra
off-shell (again this only appears in the anticommutator of two supersymmetries). 
The bottom line is that the softness of the algebra is manifested only in the structure functions~$f_{AB}{}{}^{\mu}$,
$f_{AB}{}{}^{C}$ with~$A,B$ both corresponding to the fermionic transformations. 
This is explicitly illustrated for the case of~$d=4$, $\CN=2$ conformal supergravity in the algebra 
of transformations~\eqref{algebra1} with~\eqref{CGCT} and \eqref{parameters1}.

In addition to the fields of supergravity, we introduce, for each of the local symmetries, a ghost system consisting 
of ghosts~$c$, anti-ghosts~$b$, and Lagrange multiplier~$B$. The ghost~$b$ and anti-ghost~$c$ for bosonic 
(fermionic) gauge symmetries are fermionic (bosonic), and the Lagrange multiplier~$B$ is bosonic (fermionic).
Now we write the transformations of all the fields under the equivariant supercharge~$\qeq$, following the prescription of 
the previous section. 
We choose a background~$\mathring{\phi}$ which admits some rigid supersymmetry and a corresponding Killing spinor, and 
set all the background value of the ghost fields to be zero except for the ghost of the local supersymmetry variation. 
The equation \eqref{isometry} 
\be
\mathring{c}^A R(\mathring{\phi})_A{}^{i} \= 0 \,\nn 
\ee
is simply the condition that the background has a fermionic isometry, i.e.~a rigid supersymmetry, and the 
corresponding parameter $\mathring{c}^A$ is simply the corresponding Killing spinor. Here we assume that we have 
a non-compact background so that all the isometry parameters are non-normalizable, otherwise we would need 
an additional gauge fixing procedure by introducing ghost for ghosts.  

In this situation, the deformed BRST  transformation given in (\ref{eq:new-brst2again}),~(\ref{eq:new-brst4again}),~(\ref{bBtrans}) and~(\ref{bBtrans}) is
\bea  \label{ModifieBRST2}
\deq  \,\widetilde{\phi}^i & \= & \cL_{\Lambda c^\mu}\phi^i+ \Lambda (\mathring{c}+ c)^A R_A{}^i 
(\phi)\,,\nn \\[0.5mm]
\deq  \,c^\mu & \= & \Lambda {c}^\nu \partial_\nu c^\mu-\half (\mathring{c} + c)^B \Lambda (\mathring{c}+c)^A f_{AB}{}^{\mu}({\phi})+\half \mathring{c}^B \Lambda \mathring{c}^A f_{AB}{}^{\mu}(\mathring\phi) \,,  \nn \\
\deq  \,{ c}^A & \= &\Lambda c^\mu \partial_\mu (\mathring{c}+c)^A-\half (\mathring{c}+c)^C \Lambda(\mathring{c}+ c)^B f_{BC}{}^{A}({\phi})+\half \mathring{c}^C \Lambda\mathring{c}^B f_{BC}{}^{A}(\mathring{\phi}) \,, \nn \\[0.5mm]
\deq \, {b}_\mu & \= &\Lambda {B}_\mu \,, \nn \\[0.5mm]
\deq  \,{ b}_A & \= &\Lambda {B}_A\,,~~~ \\
\deq \, {B}_\mu & \= &\cL_{ \frac{1}{2} \mathring{c}^B \Lambda \mathring{c}^A f_{AB}{}^{\mu}(\mathring\phi)}{b}_\mu+\half \partial_\mu    
\Bigl( \mathring{c}^C\Lambda \mathring{c}^B f_{BC}{}^{A}(\mathring{\phi})\Bigr){b}_A\,,~~~~~~~~~~~~~~~~~\nn \\
\deq {B}_A & \= &\cL_{\frac{1}{2} \mathring{c}^B \Lambda \mathring{c}^A f_{AB}{}^{\mu}(\mathring\phi)}{b}_A+\half  \mathring{c}^L \Lambda \mathring{c}^C f_{CL}{}^{B}(\mathring{\phi}) f_{BA}{}^{\mu}(\phi){b}_{\mu} \cr 
&& \qquad \qquad \qquad \qquad \qquad  \qquad \qquad + \half  \mathring{c}^M \Lambda \mathring{c}^L f_{LM}{}^{B}(\mathring{\phi}) 
f_{BA}{}^{C}(\phi){b}_{C}\,. \nn
\eea
Now we express the equivariant cohomology without the formal  grassmann parameter $\Lambda$, by defining  $\deq= \Lambda \qeq$.
Since we only have background ghosts for supersymmetry, the surviving background bilinears are the Killing vector 
\be \label{KillingV}
\mathring{v}^\mu \;:=\; \half \mathring{c}^B  \mathring{c}^A f_{AB}{}^{\mu}(\mathring{\phi}) \,,
\ee
and the parameters for bosonic transformations
\be \label{BosonicPara}
\mathring\varepsilon_3^A \;:=\; \half  \mathring{c}^C \mathring{c}^B f_{BC}{}^{A}(\mathring{\phi})\,.
\ee

Now, recalling the discussion after Equation~\eqref{Nalgebra} 
that $f_{BA}{}^{\mu}(\phi)=0$ and that $f_{BA}{}^{C}({\phi})$ is constant whenever 
the index $B$ labels a bosonic symmetry transformation, we find that some of the structure functions in the 
transformations~\eqref{ModifieBRST2} are actually constant. 
A direct calculation of the various commutators results in the algebra:
\be \label{equiCoho}
\qeq^2 \= \cL_{\mathring{v}}  + \sum_{A\in\text{bos}} \delta_A(\mathring\varepsilon_3^A)\,,
\ee
where the sum in the second term is now over all bosonic symmetries except general coordinate transformations.

It is worth re-emphasising that the deformed BRST transformations~\eqref{ModifieBRST2} are consistently defined around 
an arbitrary supersymmetric background. The consequent algebra~\eqref{equiCoho} depends on the choice of 
background through its rigid symmetry parameters. On specializing to a flat background, we recover the algebra discussed  
in~\cite{Baulieu:2012jj,Bae:2015eoa,Imbimbo:2018duh}.  
%

\subsection{Matter multiplets coupled to supergravity \label{sec:mattersugra}}

The general formalism explained in the previous section can also be applied in the same manner when 
matter multiplets are coupled to supergravity. Many such examples of such constructions have been 
discussed recently (see e.g.~the review collection~\cite{Pestun:2016zxk}).
In this subsection, we show that our general formalism gives a uniform explanation for the various constructions.  

Suppose a matter multiplet is accompanied by internal gauge symmetry $G$ which we take to be generic 
non-abelian Lie group. Then  the superconformal symmetry gets  the central extension; in general,  
anti-commutation of two supercharges $Q$ generates the internal gauge symmetry $G$ with field dependent 
parameter. Thus the structure functions  are enlarged to include the internal gauge algebra, 
$\{{f}_{BC}{}^{A}(\phi)\} \rightarrow \{{f}_{BC}{}^{A}(\phi)\,,{f}_{BC}{}^{I}(\phi)\,,{f}_{J K}{}^{I}\}$,  
where $I\,, J\,, K$ are the gauge index and ${f}_{JK}{}^{I}$ is constant.  In addition to  the matter 
multiplet $\{\phi_m^i \}$, we include the ghost multiplet $\{c^I\,, b_I\,, B_I \}$ of the internal gauge 
symmetry~$G$ to the Weyl multiplet and its ghost multiplets. As in the previous section, we use the 
deformed BRST transformation as in~(\ref{ModifieBRST2}) to get the algebra~(\ref{equiCoho}).

In order to consider the matter fields on rigid supergravity background,  we suppress all the quantum fluctuations 
of the Weyl multiplet and its ghost fields and set them to their background values. Thus we have
\bea\label{QeqQFT}
\qeq \widetilde\phi_m^i  & \= & \mathring{c}^A R_A {}^i(\mathring\phi + \widetilde\phi_m)
+ c^I R_I{}^{i}(\mathring{\phi}+\widetilde{\phi}_m)\nn \\
\qeq c^{I} &\=& -\half \mathring{c}^C \mathring{c}^B ( f_{BC}{}^{I}(\mathring\phi+ \widetilde{\phi}_m)- f_{BC}{}^{I}
(\mathring{\phi}))+ \half c^K  c^J f_{J K}{}^{I}\nn\\
\qeq b_{ I} &\=&  B_{ I} \\
\qeq B_{ I} &\=&  \cL_{\frac{1}{2} \mathring{c}^B  \mathring{c}^A f_{AB}{}^{\mu}(\mathring\phi)}{b}_{I}+ 
\half  \mathring{c}^{\,B}  \mathring{c}^{\,A} f_{AB}{}^{J}(\mathring{\phi}) f_{J I}{}^{K}{b}_{K}\,,\nn
\eea
and the algebra closes equivariantly to
\be
\qeq^2 \= \cL_{\mathring{v}} +\sum_{A\in\text{bos} }\delta_A(\mathring{\varepsilon}_3^A) + \delta_G(\mathring{a})\,,
\ee
where the parameters~$\mathring{v}^\mu$ and  $\mathring{\varepsilon}_3^A$ are the Killing vector~(\ref{KillingV}) and 
rigid bosonic symmetry parameters~(\ref{BosonicPara}), 
respectively, 
and the $\mathring{a}^I$ is rigid parameter for the internal gauge group $G$ defined as
\be\label{RigidGparameter}
\mathring{a}^I\=\frac{1}{2}  \mathring{c}^{\,B}  \mathring{c}^{\,A}f_{AB}{}^{I}(\mathring{\phi})\,.
\ee

We now illustrate the simple example of an abelian vector multiplet coupled to $\cN=2$ supergravity background that 
we use in the following. 
The vector multiplet consists of a vector field $A_\mu$, a scalar $X$, two gaugini $\lambda^i$ which form an $SU(2)$ 
doublet of chiral fermions, and the auxiliary scalars $Y^{ij}$ which form an $SU(2)$ triplet. 
The algebra that is used for localization is that of a rigid supersymmetry $Q^2$ 
which squares to bosonic symmetries with field dependent parameters:
\be
Q^2 \= \cL_{\mathring{v}} + \mbox{Gauge}(a) \,,
\ee
where we have assumed that there are no other bosonic transformations on the right-hand side, just to make 
the discussion simpler.  
Here, $a$ is the $U(1)$ gauge parameter given by 
$a =-\mathring{v}^\mu A_\mu -2 \i  (\bar{\mathring{c}}_{i-}\mathring{c}_-^i X+\bar{\mathring{c}}_{i+}\mathring{c}_+^i \bar{X})$. 
Note that this includes the background value as well as fluctuation of fields. 
In order to get a rigid symmetry algebra, one introduces the ghost system $(c, b, B)$ for the $U(1)$ gauge symmetry,
and uses the combination~$\wh Q = Q +\qb$. 
In this case one has to additionally work out the transformations of~$Q$ on the ghost system demanding  
consistency of the algebra (see~\cite{Baulieu:1988xs,Pestun:2007rz,Hama:2012bg,David:2016onq} for details of this
procedure in some examples).

Our formalism above systematizes this procedure, and the transformation rules of~$\wh Q$ are precisely those 
of~$\qeq$. The transformations of the rigid supersymmetry~$Q$ correspond to the terms  
involving~$\mathring{c}$, and the other terms correspond to the BRST transformation $\qb$.\footnote{The constant gauge 
transformation parameter 
$\mathring{a}^I= \frac{1}{2}  \mathring{c}^B  \mathring{c}^A f_{AB}{}^{I}(\mathring{\phi}) $ in \eqref{RigidGparameter} 
corresponds to the parameter
$a_0$ that appears in (4.12) of~\cite{Pestun:2007rz} or (4.9) of~\cite{Hama:2012bg}. 
In the (\ref{QeqQFT}), it  naturally appears as a  part of the rigid supersymmetry $Q$ transformation of the ghost fields .  
A difference is that  since we do not consider the zero 
mode of the ghost fields, the multiplet of ghost for ghost is absent. i.e.  $ \widetilde{a}_0= \widetilde{c}_0= c_0 = b_0=0$ in~\cite{Pestun:2007rz}. } 
In this case one obtains
\begin{eqnarray}\label{Qeqghost}
&&\qeq \, c\=-\widetilde{a}\,,~~~~~~~~~~~~\qeq \, \widetilde{a}\=-\cL_{\mathring{v}}c\,,\\
&&\qeq \, b\=B\,,~~~~~~~~~~~~~~\qeq \, B\=\cL_{\mathring{v}}b\,,\nn
\end{eqnarray}
in agreement with the construction of the combined cohomology~$\wh Q$ in each case. 
The algebra closes to bosonic symmetries with field independent rigid parameters,
\be
\qeq^2\= \cL_{\mathring{v}} + \mbox{Gauge}(\mathring{a}) \,,
\ee
as can be read off directly from~(\ref{QeqQFT}).

\section{Twisted fields and algebra of $\CN=2$ conformal supergravity \label{sec:twisted}}

In this section we implement the twisting procedure described above on all the fields of the~$\CN=2$ supergravity (Weyl)
multiplet. We then classify all the twisted fields as representations of the supersymmetry 
algebra~\eqref{equiCoho}. 
This representation, 
called the \emph{cohomology complex}, is of the form $(\Phi\,, \qeq \Phi\,,\Psi\,, \qeq \Psi)$. 
Here~$\Phi$ and~$\Psi$ denote the collection of some of the bosons and fermions, respectively, of the theory 
which we shall call~\emph{elementary}. 
The rest of the bosonic and fermionic fields are in the collections $\qeq \Psi$ and~$\qeq \Phi$, respectively. 
We can think of this procedure as a change of variables 
in the (matter+ ghost) field space from the fields labelled as usual under local Lorentz indices to a set of fields 
that are paired up under the operator~$\qeq$. This change of variables will be very useful when we compute 
the functional integral using localization, as the algebra~\eqref{equiCoho} is then manifestly satisfied on these variables
removing any issues caused by gauge choices.

In order to achieve such a classification we need, firstly, an operator~$\qeq$ with a well-defined off-shell action in the 
theory. This is precisely what we achieved in the previous sections for~$\CN=2$ supergravity around any supersymmetric
background that admits a Killing spinor~$\ve^i$, we shall refer to~$\qeq$ as the supercharge from now on, and 
the transformations as supersymmetry transformations from now on. The next step is to twist the various fermionic fields, 
i.e.~construct linear combinations with the Killing spinor so as to obtain a set of fields with purely bosonic quantum numbers. 
Having done that the problem reduces to tracking the supersymmetry transformations on all the fields and classifying 
them into the four sets listed above. This classification of course only respects the superalgebra~\eqref{equiCoho} and, in particular,
the local Lorentz components of the same field can end up in different sets.

We reorganize the variables through the following procedure. We consider a local change of variables. 
We also demand that this change of variables is invertible, as otherwise the functional integration measure would 
be singular.
\begin{enumerate} 
\item We choose a particular twisting of all the spinorial fields, and make sure that it is invertible. 
The way of twisting may not be unique, but the following procedure will ensure if our choice of twisting 
is good for the cohomological classification.
\item  We start with a given component, say $\phi_R$, of a bosonic field~$\phi$ in some representation~$R$ 
of the gauge group, and consider the variation~$\qeq \phi_R$ which is clearly in the same 
representation, and may be a composite combination of bosonic fields and the twisted fermionic fields with 
coefficients consisting of bilinears of the Killing spinor~$\ve^i$. 
\item We find a term where the twisted fermionic field $\psi_R$ in the same 
representation as the boson $\phi_R$ linearly appears.
The constraints we impose are that this fermionic field should not contain derivatives---otherwise the change of 
variables will not be invertible (as the constant modes will not be present)---and that the coefficient of this term 
should be regular everywhere for the invertibility.    
If we can find such a~$\psi_R$, then 
 we classify~$\phi_R$ as an elementary bosonic variable 
in $\Phi$ and $\qeq \phi_R$ in $\qeq \Phi$.
We may exclude the $\psi_R$ from the set~$\Psi$ of elementary fermionic variables.
\item In the same way, we find the fermionic variables in $\Psi$ and the corresponding bosonic 
variables in $\qeq \Psi$. 
\item Keep the process going until all the variables are classified. If we fail, then we reconsider the other way of twisting.
\end{enumerate}
This procedure yields a consistent set of twisted variables which smoothly fall into representations of~$\qeq$.  
The nature of the change of variables from the original quantum field variables to the cohomological variables is of the 
form linear transformation + non-linear transformation. Here the coefficients of the linear term always include the 
background spinors, while the non-linear terms can be thought of as fluctuations. Thus, at least for small fluctuations,
the Jacobian is a constant. 
This is one of the big advantages of the background field method, and is an important difference with the discussion 
in~\cite{Baulieu:2012jj,Bae:2015eoa,Imbimbo:2018duh}. 
In the applications that follow, we assume that this is the case in the full transformation and 
there is no Gribov-type singularity.

We now use these ideas to classify the cohomology complex of~$\CN=2$ supergravity. 
We begin by reviewing the simpler and known case of the vector multiplet fields~\cite{Baulieu:1988xs} 
to set up the formalism and notations. Our conventions for spinors and gamma matrices are presented in Appendix~\ref{Gamma}.

\subsection{Vector multiplet}
The~$\CN=2$ vector multiplet~$\bigl(A_{\mu}\,,  X\,, \bar{X}\,, \l^{i}\,, Y^{ij}  \bigr)$ consists of a vector field~$A_\mu$,
two  scalars~$X$ and $\bar{X}$, two gaugini~$\l^i$ which form an $SU(2)$ doublet of chiral fermions, and the 
auxiliary scalars~$Y^{ij}$ which form an $SU(2)$ triplet. 
The vector field~$A_\mu$ is a gauge field for~$U(1)$ gauge symmetry, and correspondingly we introduce 
the ghost system~$(b,c,B)$. The ghost~$b$ and anti-ghost~$c$ are fermionic, and the Lagrange multiplier~$B$ is bosonic.

First we write the spinorial gaugini fields~$\l^i$ 
in terms of bosonic variables by projecting against the 
fixed Killing spinors~$\ve^i$ using ($\gamma_5\ve^i\,,\gamma^{\mu}{\ve}^i\,, \epsilon_{ij}\ve^j)$ 
as a basis. The resulting twisted variables are
\be 
\label{Cohovariablegauge}
\lambda \= \bar\ve_{i}\gamma_5\lambda^{i}\,,\qquad 
\lambda_{\mu} \= {\bar\ve}_{i}\,\gamma_{\mu}\,\lambda^{i}\,,\qquad 
\lambda^{ij} \= -2\,\ve^{(i} \,C\lambda^{j)}\,, 
\ee
where the matrix~$C$ is the charge conjugation matrix, and the inverse relation is
\be
\label{inverserelationgauge}
\lambda^{i} \=  (\bar\ve_{j}\,\ve^{j})^{-1}
\bigl(\gamma_5\ve^{i} \lambda + \gamma^{\mu}{\ve}^{i}\,\lambda_{\mu} + \epsilon_{jk}\,\ve^{k}\,\lambda^{ij} \bigr) \,.
\ee
Here we assume that the coefficient $\bar\ve_{j}\,\ve^{j}$ is non-singular, which ensures the invertibility of the twisting, 
so that  the $8$ gaugini degrees of freedom are 
now encoded in the bosonic coefficients $(\lambda\,,\lambda_{\mu}\,,\lambda^{ij})$.
Here, we could also choose a different twist  using  ($\ve^i\,,\gamma^{\mu}\gamma_5{\ve}^i\,,
\epsilon_{ij}\gamma_5\ve^j)$, but we shall not work it out. As we will see that the choice \eqref{inverserelationgauge} 
reads to a consistent cohomological classification.  

We start with the gauge field~$A_\mu$. The variation of the quantum fluctuation
of~$A_\mu$ is
\be 
\qeq \, \widetilde{A}_\mu \= \l_\mu +\partial_\mu c\,, 
\ee
Here the twisted variable $\l_\mu$ appears without derivative, and 
so~$A_\mu$ belongs to~$\Phi$ and~$\l_\mu$ is excluded from~$\Psi$. 
Indeed, the variation of $\l_\mu$ 
\be
\qeq \l_{\mu} \= \CL_{v} \, \widetilde{A}_\mu + \partial_{\mu}a
\ee
with 
\be \label{defa}
a\; := -v^\mu {A}_\mu -\i {X}_1 ({\bar\ve}_i{\ve}^i ) -{X}_2({\bar\varepsilon}_i\gamma_5{\varepsilon}^i) \,,
\ee
does not contains any term without derivatives of other bosonic variables. Here we let~$X_1 := X+\bar{X}$ and~$X_2:= -\i (X-\bar{X})$. 
Next, we consider the variation of the quantum fluctuation of~$X_2$:
\be
\qeq \, \widetilde{X}_2 \= \l \,.
\ee
It does not contain any derivatives, and thus the variable~$\widetilde{X}_2$ belongs to~$\Psi$ and~$\l$ 
is excluded from $\Psi$\footnote{If we started from $\widetilde{X}_1$, then we would get $\qeq \widetilde{X}_1 = \i (\bar\ve_j \ve^j)^{-1}\bigl(\bar\ve_i \gamma_5\ve^i \l  + \bar\ve_i \gamma^\mu \ve^i \l_\mu \bigr)$ which has a singular coefficient $\bar\ve_i \gamma_5\ve^i$, and thus the change of variable $(\l\,,\l_\mu \,, \l^{ij})\rightarrow (\qeq \widetilde{X}_1\,, \qeq \widetilde{A}_\mu\,, \l^{ij})$ would be singular. }. The remaining twisted gaugino field~$\l^{ij}$ varies into
\bea \label{Qlambdaij}
&&\qeq \, \lambda^{ij} \= \; \bar\ve_{k}\ve^{k}\,Y^{ij}+ 2 {\ve}^{(i}C\gamma^{\mu}\ve^{j)}\partial_{\mu}X_2 
\\
&&~~~~~~~~~
+\ve_+^{(i} C\gamma^{ab}\ve_+^{j)}\left[F^{-}_{ab}-\frac{1}{8} ( X_1- \i X_2)T^-_{ab}\right]+{\ve}_-^{(i} 
C\gamma^{ab}{\ve}_-^{j)}\left[F^{+}_{ab}-\frac{1}{8}  (X_1 +\i X_2) {T}^+_{ab}\right]\,.\nonumber
\eea
Since the auxiliary field~$Y^{ij}$ appears without derivative and with a regular coefficient $\bar\ve_k\ve^k$.
the field~$\l^{ij}$ belongs to~$\Psi$ and $Y^{ij}$ does not belong to~$\Phi$. 

The bosonic variable~$X_1$ is not yet classified. For this we look at the variation of ghost fields which were already 
presented in (\ref{Qeqghost}):
\begin{eqnarray}
&&\qeq c\=-\widetilde{a}\,, \qquad  \qquad \qeq \widetilde{a}\=-\cL_{\mathring{v}}c\,, \nn\\
&&\qeq b\=B\,, \qquad  \qquad \; \qeq B\=\cL_{\mathring{v}}b\,. \nn
\end{eqnarray}
From the expression~\eqref{defa}, we see that the field~$\widetilde{a}$ includes the field~$X_1$ without derivative 
and with non-singular coefficient~$\bar\ve_i \ve^i$. Thus $c$ belongs to $\Psi$ and it is natural that $X_1$ is not part of $\Phi$. 
Finally, the classification of the anti-ghost and the auxiliary field is trivial. The $b$ varies into $B$ with no derivative. 
Thus $b$ belongs to $\Psi$ and $B$ is not in $\Phi$.
From these transformation rules, we see that the cohomological variables for the bosons and fermions variables are 
organized as in Table~\ref{VectorMultcharges}.
{\begin{table}[h] 
\begin{center}
\begin{tabular}{|c|c|}
\hline
$\Phi$ & $\Psi$ \\
\hline
~$\widetilde{A}_\mu\,, \widetilde{X}_{2}~$&~ $\l^{ij}\,, b\,,c$   ~\\
\hline
\end{tabular}
\caption{The elementary variables (5 bosons and the 5 fermions) in the cohomological representation of 
the vector multiplet fields, including the ghost system. The rest of the fields are in $\qeq$ variations of the elementary variables.} 
\label{VectorMultcharges}
\end{center}
\end{table}}
This above discussion was simply a review of known results~\cite{Baulieu:1988xs,Pestun:2007rz,Hama:2012bg}, 
which we went through in order to explain our systematics. Our real interest is of course in the Weyl multiplet, to 
which we turn now in order to achieve a similar classification using these ideas.

\subsection{Weyl multiplet \label{sec:Weyl}}

The independent physical fields of the Weyl multiplet consist of~$24+24$ independent degrees of freedom,  
as reviewed in Appendix~\ref{ConfGravity}. We collect them in Table~\ref{WeylFields}. 
Here we are interested in the off-shell counting of the degrees of freedom 
in a covariant manner, i.e.~without taking into account the redundancies due to gauge transformations. 
This gives a count of 
a total of~43 bosonic and~40 fermionic degrees of freedom. 
This mismatch, as we have discussed, is due to the gauge symmetries not commuting with supersymmetry, and 
it will be cured by the addition of ghosts. 
Thus we introduce, for each of the local symmetries, a ghost system consisting of ghosts~$c$, anti-ghosts~$b$, 
and Lagrange multiplier~$B$. The ghost~$c$ and anti-ghost~$b$ for bosonic (fermionic) gauge symmetries 
are fermionic (bosonic), and the Lagrange multiplier~$B$ is bosonic (fermionic).
These are presented in Table~\ref{WeylGhosts}. 
Together, the matter and ghost fields of the Weyl multiplet consist of $94+94$ degrees of freedom.

{\begin{table}[h] 
\begin{center}
\begin{tabular}{|c|c|c|}
\hline
Local symmetry &Gauge fields & Degrees of freedom \\
\hline
g.c.t &  $e_{\mu}^{a}$  & 16B \\
Dilatation $D$&  $A^D_{\mu}$  & 4B \\
Sp. conf. $K^a$&  $f_{\mu}^{a}$ & composite \\
Lorentz $M_{ab}$&  $\w_{\mu}^{ab}$ & composite \\
$SO(1,1)_R$&  $A^R_{\mu}$  & 4B\\
$SU(2)_R$&  $\CV_{\mu \, j}^{\, i}$ & 12B \\
\hline
$Q$-susy &  $\psi_{\mu}^{\,i}{}$ & 32F\\
$S$-susy &  $\phi_{\mu}^{\,i}{}$ & composite \\
\hline
\hline
~ & Auxiliary fields & Degrees of freedom \\
\hline
&  $ 
T^{\pm}_{ab}$  & 6B \\
&  $D$ & 1B \\
&  $\chi^{\,i}$  & 8F \\
\hline
\end{tabular}
\caption{The 43 bosonic(B) and 40 fermionic(F) matter fields of the Weyl multiplet.} 
\label{WeylFields}
\end{center}
\end{table}}
%

{\begin{table}[h] 
\begin{center}
$\begin{tabular}{|c|c|c|}
\hline
Local symmetry &Ghosts & Degrees of freedom \\
\hline
g.c.t &  $(c^\mu,b_\mu,B_\mu)$  & 8F \quad 4B \\
Dilatation $D$&  $(c_D,b_D,B_D)$  & 2F \quad 1B \\
Sp. conf. $K^a$&  $(c_K^{~a}\,,b_K^{~a}\,,B_K^{~a})$ & 8F \quad 4B \\
Lorentz $M_{ab}$&  $(c_M^{ab}\,,b_M^{ab}\,,B_M^{ab})$ & 12F \quad 6B\\
$SO(1,1)_R$&  $(c_R,b_R,B_R)$  & 2F \quad 1B\\
$SU(2)_R$&  $(c_R^{\;i}{}_{\,j}\,,b_R^{\;i}{}_{\,j}\,,B_R^{\;i}{}_{\,j})$ & 6F \quad 3B \\
\hline
$Q$-susy &  $(c_Q^{\,i}{}\,,b_Q^{\,i}{}\,,B_Q^{\,i}{})$ & 16B \quad 8F\\
$S$-susy &  $(c_S^{\,i}{}\,,b_S^{\,i}{}\,, B_S^{\,i}{})$& 16B \quad 8F  \\
\hline
\end{tabular}$
\caption{The 51 bosonic(B) and~54 fermionic(F) ghosts of the Weyl multiplet.} 
\label{WeylGhosts}
\end{center}
\end{table}}

We now present the details of the twisting and the representation of these fields as pairs under the supercharge~$\qeq$.
The twisted fermionic variables for~$\psi_\mu^i\,, \chi^i\,, c_Q^i\,,c_S^i$ are
\bea  \label{WeylFieldsInvTwist}
\psi_{\mu} & \= & \bar\ve_{i}\gamma_5\psi_\mu^{\,i} \,, \qquad \;
\psi_{\mu}{}^{a} \=  {\bar\ve}_{i}\gamma^{a}\psi^{\,i}_{\mu}\,, \qquad \; \; \psi_{\mu}^{ij} \= - 2 \ve^{(i}C\psi_{\mu}^{j)}\,,\\
\chi & \= & \bar\ve_{i}\chi^{i} \,,\qquad \qquad  \chi^{a} \=  {\bar\ve}_{i}\gamma_5\gamma^{a}\chi^i\,,\qquad  
\chi^{ij} \= - 2 \ve^{(i}C\gamma_5\chi^{j)}\,,\\
c_S & \= & \bar\ve_{i}\gamma_5 c_S^{\;i} \,,\qquad \quad c_S^{~a} \=  {\bar\ve}_{i}\gamma^{a}c_{S}^{\;i}\,,\qquad \quad 
c_S^{ij} \= - 2 \ve^{(i}C c_S^{\;j)}\,,\\
c_Q & \= & \bar\ve_{i}\gamma_5 c_Q^{\;i} \,,\qquad \; c_Q^{~a} \=  {\bar\ve}_{i}\gamma^{a} c^{\;i}_{Q}\,,\qquad \quad 
c_Q^{ij} \= - 2 \ve^{(i}Cc_Q^{\;j)}\,,
\eea
with inverse relations:
\bea \label{WeylFieldsTwist}
\psi_{\mu}^{\,i} &\= &(\bar\ve_i\ve^i )^{-1}\bigl(\psi_{\mu}\gamma_5\ve^i +\psi_{\mu}^{~a}\gamma_{a}{\ve}^i + \psi_{\mu}^{ij}\,\epsilon_{jk}\ve^k\bigr)\,,\\
\chi^i &=&(\bar\ve_i\ve^i )^{-1}\bigl(\chi\ve^i  + \chi^a\gamma_a \gamma_5\ve^i  +\chi^{ij}\,\epsilon_{jk}\gamma_5\ve^k \bigr)\,,  \\
c_S^{\;i} &=&(\bar\ve_i\ve^i )^{-1}\bigl(c_S \gamma_5\ve^i +c_S^{~a}  \gamma_a \ve^i +c_S^{ij}\,\e_{jk}\ve^k\big)\,,\\
c_Q^{\;i} &=&(\bar\ve_i\ve^i )^{-1}\bigl( c_Q  \gamma_5\ve^i +c_Q^{~a}  \gamma_a \ve^i +c_Q^{ij}\,\e_{jk}\ve^k\big)\,.
\eea
The spinorial anti-ghosts~$b_Q$, $b_S$ and Lagrange multipliers~$B_Q$, $B_S$ corresponding to the fermionic transformations 
can be twisted in the same way as the ghosts. 

The classification of the cohomological variables of the Weyl multiplet is a little more involved than those of the 
vector multiplet, but follows exactly the same general principles. 
We recall the general definition of $\qeq$ given in \eqref{ModifieBRST2}. The details can be read off from the algebra,
which we present in Appendices~\ref{superconfalg} and \ref{weylmultiplet}. 
As in the previous subsection we focus on terms that are linear in the fields with no derivatives and with 
non-singular coefficient. This allows us to go through the whole multiplet and classify the various fields. In this 
discussion below, we use ellipses to denote other terms that appear in the variations. 
Once we finish the full classification, we present all the detailed fields variations.

We begin with the defining field of the Weyl multiplet, namely the vielbein~$e_\mu{}^a$. The variation is
\bea
&&\qeq \,  \widetilde{e}_{\mu}^{~a}\=\psi_{\mu}^{~a} +\cdots 
\eea
Since the gravitino twisted variable $\psi_\mu{}^a$ appears linearly without derivative, we classify $\widetilde{e}_\mu{}^a$ 
into $\Phi$ and exclude $\psi_\mu{}^a$  from $\Psi$. Now consider the other gravitino twisted variables:
\bea\label{Qpsimu}
\qeq \, \psi_\mu &\=& -  c_{Sa}\mathring{e}_{\mu}^a+\widetilde{A}^R_{\mu}\bar\ve_i\ve^i  
+\cdots\,,\\
\qeq \, \psi_\mu^{ij} &\=& \widetilde{\cV}_\mu{}^{(i}{}_{k}\e^{j)k}\, \overline{\ve}_l\ve^l+\cdots\,.
\eea
As the right-hand sides contain pure bosonic variables in the same representation, 
we classify~$\psi_\mu$ and~$\psi_\mu^{ij}$ into $\Psi$ and exclude~$\widetilde\cV_{\mu}{}^i{}_{j}$ from $\Phi$. 
In the first equation above, it is not immediately clear which one among the $c_{S a}$ and $A^R_\mu$ 
can be excluded from $\Phi$. However, we can exclude $c_{S a}$ from $\Phi$ by observing that the 
variation of $\widetilde{A}^R_\mu$ gives
\bea
&&\qeq \, \widetilde{A}^R_\mu \= - \chi_a \mathring{e}_{\mu}{}^a +\cdots\,.
\eea
Thus $\widetilde{A}^R_\mu$ belongs to $\Phi$ and $\chi^a$ 
can be excluded from $\Psi$. 

Now consider the other auxiliary fermion twisted variables $\chi$ and $\chi^{ij}$.  
Since variation of the $\chi$ gives the auxiliary scalar $\widetilde{D}$ as
\bea
&&\qeq \, \chi \=\widetilde{ D}\,\bar{\ve}_i \ve^i+\cdots\,,
\eea
we put $\chi$ into $\Psi$ and exclude $\widetilde{D}$ from $\Phi$. We can find the $\chi^{ij}$  from the variation of the 
tensor field $\widetilde{T}_{ab}^{+}$ or $\widetilde{T}_{ab}^-$ as
\be \label{QT}
\qeq \, \widetilde{T}_{ab}^{\pm} \= 4\i \, (\bar\ve_i\ve^i)^{-1} \,  \ve_{\mp}^{(i}C \, \gamma_{ab} \, \gamma_5 \, \ve_{\mp}^{j)} \,\chi_{ij} \,+\cdots\,,
\ee
where $\chi_{ij}:= \e_{ik}\e_{jl} \, \chi^{kl}$. At present the mapping looks nontrivial because 
the fields have different representation under the local Lorentz and R-symmetry group 
$SU(2)_{+} \times SU(2)_{-} \times SU(2)_R$ :  
while the $\widetilde{T}_{ab}^{+}$ and $\widetilde{T}_{ab}^-$ have representation $(1,3,1)$ and $(3,1,1)$,  
the $\chi_{ij}$ has $(1,1,3) $. The right hand side of~\eqref{QT} provides the twisting procedure 
such that the representation of~$\chi_{ij}$ is converted to the representation of the correct combination of~$\widetilde{T}_{ab}^-$ 
and~$\widetilde{T}_{ab}^+$ depending on the point of the manifold. At the point where $\ve_-^i =0$  the $\widetilde{T}_{ab}^-$ 
maps to~$\chi_{ij}$ , and at the point where $\ve_+^i=0$ the $\widetilde{T}_{ab}^{+}$  maps the~$\chi_{ij}$. 
Therefore one of the~$\widetilde{T}_{ab}^{\pm}$ belongs to $\Phi$ and the $\chi^{ij}$ is excluded from $\Psi$. 
The other one of the $\widetilde{T}_{ab}^{\pm}$ can be found from the variation of the ghost for Lorentz symmetry as
\be
\qeq \, c^{ab} \= -\frac{1}{4}\i\bigl( \bar\ve_{i+}\ve_+^i \widetilde{T}^{ab +} + \bar\ve_{i-}\ve_-^i \widetilde{T}^{ab-}\bigr) 
-(\bar\ve_i \ve^i)^{-1}\,\ve^{(i}C\gamma^{ab}\gamma_5\ve^{j)}\,c_{S ij}+ \cdots\,.
\ee
This variation in fact includes the ghost for the $S$ symmetry $c_S^{ij}$ as well. 
Again, by the twisting procedure,  at the point where $\ve_-^i =0$  the $c^{ab}$ maps to $\widetilde{T}_{ab}^{+}$ 
and $c_S^{ij}$, and at the point where $\ve_+^i=0$ the $c^{ab}$ maps to $\widetilde{T}_{ab}^{-}$ and $c_S^{ij}$. 
Thus the $c^{ab}$ belongs to $\Psi$ and the other one of $\widetilde{T}^{\pm}_{ab}$  and $c_S^{ij}$ can be 
excluded from $\Phi$. 

Consider the yet unclassified twisted variable for the $S$ supersymmetry ghost 
field~$c_S$. From the $\{Q,S\}$ algebra \eqref{algebra2}, we find
\be
\qeq \, c_D \= c_S+\cdots \,.
\ee
Thus $c_D$ belongs to $\Psi$ and $c_S$ can be excluded from $\Phi$. Now from the $[A^R, Q]$ 
and $[V^R_\Lambda, Q]$ algebra, we read off
\bea
\qeq \, c_Q & \= & -\frac{1}{2} c_R \,\overline{\ve}_i\ve^i+\cdots\,,\\
\qeq \, c_Q^{\;ij} & \= & c_{R}^{\;ij} \, \overline{\ve}_k\ve^k+\cdots\,,
\eea
where $c_R^{\;ij}:= c_R^{\;(i}{}_{k}\e^{j)k}$. Thus the ghost for supersymmetry $c_Q$ 
and $c_Q^{\;ij}$ belong to $\Phi$ and $c_R$ and $c_R^{\;ij}$ can be excluded from $\Psi$. 
The rest of the supersymmetry ghost $c_Q^{\;a}$ can be found from the variation of the translation ghost $c^{\mu}$,
\be
\qeq \, c^{\mu} \= -2\,c_Q^{\,a} \,\mathring{e}_{a}{}^{\mu}+\cdots\,.
\ee
Thus the $c^\mu$ is in $\Psi$ and $c_Q^{\;a}$ is excluded from $\Phi$.

Finally consider the transformation of $\widetilde{A}^D_\mu$: 
\be
 \qeq \, \widetilde{A}^D_\mu \= c_{Ka}\mathring{e}^a_{\mu}+\cdots\,.
\ee
Thus we finish the classification of matter and ghost fields by putting the $\widetilde{A}_\mu^D$ into $\Phi$ 
and excluding $c_K^{\;a}$ from~$\Psi$. 

The classification of all the anti-ghosts $b$ and the Lagrange multiplier fields $B$ is straightforward since they 
form a closed multiplet under the $\qeq$ by themselves. The algebra takes the form
\be
\qeq b \= B\,, \qquad \qeq B \= \cL_{\mathring{v}} b +\cdots\,,
\ee
where the rest of the terms in the variation of~$B$ are given in~\eqref{ModifieBRST2}. 
Since $B$ linearly appears without derivative on the right-hand side, we classify the anti-ghost~$b$ in~$\Phi$ (or~$\Psi$) 
and the Lagrange multiplier~$B$ into $\qeq\Phi$ (or $\qeq\Psi$) when~$b$ is bosonic (or fermionic).

The final classification of all the Weyl multiplet fields is presented in Table~\ref{WeylPairing1}.

{\begin{table}[h] 
\begin{center}
$\begin{tabular}{|c|c|c|}
\hline
$\Phi$ & $\Psi$  \\
\hline
$\widetilde{e}_{\mu}^{\,a}\,, \widetilde{A}^R_{\mu}\,,\widetilde{A}^D_\mu\,,$ $\widetilde{T}^{+/-}_{ab}$& ~$\psi_\mu\,,\psi_{\mu}^{\;ij}\,,\chi\,,$~ \\
$c_Q$\,, $c_Q^{ij}$\,, & $ c^{\mu}\,, c_M^{\,ab}\,,c_D\,,$   \\
$b_{Q}$\,, $b_{Q}{}_a$\,, $b_{Q}^{\;ij}$\,, & $ b_\mu\,,b_{M}^{\;ab}\,,b_{D}\,,$   \\
$b_{S}$\,, $b_{S}{}_a$\,, $b_{S}^{\;ij}$&  $b_{K}^{\;a}\,,b_{R}\,,b_{R}^{\;i}{}_{j}$  \\
\hline
\end{tabular}$
\caption{The elementary variables (47 bosons + 47 fermions)  in the cohomological representation of the 
Weyl multiplet fields and ghosts. Here $\widetilde{T}^{+/-}_{ab}$ refers appropriate combination of $\widetilde{T}^{+}_{ab}$ and $\widetilde{T}^{-}_{ab}$ depending on the spacetime point as explained around \eqref{QT}. } 
\label{WeylPairing1}
\end{center}
\end{table}}

Now we turn to the full transformation rules. It will be useful to define the following field-dependent parameters:
\bea
v^\mu &\=& v^a e_{a}{}^\mu= \bar\ve_{ i} \gamma^a \ve^i e_a{}^{\mu}\,,\nn \\
\ve^{ab} &\=& -v^\mu \omega_\mu^{ab}+\i\frac{1}{4}\bar\ve_{i+ }\ve_{+}^{i}T^{ab+}+\i\frac{1}{4}\bar\ve_{i- }\ve_{-}^{i}T^{ab-} 
+\bar\ve_{i }\gamma^{ab}\gamma_5\eta^{i}\,,\nn \\
\ve_D &\=&  -v^\mu A_{\mu}^D - \bar\ve_{i }\gamma_{5}\eta^{i}\,,\\
\ve_K^{~a} &\=&  -v^\mu f_\mu^a -\i \frac{1}{4}\bar\ve_{i+ }\ve_{+}^{~i}D_{b}T^{ab+}-\i\frac{1}{4}\bar\ve_{i- }\ve_{-}^{~i}D_bT^{ab-}
-\frac{3}{4}v^a D +\frac{1}{2} \bar\eta_{i }\gamma^{a}\eta^{i}\,, \nn \\
\ve_Q^{~i} &\=& -\frac{1}{2}v^\mu \psi_\mu^i \,, \qquad \qquad
\ve_S^{~i} \=  -\frac{1}{2}v^\mu \phi_\mu^{i}+\frac{3}{2}\bar\ve_{j+} \ve_+^j  \chi_{-}^i-\frac{3}{2}\bar\ve_{j-} \ve_-^j  \chi_{+}^i\,, \nn\\
\ve_R  &\=&  -v^\mu A_\mu^R + \bar\ve_i \eta^i\,,\nn 
\\
\ve_R{}^i{}_{j}  &\=&  \frac{1}{2}v^\mu \cV_\mu{}^i{}_{j}+ 2 \bar\ve_j \gamma_5 \eta^i -\delta^i_j \bar\ve_k \gamma_5 \eta^k\,. \nn
\eea
These bilinears are all constructed out of the background values of the Killing spinors for supersymmetry~$\ve^{i}$ and 
conformal supersymmetry~$\eta^{i}$. The field-dependence of the gauge algebra occurs because of the field-dependence 
of these bilinears. We see here that, indeed, it only occurs in the anticommutators of two fermionic transformations,
consistent with the discussion in Section~\ref{sec:sugra}.

Our goal now is to write down the transformation rules for all the (elementary) variables of the theory which we classified above. 
The transformations should be expressed in terms of the twisted variables which we also defined above. It turns out that it is 
easier to actually write out the equivalent transformation rules in terms of the original variables for a couple of reasons.
Firstly, some of the equations involve derivatives, and the derivative operation does not commute with the twisting as our Killing 
spinors are not constant. This of course can be overcome if we express everything in terms of covariant derivatives---which 
do kill the Killing spinors---and write out the non-covariant terms with connections. The second (and real) reason we use 
the original fields stems from the non-linearity of supergravity. Almost all terms on the right-hand side of the gauge 
variations involve at least bilinears of fields, if not higher powers. To rewrite the bilinears we have to insert a 
spinorial basis that we used to twist, which is 8 dimensional. Doing so makes the equations much longer. 
Therefore we present all the equations in terms of the original field variables. In any discussion of the twisted theory, 
one should use the twisted variables presented in~\eqref{WeylFieldsTwist}, \eqref{WeylFieldsInvTwist}, which is a 
linear transformation of the original fields. 

We write some of the transformation rules below to illustrate their form, and record the full list in Appendix~\ref{sec:transformations}. 
We begin with the transformation rules of the $(b, B)$ ghost fields for general coordinate transformations, Lorentz transformations, and supersymmetry:
\be
\begin{array}{ll}
\qeq b_\mu \= B_\mu \,, ~~~~&\qeq B_\mu \=\cL_{\mathring{v}}b_\mu +\partial_\mu \mathring{\ve}^{ab}b_{ab}
+\partial_\mu \mathring{\ve}_{D}b_{D}+\partial_\mu \mathring{\ve}_K^{a}b_{Ka}+\partial_\mu \mathring{\ve}_R b_R
+\partial_\mu \mathring{\ve}_R^{ij}b_{Rij} \,,\\
\qeq b_{ab}\= B_{ab}\,,& \qeq B_{ab}\=\cL_{\mathring{v}}b_{ab}+\mathring{\ve}_{a}{}^{c}b_{cb}
+\mathring{\ve}_{b}{}^{c}b_{ac}+\mathring{\ve}_{K[b}b_{K a]}\,, \\
\qeq b_Q^{~i}\= B_Q^{~i}\,,& \qeq B_Q^{~i}\= \cL_{\mathring{v}}b_{Q}^{~i}+ \frac{1}{4}\mathring{\ve}^{ab}\gamma_{ab}b_{Q}^{~i} 
+\mathring{\ve}^i{}_{j}b_{Q}^{~j}+\frac{1}{2}\mathring{\ve}_D b_{Q}^{~i}+\frac{1}{2}\mathring{\ve}\gamma_5 b_{Q}^{~i}
+ \mathring{\ve}_K^{~a}\gamma_a\gamma_5 b_{S}^{~i}\,.
\end{array}\ee
The rest of the~$(b, B)$ transformations follow a similar pattern and are presented in \eqref{bBtransFull}.
The transformation rules of $c$ ghost fields for the same transformations 
are as follows (here $\widetilde{e}_a^{~\mu}:= e_a^{~\mu}- \mathring{e}_{a}^{~\mu}$):
\be 
\begin{array}{ll}
\qeq c^\mu &\= - 2\, \bar{\ve}_i \gamma^a c_Q^i \,\mathring{e}_{a}^{~\mu } - 2\, \bar{\ve}_i \gamma^a c_Q^i \,\widetilde{e}_{a}^{~\mu }  
- \bar\ve_i \gamma^a \ve^i \widetilde{e}_a^{~\mu}+c^\nu\partial_\nu c^\mu- \bar{c}_{Qi}\gamma^a c_{Q}e_{a}^{\,\mu}\,,\\
 \qeq c^{ab}&\= -\i \frac{1}{4}{\bar\ve}_{i+ }\ve^{i}_{+ }\widetilde{T}^{ab+}- \i \frac{1}{4}{\bar\ve}_{i- }\ve^{i}_{-}\widetilde{T}^{ab-}
- {\bar\ve}_{i }\gamma^{ab}\gamma_5 c_S^i-\overline{{c}_{Q}}_{ i }\gamma^{ab}\gamma_5 \eta^i- \overline{{c}_{Q}}_{ i }\gamma^{ab}\gamma_5 c_S^i
\\
&\qquad  -\i \frac{1}{2}{\bar\ve}_{i+ }c_{Q+}^{i}T^{ab+}- \i \frac{1}{2}{\bar\ve}_{i- }c_{Q-}^{i}T^{ab-}
   -\i \frac{1}{4}\overline{c_Q}_{i+ }c^{i}_{Q+ }{T}^{ab+}- \i \frac{1}{4}{\overline{c_Q}}_{i- }c^{i}_{Q-}{T}^{ab-}
\\
&\qquad +c^\mu \partial_\mu c^{ab} + ({\bar\ve+\overline{{c}_Q}})_i \gamma^\mu (\ve+c_Q)^i \omega_\mu^{ab}- \bar\ve_i \gamma^c \ve^i \mathring{e}_{c}^{~\mu}\mathring{\omega}_\mu^{ab}+c^{ac}c_{c}{}^{b} \,,\\
\qeq c_Q^i 
& \=- \frac{1}{2}c_R\gamma_5 \ve^i+c_R^{~i}{}_{j}\,\ve^j -\frac{1}{2}c_D \,\ve^i  +\frac{1}{4}c^{ab}\gamma_{ab}(\ve+c_Q)^i  \\
&\qquad +c^\mu \partial_\mu(\ve+ c_Q)^i + \frac{1}{2}(\bar\ve+ \bar{c}_Q)_j \gamma^\mu (\ve +c_Q)^j \psi_\mu^i
- \frac{1}{2}c_R\gamma_5 c_Q^i  +c_R^{~i}{}_{j}c_{Q}^j -\frac{1}{2}c_D c_Q^i \,.
\end{array}
\ee

The rest of the $c$ ghost transformations are presented in \eqref{cCtransFull}. 
The transformation rules for the vielbein and gravitini are 
\be
\begin{array}{ll}
\qeq  \widetilde{e}_{\mu}^{~a}
	&\= \bar\ve_i \gamma^a \psi_\mu^{\;i}+ c^\nu \partial_\nu e_{\mu}^a+\partial_\mu c^\nu e_{\nu}^a+ c^{ab}e_{\mu b} 
	-c_D e_\mu^a +\overline{{c}_Q}_{i} \gamma^a \psi_\mu^{~i}\,, \\
\qeq \psi_\mu^{~i}&\= 2\cD_{\mu}(\ve+c_{Q})^i + c^\nu \partial_\nu \psi_\mu^{~i} +\partial_\mu c^\nu \psi_\nu^{~i} + \frac{1}{4}c^{ab}
	\gamma_{ab}\psi_{\mu}^{~i}-\frac{1}{2}c_D \psi_{\mu}^{~i}-\frac{1}{2}c_R \gamma_5 \psi_{\mu}^{~i}\\
& \qquad +\;c^i{}_{j}\psi_{\mu}^{~i}+\i \frac{1}{16}  T^{ab} \gamma_{ab}\gamma_\mu (\ve+c_Q)^i 
	+\gamma_\mu \gamma_5 (\eta+c_S)^i\,,\\
&\= 2\wt \cD_{\mu}\ve^i+\mathring{\gamma_\mu} \gamma_5 c_S^i+\i \frac{1}{16}  \gamma_{ab}(T^{ab}\gamma_\mu - \mathring{T^{ab}} \mathring{\gamma_\mu}) \ve^i+
2\cD_{\mu}c_{Q}^i + c^\nu \partial_\nu \psi_\mu^{~i} +\partial_\mu c^\nu \psi_\nu^{~i}  \\
& \qquad+ \frac{1}{4}c^{ab}
	\gamma_{ab}\psi_{\mu}^{~i} -\frac{1}{2}c_D \psi_{\mu}^{~i}-\frac{1}{2}c_R \gamma_5 \psi_{\mu}^{~i}+\;c^i{}_{j}\psi_{\mu}^{~i} +
 \i \frac{1}{16}  \gamma_{ab} T^{ab}  \gamma_\mu c_Q^i 
	+\wt\gamma_\mu \gamma_5 c_S^i + \wt \gamma_\mu \gamma_5 \eta^{i} \,,  
\end{array}
\ee
where the covariant derivative~$\cD_\mu\ve^i$ is
\begin{eqnarray} 
&&\cD_\mu \varepsilon^i=(\partial_{\mu}- \frac{1}{4}\omega_{\mu ab}\gamma^{ab} +\frac{1}{2}A^D_\mu  + \frac{1}{2}A^R_\mu \gamma_5)\varepsilon^i +\frac{1}{2}\cV_\mu{}^i{}_j \varepsilon^j\,.
\end{eqnarray}
In the variation of the gravitini, we have defined~$T^{ab}=T^{ab+}+T^{ab-}$, the fluctuation of the covariant 
derivative~$\wt \cD_\mu\equiv\cD_\mu  - \mathring{\cD}_\mu$, the fluctuation of the gamma matrix~$\wt \gamma_{\mu} = \gamma_{a} \, \wt e_{\mu}^{~a}$, 
and used the fact that the~$(\ve^i, \eta^i)$ obey the background Killing spinor equation.
The rest of the transformation rules for the Weyl multiplet fields are presented in \eqref{WeyltransFull}.

We end this section with a couple of comments. Firstly, 
as an illustration of our discussion about why we use the original variables, we can look at the variation of the 
vielbein in terms of twisted variables:
\bea
&&\qeq  \widetilde{e}_{\mu}{}^{a}\=\psi_{\mu}{}^{a} + \cL_{c^\nu} e_{\mu}^{\:a}+ c^{ab}e_{\mu b} -c_D e_\mu^a \\
&&~~~~~~~~~~~~+ (\bar\ve_i\ve^i)^{-2}\bigl[-(\bar\ve_i \gamma^a \ve^i) c_{Q}\psi_\mu 
+(\bar\ve_i \gamma_5 \ve^i)  c_{Q }\psi_{\mu}^{~a}+(\ve^{i} C\gamma_5 \gamma^a \ve^{j}) c_{Q}\psi_{\mu ij}\bigr.\nonumber\\
&&~~~~~~~~~~~~~~~~~~~~~~~~~+(\bar\ve_i \gamma_5 \ve^i)c_Q^{\,a}\psi_\mu +(\bar\ve_i \gamma_b \ve^i)c_Q^{\,a}\psi_\mu^{\,b} 
+2(\bar\ve_i \gamma_b \ve^i)c_Q^{\; [b}\psi_{\mu}^{\, a]}-(\ve^i C\gamma^{a}{}_{b}\ve^j) c_Q^{\; b}\psi_{\mu ij}\nonumber\\
&&~~~~~~~~~~~~~~~~~~~~~~~~~ -(\ve^i C\gamma^a\gamma_5 \ve^j)c_{Q ij} \psi_\mu 
-(\ve^i C\gamma^a{}_{b} \ve^j)c_{Q ij}\psi_{\mu}^{\,b}+{\textstyle \half} (\bar\ve_k \gamma^a \ve^k)c_{Q ij}\psi_{\mu}^{\,ij}\bigr]\,.\nonumber
\eea
As we discussed above, we see that one bilinear term in the original variables has become ten terms in terms of the twisted variables.
Secondly, we can now explicitly see the promised linear $+$ non-linear form of the~$\qeq$-variations of the 
elementary fields. The linear part is the twisted variable which we have presented as the first term 
in the above variations. 

\section{Equivariant cohomology and black hole functional determinants \label{sec:BH}}

In this section we discuss the functional integral for the exact quantum entropy of half-BPS black holes 
in~$\CN=2$ superconformal gravity coupled to vector multiplets. Using the formalism developed above, 
we show how the functional integral reduces to an ordinary integral using supersymmetric localization filling in 
a gap in the formal derivation of the result for the graviton multiplet. We then compute the one-loop determinants 
of the deformation operator over the non-BPS fluctuations of the Weyl and vector multiplets in the localization formula.  
This determinant was computed in~\cite{Gupta:2015gga,Murthy:2015yfa} for vector and hyper multiplets using index theory. 
The symmetries of the problem combined with consistency with the on-shell 
computations at large charges~\cite{Sen:2011ba} also pinned down the determinant for the graviton multiplet. 
Here we give a first-principles calculation for the off-shell graviton multiplet, using the covariant formalism developed 
in the previous sections. As part of this calculation we need to deal with the subtleties of the so-called boundary modes
first discussed in~\cite{Gupta:2012cy}. To this end we develop a treatment of the boundary modes 
consistent with our formalism based on supersymmetry.

\subsection{Review of exact quantum entropy of BPS black holes}

The underlying theory we consider is $\CN=2$ superconformal gravity coupled to a number of matter multiplets
that we discussed in Sections~\ref{sec:sugra} and \ref{sec:mattersugra}.  
This theory has extra fields that transform under gauge transformations compared to the physical 
fluctuating fields around the black hole. As in any gauge theory, in order to make contact with the 
physics (in this case, of the black hole), one has to consider gauge-invariant combinations. 
In particular, we consider the Weyl multiplet coupled to~$\nv+1$ vector 
multiplets, labelled by~$I=0,\cdots \nv$, and one hyper multiplet. 
Of these, one vector multiplet and one hyper multiplet act as the so-called 
compensating multiplets, and can be gauged away if required. 

This theory has a black hole solution which preserves 4 out of 8 supercharges. The near-horizon 
configuration is a fully supersymmetric solution in its own right. The geometry is~$AdS_{2} \times S^{2}$
with equal and opposite scalar curvatures.
The near-horizon configuration has an $SL(2) \times SU(2)$ bosonic symmetry, 
the two factors acting on the~$AdS_{2}$ and~$S^{2}$ parts respectively. Each gauge field 
has a fixed electric and magnetic field strengths consistent with the bosonic symmetry, and 
constant scalars. The above bosonic symmetries together with the eight supersymmetries 
form an~$SU(1,1 | 2)$ superalgebra. The curvatures, fields strengths, and the scalar values 
are all fixed by the attractor equations, or equivalently, by the supersymmetry equations.

The problem of computing the exact quantum entropy of the original black hole was proposed in~\cite{Sen:2008vm} 
as the computation of the functional integral of the gravitational theory whose fields~$\phi_\text{sugra}$ asymptote 
to the near-horizon background just discussed:
\be \label{qef}
\exp \bigl( \SBH (q, p) \bigr) \; \equiv \; W (q, p) \= \int_{\rm{AdS}_2} [D\phi_\text{sugra}]  \, 
\exp\Bigl( -i \, q_I \oint_{\tau}  A^I  - S_\text{sugra}(\phi_\text{sugra}) \Bigr)  \, . 
\ee 
There are various infra-red divergences that arise from the infinite volume of~$AdS_{2}$, which are 
taken into account by appropriate counterterms. 

The idea of solving this integral exactly by localization methods was put forward 
in~\cite{Banerjee:2009af},~\cite{Dabholkar:2010uh}, which we review briefly below. This endeavor is different at a conceptual level from 
using localization to solve functional integrals in quantum field theory as there is no good a priori 
definitions because of the usual UV problems of gravity. Nevertheless, treating it as a formal object 
which is consistent with supersymmetry, the idea of~\cite{Dabholkar:2010uh,Dabholkar:2011ec} 
was to reduce it to a sensible integral which can then be compared to microscopic string theory. 
Even with this philosophy, we have to deal with the question posed in the introduction, namely what is a 
good choice of supercharge with which to localize. 
The route pursued in~\cite{Dabholkar:2010uh,Dabholkar:2011ec} is to choose the attractor solution 
as a background and use one of the supersymmetries of this background, and hope that all the 
gauge-invariances of the supergravity theory are consistently fixed in the quantum theory. 

We can now give a more systematic treatment of the Weyl multiplet and the gauge-fixing procedure in the 
quantum theory using the formalism developed in the previous sections. We write down a symmetry 
generator~$\qeq$ as in Section~\ref{sec:sugra} coming from the supersymmetry variations of the classical 
attractor background, and promote it to a covariant operator in the full quantum theory including the ghosts 
for all the gauge symmetries. According to the discussion in~\cite{dWMR} we should consider the original 
gauge-fixed functional integral~$Z$ using the action 
\begin{equation} \label{eq:BRST-action}
S_\text{sugra} \=  \int  \mathrm{d}^{4} x \,  \Bigl( \mathcal{L}^\text{phys}_\mathrm{sugra} 
- \delta_{\mathrm{eq}} \bigl( b_\alpha    \, F^\alpha \bigr) \Bigr) \,,
\end{equation}
where the gauge-fixing conditions~$F^\a$ are assumed to completely fix all the gauge invariances of the 
theory.\footnote{In the black hole context we will adopt a covariant gauge as in~\cite{Banerjee:2009af}.}

In order to localize, one begins by choosing a Killing spinor in the background attractor geometry, which we present in the appendix \ref{KillingS},  that 
generates a fermionic symmetry obeying the algebra
\be \label{specificQ}
\qeq^{2} \=  L_{0} - J_{0} \, , 
\ee
where~$L_{0}$ and~$J_{0}$ are the Cartan generators of the~$SL(2)$ and the~$SU(2)$ algebras, respectively. 
Next, one deforms the action as
\be \label{deformact}
S_\text{sugra} = S(0) \to S(\l) = S(0) + \lambda \, \qeq \, \CV \,, 
\ee 
with
\be\label{deformactV}
\qquad \CV \= \int \, d^{4} x \, 
\sqrt{\mathring{g}}\, \sum_\psi \, \overline{\psi} \, \qeq \psi  
\ee
summed over all the physical fermions of the theory.  Since~$L_0-J_0$ is a compact~$U(1)$ isometry,
this deformation obeys the condition~$\qeq^2 \CV=0$. This leads to the result that the functional integral 
reduces to an integral over the critical points of the deformation term, weighted by the original action 
times a one-loop determinant of the deformation action~$\qeq \, \CV$.
The critical points are given by the localization equations
\be
\qeq \,\psi \= 0  \,, \qquad \text{for all physical fermions} \; \psi \, ,
\ee
to be solved along with the gauge conditions~$F^\a =0$.

The variables for these localization equations 
are the metric and matter fields, as well as the bosonic ghosts for supergravity. In other words, 
the problem reduces to finding all metric and gauge field configurations which asymptote to the attractor 
background and admit some supercharge that asymptote to the background supercharge~$\qeq$. 
It was shown in~\cite{Gupta:2012cy} that the solution to this problem\footnote{This is true modulo an 
assumption in~\cite{Gupta:2012cy} regarding the~$SU(2)_R$ gauge field which can probably be removed 
upon coupling to charged hyper multiplets and repeating the localization calculation.}  was parameterized 
by an~$\nv+1$-dimensional manifold, whose points label the off-shell BPS fluctuations of the scalar field 
in each vector multiplet in a gauge of~$\sqrt{g}=\sqrt{g_0}$ where $g_0$  is for the $AdS_2 \times S^2 $ metric with unit radius.

The localizing manifold is thus labelled by~$(\nv+1)$ real parameters~$\{\phi^{I}\}$, $I=0,\cdots, \nv$. 
The result~\cite{Dabholkar:2010uh} of evaluating the functional integral~\eqref{qef} is 
\begin{equation} \label{LocIntegral}
 W^\text{pert} (q, p) = \int_{\mathcal{M}_{Q}}  \, \prod_{I=0}^{\nv} d\phi^{I} \, \exp\Big(- \pi  \, q_I  \, \phi^I 
 + 4 \pi \, \Im{F \big((\phi^I+\i p^I)/2 \big)} \Big)
 \, Z^{\qeq\CV}_\text{1-loop}(\phi^{I})\, ,
\end{equation}
where~$F$ is the holomorphic prepotential of the supergravity theory (which can contain terms with 
arbitrary derivatives). 
The superscript ``pert'' indicates that this is an all-order perturbation theory result around the attractor configuration. 
There may be additional non-perturbative contributions, for example from orbifold 
configurations~\cite{Banerjee:2008ky, Murthy:2009dq, Dabholkar:2014ema}.

The problem thus reduces to evaluating the one-loop determinant in the expression~\eqref{LocIntegral}. 
It was argued in~\cite{Murthy:2015yfa} that since there is only one scale set by~$e^{-\CK}:= \,  -i(X^I \,\bar{F}_I   - \bar{X}^I F_I)$ in the localization background, where $\CK$ is called  K\"ahler potential and  $F$ is the holomorphic prepotential, 
the functional determinant will have the symplectically invariant form (ignoring infinite constants):
\be \label{def1loop}
Z_\text{1-loop}(\phi^{I}) \= \exp \bigl(-a_{0} \,\CK(\phi^{I}+\i p^I ) \bigr) \,.
\ee
The number~$a_{0}$ receives contributions from each multiplet of the~$\CN=2$ supergravity theory:
\be \label{defa0}
a_{0} \= a_{0}^\text{grav} \+  (\nv + 1) \, a_{0}^\text{vec} \+  \nh \, a_{0}^\text{hyp} \, , 
\ee
where~$(\nv+1)$, $\nh$ are the number of  vector and hyper multiplets in the off-shell theory, 
respectively.\footnote{Any other multiplets like spin~$3/2$ multiplets will also contribute linearly.}
When all the electric and magnetic charges of the black hole scale equally to be very large, 
we can do a saddle-point analysis of the integral~\eqref{LocIntegral} to obtain
\be \label{leadlog}
\SBH \= \frac{A_{H}}{4} + a_{0} \log A_{H} \+ \cdots \, .
\ee
The number~$a_0$ was calculated for vector and hyper multiplets in~\cite{Gupta:2015gga,Murthy:2015yfa} to be
\be
a_{0}^\text{vec}\=-a_{0}^\text{hyp}\=-1/12 \,.
\ee
We now move on to compute the number~$a_{0}$ for the Weyl multiplet, after 
first reviewing the fixed-point formula for the computation of the determinant.

\subsection{Functional determinants from a fixed point formula \label{sec:determinant}}

An elegant formalism to compute the one-loop determinant was given 
in~\cite{Pestun:2007rz,Hama:2012bg, LeeSJ, Hosomichi:2015jta}.
The idea is to first organize all the fluctuating fields of the theory into cohomological variables, 
i.e.~representations of the form~$(\Phi\,, \qeq \Phi\,,\Psi\,, \qeq \Psi)$ of the equivariant algebra~$\qeq^2= H$.
This is exactly what we achieved in Section~\ref{sec:twisted}
for the case of supergravity, where we arranged all the fields 
as elementary bosons~$\Phi$ and fermions~$\Psi$ and their respective~$\qeq$-partners.

The supercharge~$\qeq$ pairs up the fields algebraically at each point in space, and therefore 
all the contribution to the superdeterminant can be understood as a mismatch between the elementary bosons 
and elementary fermions, which is kept track by the operator~$D_{10}: \Phi \to \Psi$. 
This follows from an algebraic analysis which we repeat below because there are subtleties when 
we apply it to our problem. We begin by writing the~$\qeq \CV$ action as follows,
\begin{eqnarray}\label{QVaction1}
\CV&\=&\int \, d^{4} x \, 
\sqrt{\mathring{g}}\,   \Biggl[(\qeq \Phi\,, \Psi) \, \Biggl( \, \begin{matrix}D_{00}&D_{01}\\D_{10}&D_{11}\end{matrix} \, \Biggr) 
\Biggl( \, \begin{matrix}\Phi\\\qeq\Psi \end{matrix} \, \Biggr) \Biggr]\,\\
\Rightarrow \qeq \CV&\=&\int \, d^{4} x \, \sqrt{\mathring{g}}\,\Biggl[ (\Phi\,, \qeq\Psi)\,K_b \, \Biggl( \, \begin{matrix}\Phi\\\qeq\Psi \end{matrix} \, \Biggr)
+ (\qeq\Phi\,, \Psi)\,K_f\, \Biggl( \begin{matrix}\qeq\Phi\\\Psi \end{matrix} \Biggr) \Biggr]\,, \label{QVaction2}
\end{eqnarray}
where
\begin{eqnarray}
&&K_{b}\=  \Biggl( \,\begin{matrix}-H&0\\0&1\end{matrix} \, \Biggr)  \Biggl( \,\begin{matrix}D_{00}&D_{01}\\D_{10}&D_{11}\end{matrix}\, \Biggr)  
+ \Biggl( \,\begin{matrix}D_{00}^{T}&D_{10}^{T}\\D_{01}^{T}&D_{11}^{T}\end{matrix} \, \Biggr)   \Biggl( \,\begin{matrix}H&0\\0&1\end{matrix}\, 
\Biggr)  \,,\label{Kb} \\
&&K_{f}\= \Biggl( \,\begin{matrix}1&0\\0&-H\end{matrix} \, \Biggr)  \Biggl( \,\begin{matrix}D_{00}^{T}&D_{10}^{T}\\D_{01}^{T}&D_{11}^{T}\end{matrix} \, \Biggr)  
-\Biggl( \,\begin{matrix}D_{00}&D_{01}\\D_{10}&D_{11}\end{matrix} \, \Biggr) \Biggl( \, \begin{matrix}1&0\\0&H\end{matrix} \, \Biggr)  \,.\label{Kf}
\end{eqnarray}
It is clear from these expressions that  
\be\label{KbKfrelation}
\Biggl( \,\begin{matrix}1&0\\0&-H\end{matrix}  \Biggr) K_{b} \= K_{f} \Biggl( \,\begin{matrix}H&0\\0&1\end{matrix} \, \Biggr)\,,
\ee
and therefore the ratio of determinants of the fermionic 
and bosonic kinetic operators in~$\qeq \CV^\text{eq}$ reduces, up to a sign, to the 
ratio\footnote{The last ratio is well-defined in that the modes with zero eigenvalue of $H$ do not contribute to it. As can be seen from Equations~\eqref{Kb}, \eqref{Kf}, the determinant of the kinetic operator on $H=0$ modes is the square of determinant of~$D_{10}$  equally for both bosons and fermions. Further, this determinant is non-zero as the modes under consideration are orthogonal to the localisation locus. Therefore the determinant for those modes is completely cancelled between bosons and fermions. }
\be \label{detratio}
Z^{\qeq\CV}_\text{1-loop} \= \sqrt{\frac{\det_{} K_f}{\det_{} K_b}} \= \sqrt{\frac{\det_{\Psi} H}{\det_{\Phi} H}} \,.
\ee
Now, the operator~$D_{10}$ pairs up the elementary bosons and fermions, and therefore any mode which is not 
in the kernel or cokernel of~$D_{10}$ does not contribute to this ratio.
Thus the ratio of determinants on the right-hand side can thus be computed from the knowledge of the index
\be \label{indD10}
\text{ind} (D_{10})(t) \; := \; \Tr_\text{Ker$D_{10}$}  \, e^{tH} - \Tr_\text{Coker$D_{10}$}  \, e^{t H} \,.
\ee
Writing the index as a series,
\be \label{indexser}
\text{ind} (D_{10})(t) \= \sum_{n} a(n) \, e^{\i  \lambda_{n} t} \, , 
\ee 
we can read off the eigenvalues~$\l_{n}$ of~$H$, as well as their indexed degeneracies~$a(n)$, and 
the ratio of determinants in \eqref{detratio} is
\be \label{detratio1}
Z_\text{1-loop}\= \prod_{n} \, \l_{n}^{-\half a(n)} \, ,
\ee
where the infinite product is regulated in a suitable manner.  

Our computation thus reduces to the computation of the equivariant index~\eqref{indD10}, with respect 
to the action of~$H$. This can be done in an elegant manner using the Atiyah-Bott fixed-point 
formula~\cite{Atiyah:1974}, which says that it reduces to the quantum-mechanical modes at the 
fixed points of the manifold under the action of~$H$. 
Denoting this action by $x \mapsto \wt x = e^{tH} x$ we have
\be \label{ABFormula}
{\rm ind}(D_{10}) \= \sum_{\{x \mid \wt x = x\}}  
\frac{{\rm Tr}_{\Phi} \, e^{tH} \, - \, {\rm Tr}_{\Psi}\,e^{tH}}{{\rm det}(1-\partial \wt x/\partial x)} \,.
\ee
We therefore simply need to compute the charges of the various 
modes under this rotation, which can be read off from our presentation of the twisted variables in Section~\ref{sec:twisted}.

Our goal now is to compute the one-loop determinant in~\eqref{LocIntegral} and, in particular, the number~$a_{0}$ defined 
in~\eqref{defa0} for the Weyl multiplet. We will do so using the fixed-point formula outlined above, but 
before doing so we remind the reader that there are some caveats and subtleties in applying the formula to the black hole problem, 
as discussed in~\cite{Gupta:2015gga,Murthy:2015yfa}. 
The main issue is that we are in a non-compact space and we should be careful about 
the boundary conditions on the various fields. These issues have been addressed in similar contexts in 
\cite{David:2016onq, Assel:2016pgi, David:2018pex}. In particular, it was shown in~\cite{David:2018pex} that 
normalizable boundary conditions are not always compatible with supersymmetry, even for scalar multiplets. 
In our analysis, we would like to have a set of boundary conditions consistent with supersymmetry. In order to achieve 
this we impose normalizable boundary conditions for all the elementary cohomological variables. Supersymmetry
then requires that a mode~$\phi$ and its superpartner~$\qeq \phi$ have the same boundary conditions.
Here we do not explicitly construct these boundary conditions---this is an important issue that needs to be addressed---but 
our results seem to be consistent with their existence.
Another technical caveat is that we need to show that the~$D_{10}$ operator in the black hole context is 
transversally elliptic with respect to the action of~$H$.  
We postpone the details of these issues to future work.
However, there is one important subtlety for the black hole problem that may affect the answer crucially, which 
is the existence of the so-called boundary modes~\cite{Banerjee:2010qc,Sen:2011ba}, we now turn to a detailed 
discussion of this matter.\footnote{We thank the referee for emphasizing their importance, which led us to include 
the following subsection in the present version of the paper.}

\subsection{Boundary modes and their effect on the 1-loop determinant}

Boundary modes are normalizable modes of gauge fields that are formally pure gauge but whose 
gauge parameters are not normalizable. 
For example there are normalizable modes of the 1-form gauge field~$A_\mu^{\text{bdry}}=\p_\mu \Lambda$
with non-normalizable~$\Lambda$.
These modes are not gauge redundancies and should be considered 
physical degrees of freedom because we have assumed a normalizable boundary 
condition on all the elementary cohomological variables, which includes the~$c$ ghost fields. 
The presence of these boundary modes makes the functional integral~\eqref{qef} 
ill-defined. This is because the gauge fields only appear in the physical action through the field strengths,  
which vanish for these modes. These modes also respect the covariant gauge condition that we adopt, 
so that they are genuine zero modes of the action in~\eqref{qef}. 
Further, as we will see below, the~$\qeq \CV$-deformation
also vanishes when evaluated on these modes. The deformed functional integral thus remains ill-defined. 
Therefore we must remove these modes from the naive computation and consider their effect separately.
In our treatment we continue to denote the critical points of the~$\qeq \CV$ action in the space of bulk modes 
as the ``localization manifold", and take into account the effect of the boundary modes in the one-loop determinant,
that is,
\be\label{factorizedZ}
Z^{\qeq\CV}_\text{1-loop}\; := \; Z^{\text{bdry}}  \, Z'^{\qeq\CV }_\text{1-loop } \,,
\ee
where the two terms on the right-hand side denote the contribution to the one-loop determinant by the 
boundary modes and the quadratic fluctuations of the bulk modes, respectively.

The main reason for the subtlety concerning the boundary modes in our formalism 
is that $\qeq$ is nilpotent on these modes instead of obeying the 
equivariant algebra~$\qeq^2= H$. We can check this explicitly using the definition of the cohomological variables and 
by noting that the normalizable boundary condition on the ghost field $c$ implies that the boundary mode 
cannot be written as a covariant derivative acting on a ghost mode.  
For example, the $1$-form gauge field~$A_\mu$ generically obeys~$\qeq A_\mu = \lambda_\mu+\p_\mu c$ but 
for the boundary mode the second term is absent, and therefore we have 
\be \label{nilpot}
\qeq^2 A_\mu^{\text{bdry}}=v^{\nu}F^{\text{bdry}}_{\nu\mu}=0\,.
\ee 
This fact plays an important role below. 

Now we look at the details of the deformation action. We use the 1-form gauge field with boundary modes~$A_\mu^{\text{bdry}}$ 
as an example. The quantity~$\CV$ in~\eqref{QVaction1} actually vanishes for these boundary modes because it is built out of 
field strengths---this can be seen easily from~\eqref{deformactV} and the supersymmetry variation of the gaugino in 
Appendix~\ref{App:vecmult}---which vanishes for the boundary modes as they are pure gauge. 
This implies that~$\qeq \CV$ also vanishes 
for the boundary modes, and we have to treat it separately as mentioned above.
Now, the fact that~$\CV$ vanishes for the boundary modes implies that~$\qeq \CV$ also 
vanishes for the superpartner~$\qeq A_\mu^{\text{bdry}}$, so it would seem that these are also 
zero modes of the deformation action.
In fact, these modes can be lifted in a supersymmetric manner 
by choosing the relevant term in~\eqref{QVaction2} directly 
as the definition of the deformed action for the fermionic superpartners of the boundary modes.  
Denoting the set of bosonic boundary modes by~$\Phi^{\text{bdry}}$, we have 
\be\label{bdryQV}
\qeq\cV|^{\text{bdry}}_{\text{fermion}} 
\=  -(\qeq\Phi^{\text{bdry}}) \,  H \, (\qeq\Phi^{\text{bdry}})\,,
\ee
where the corresponding kinetic operator is nothing but the left-upper block diagonal part~$D_{00}$ of the fermionic kinetic operator in \eqref{Kf}. 
Now, with this definition it is not obvious that the deformation action is $\qeq$-exact. However, one can easily check that it is the case. 
Indeed we have
\be \label{bdryQV2}
\qeq\cV|^{\text{bdry}}_{\text{fermion}}\= \qeq\left[\Phi^{\text{bdry}} \, H \, (\qeq\Phi^{\text{bdry}})\right]\,,
\ee
by using the nilpotency of the boundary modes. 
(We can also regard this as replacing~$D_{00}$ in~\eqref{QVaction1}  by~$H$ for these modes.) 
Note, in particular, that this does not lift the bosonic boundary modes themselves because of the nilpotency condition~\eqref{nilpot}.
The above arguments were made for the example of the~$1$-form gauge field, but it applies to any bosonic boundary mode. 
We will therefore take the action~\eqref{bdryQV}, or equivalently~\eqref{bdryQV2}, for the  fermionic partners of 
all bosonic boundary modes~$\Phi^{\text{bdry}}$.

Now we turn to the determinant~\eqref{factorizedZ}. For the moment we assume that there are no fermionic zero modes. 
Although the bulk part~$Z'^{\qeq\CV }_\text{1-loop }$ 
does not include the determinant over the bosonic boundary modes $\Phi^{\text{bdry}}$, it does include the 
determinant over their partners~$\qeq \Phi^{\text{bdry}}$ because these partners are not zero modes of the~$\qeq \CV$ 
action by our construction. Now we want to reduce the ratio of the determinant of bosons and fermions 
similarly to~\eqref{detratio} using the relation \eqref{KbKfrelation}. 
But here we should note that the relation \eqref{KbKfrelation} makes sense only when the operators on the left- and 
right-hand side of the equation act on the space excluding the boundary modes $\Phi^{\text{bdry}}$ as well as their 
partners~$\qeq \Phi^{\text{bdry}}$. 
Therefore a reduction similar to~\eqref{detratio} happens after splitting the determinant of $K_f$  into the 
determinant over $\qeq \Phi^{\text{bdry}}$ and the rest of the modes, so that we obtain
\be \label{detratio2}
Z'^{\qeq\CV }_\text{1-loop} \= \sqrt{\frac{\det_{} K_f}{\det'_{} K_b}}\= \sqrt{{\det}_{\qeq \Phi^{\text{bdry}} }H} \,\sqrt{\frac{\det'_{} K_f}{\det'_{} K_b}}
\= \sqrt{{\det}_{ \Phi^{\text{bdry}} }H} \,\sqrt{\frac{\det_{\Psi} H}{\det'_{\Phi} H}} \,.
\ee
For the last equality, we use that the determinant of $H$ over $\qeq \Phi^{\text{bdry}}$ is equal to the determinant over~$\Phi^{\text{bdry}}$.
We can write the logarithm of this determinant in an integral representation as follows,
\begin{eqnarray}
\log Z'^{\qeq\CV }_{\text{1-loop}}&\=& -\frac{1}{2}\int_\epsilon^{\infty} \, \frac{dt}{t}\left( \Tr_{ \Phi^{\text{bdry}}}e^{t H} + \Tr_{\Psi
}e^{tH} - \Tr'_{\Phi}e^{tH}\right) \nn \\
&\=& -\int_\epsilon^{\infty}\frac{dt}{t} \, \Tr_{\Phi^{\text{bdry}}}e^{t H} -\frac{1}{2}\int_\epsilon^{\infty}\frac{dt}{t}\left( \Tr_{\Psi
}e^{tH} - \Tr_{\Phi}e^{tH}\right)   \nn \\ 
&\; \equiv \;& -\int_\epsilon^{\infty}\frac{dt}{t} \, \nbd^\Phi+\frac{1}{2}\int_\epsilon^{\infty}\frac{dt}{t} \, \text{ind} (D_{10})(t) \,.\label{n0plusIndbos}
\end{eqnarray}
To reach the second line, we add and subtract half the trace over $\Phi^{\text{bdry}}$  
so that the trace in the second term is now over the complete normalisable function space. 
In reaching the third line we have denoted the difference of the traces in the full spaces~$\Phi$ 
and~$\Psi$ as~$\text{ind}(D_{10})$.
Here we have defined the number of boundary modes by 
\be \label{defnbd}
\nbd^\Phi \; := \;  \Tr_{\Phi^{\text{bdry}}}e^{tH} \Big{|}_{t^0} \,,
\ee
where the notation~$|_{t^0}$ means that we pick the constant ($t^0$) term in a Laurent expansion around zero. 
This definition will pick out the~$\epsilon$-independent term in the first integral,
similar to the bulk calculation.  
Further, the traces involved in this definition will turn out to be actually regular around~$t=0$ for every 
field~$\phi$ with a boundary mode,
and therefore we can replace the above definition by~$\nbd^\phi\=\lim_{t\rightarrow 0} \Tr_{ \phi^\text{bdry}}e^{tH}$.
This is simply a regulated version of~$\nbd^\phi\= \Tr_{ \phi^\text{bdry}}1$, which justifies the terminology ``the number of boundary modes".
This regulator is not exactly the same as the one used in the on-shell calculation~\cite{Sen:2011ba}, and is 
more suited to our off-shell localization calculation. However, as we shall see, our actual answers for~$\nbd$ agree with the on-shell values
of the ``the number of zero modes" defined in~\cite{Sen:2011ba}.

So far we have discussed elementary bosons. There are three more types of modes in our complex, namely non-elementary
bosons, elementary fermions, and non-elementary fermions. There are gauge fields, and corresponding boundary modes, 
in each of these spaces, to which we turn now one by one. 
We begin with non-elementary bosons. The non-elementary bosons are generally combinations of auxiliary fields and derivatives
of elementary fields. For example, in the vector multiplet we have the combination~\eqref{Qlambdaij} of~$Y^{ij}$ and the field strength. 
In this simplest example, it is well-known that the quadratic term in the Lagrangian for~$Y^{ij}$ is non-propagating. This is 
in fact more general, and we can check that the kinetic term~$D_{11}$ for the non-elementary bosons in the action~\eqref{QVaction2}
is simply 1. This means that even when there are gauge fields like the auxiliary~$SU(2)$ gauge field~$\CV^i_{\mu j}$ for the Weyl 
multiplet, they are not zero modes of the~$\qeq \CV$ action. So in this case there is no modification to our regular treatment.

Next we move to the boundary modes coming from fermionic gauge fields. We shall call such elementary fermions~$\Psi^{\text{bdry}}$
and composite fermions~$\qeq \Phi^{\text{pre-bdry}}$. The corresponding superpartners are~$\qeq \Psi^{\text{bdry}}$ and~$\Phi^{\text{pre-bdry}}$,
respectively.
As in the case of the bosonic gauge field~\eqref{nilpot}, the supercharge~$\qeq$ squares to the field strength of 
the fermion, which is zero for pure gauge modes. Thus~$\qeq$ is nilpotent on the boundary modes. 
If we take the deformation action defined  by~$\qeq \CV$ with~$\CV$ given by~\eqref{QVaction1}, we find that the 
kinetic terms of the fermionic boundary modes are:
\be
\Psi^{\text{bdry}} \, \qeq^2 \, \Psi^{\text{bdry}} \,, \qquad \bigl( \qeq \Phi^{\text{pre-bdry}}\bigr) \qeq^2 \bigl( \qeq \Phi^{\text{pre-bdry}} \bigr)\,,
\ee
corresponding to the diagonal part of the lower-right block and upper-left block of \eqref{Kf}, respectively, 
which vanish because of the nilpotence of~$\qeq$.
The corresponding superpartners have the following kinetic terms, 
\be
\bigl(\qeq \Psi^{\text{bdry}}\bigr) \, 1\, \bigl(\qeq  \Psi^{\text{bdry}}\bigr)\,, \qquad \Phi^{\text{pre-bdry}} \, H^2 \, \Phi^{\text{pre-bdry}} \,,
\ee
corresponding to the diagonal part of the lower-right block and upper-left block of \eqref{Kb}, respectively, which is well-defined. 
Thus we see that we do not need any modification to the action for the fermionic boundary modes and their superpartners. 

Now an analysis similar to the one that leads to~\eqref{detratio2}, in the case that there are no bosonic zero modes, 
leads to the following determinant,
\be \label{}
Z'^{\qeq\CV }_\text{1-loop} \= \sqrt{\frac{\det'_{} K_f}{\det_{} K_b}}\= ({{{\det}_{\Phi^{\text{pre-bdry}} }H^2} } )^{-\frac12}\,\sqrt{\frac{\det'_{} K_f}{\det'_{} K_b}}
\= ({\det}_{ \Phi^{\text{pre-bdry}} }H^2)^{-\frac12} \,\sqrt{\frac{\det'_{\Psi} H}{\det'_{\Phi} H}} \,.
\ee
As before, we can write this in an integral representation as follows, 
\begin{eqnarray}
\log Z'^{\qeq\CV }_{\text{1-loop}}&\=& -\frac{1}{2}\int_\epsilon^{\infty} \, \frac{dt}{t}\left( -2\Tr_{ \Phi^{\text{pre-bdry}}}e^{t H} + \Tr'_{\Psi
}e^{tH} - \Tr'_{\Phi}e^{tH}\right) \nn \\
&\=& \frac{1}{2}\int_\epsilon^{\infty}\frac{dt}{t} \,\left( \Tr_{\qeq \Phi^{\text{pre-bdry}}}e^{t H}+ \Tr_{\Psi^{\text{bdry}}}e^{t H} \right)-\frac{1}{2}\int_\epsilon^{\infty}\frac{dt}{t}\left( \Tr_{\Psi
}e^{tH} - \Tr_{\Phi}e^{tH}\right)   \nn \\ 
&\; \equiv \;&  \frac12 \int_\epsilon^{\infty}\frac{dt}{t} \,( \nbd^{\qeq\Phi}+ \nbd^\Psi)+\frac{1}{2}\int_\epsilon^{\infty}\frac{dt}{t} \, \text{ind} (D_{10})(t) \,,\label{n0plusIndfer}
\end{eqnarray}
where we have defined the number of (elementary and composite) fermionic boundary modes as
\be \label{defnbd}
\nbd^\Psi \; := \;  \Tr_{\Psi^{\text{bdry}}} \, e^{tH} \Big{|}_{t^0} \,, \qquad \nbd^{\qeq\Phi} \; := \;  \Tr_{\qeq\Phi^{\text{pre-bdry}}} \, e^{tH} \Big{|}_{t^0} \,.
\ee

The general formula when there are both bosonic and fermionic zero modes is found by putting together the full discussion. 
We thus reach the final formula for the modified one-loop determinant:
\be \label{n0plusIndfinal}
\log Z'^{\qeq\CV }_{\text{1-loop}}
\= -\int_\epsilon^{\infty}\frac{dt}{t} \, \nbd^{\text{bos}}
+\frac12 \int_\epsilon^{\infty}\frac{dt}{t} \, \nbd^{\text{fer}}+\frac{1}{2}\int_\epsilon^{\infty}\frac{dt}{t} \text{ind} (D_{10})(t) \,,
\ee
where~$\nbd^{\text{bos}}$ and~$\nbd^{\text{fer}}$ are the total number of bosonic and fermionic boundary modes, respectively. 
In Appendix~\ref{App:bdrymodes}, we calculate the number of zero modes for the various fields in our problem.

The whole discussion above has been done based on the assumption that  the boundary modes of the bosons and 
fermions are not paired by~$\qeq$. It is easy to see that our final formula~\eqref{n0plusIndfinal} remains unchanged even 
if we relax this assumption and there is such a pairing. 
Let us start with the bosonic case and consider the determinant~\eqref{detratio2}. In this case if the fermionic mode~$\qeq\Phi^{\text{bdry}}$ is 
also a boundary mode, then there is no lifting of this mode, so that  the 
ratio~\eqref{detratio2} does not have the term~$\sqrt{{\det}_{ \Phi^{\text{bdry}} }H}$. This means that, in the trace formula~\eqref{n0plusIndbos},
the first term of the first line is absent, which implies that the first term of the second line has a factor of~$\half$ and 
thus we get $-\half \nbd^\Phi$ in the first term of the last line instead of~$-\nbd^\Phi$.  This result can be understood 
as a cancellation between the number of  bosonic boundary modes and fermionic boundary modes, i.e.
\be
-\half \nbd^\Phi \= -\nbd^\Phi +\half \nbd^\Phi \= -\nbd^\Phi +\half \nbd^{\qeq\Phi}\,.
\ee
Thus, by adding a fermionic zero mode term and subtracting a bosonic zero mode contribution, each with a factor of~$\half$, 
we reach precisely the formula~\eqref{n0plusIndfinal}.
The same analysis holds, mutatis mutandis, for the fermionic case.

\subsection{Computation of the black hole determinant in supergravity}

In this subsection we evaluate the one-loop determinant~$Z^{\qeq\CV}_\text{1-loop}$ given in~\eqref{factorizedZ}. 
As was discussed in~\cite{Gupta:2015gga,Murthy:2015yfa}, the 1-loop determinant depends on the coordinates of the localization 
manifold only through one combination of fields called $\ell$. In order to see this, we note that  the metric that enters the index theorem calculation 
should be the physical metric, whose kinetic term is given by the Einstein-Hilbert Lagrangian. In terms of the 
metric~$g_{\mu\nu}$ and the scalar fields~$X^I$ that enter the action of~$\CN=2$ 
supergravity~\cite{deRoo:1980mm,deWit:1980tn,deWit:1984px}, the physical metric is the 
composite~$e^{-\CK (X^{I})} g_{\mu\nu}$. 
The~$AdS_{2} \times S^{2}$ line element is thus given by 
\bea \label{metricreal}
&& ds^2 \=\ell^2 \bigl(d\eta^2 + \sinh^2\eta \, d\tau^2 \bigr) + \ell^2\bigl(d\psi^2 + \sin^2\psi \, d\phi^2 \bigr)\,, 
\eea
where~$\ell$ is the overall physical size of the~$AdS_{2} \times S^{2}$ metric governed by the above 
field-dependent physical metric. 
The calculation is simplified by going to complex coordinates in which the metric is
\begin{equation}\label{metriccomplex}
    ds^2 \= \ell^{2} \biggl( \frac{4 dw d\bar w}{(1- w \bar w)^2} + \frac{4 dz d\bar z}{(1 + z \bar z)^2} \biggr) \, .
\end{equation}
At the fixed points, i.e.~the center of~$AdS_{2}$, the overall size is given by~$\ell^{2} = e^{-\CK (\phi^{I}+\i p^I)}$.

The one-loop determinant~$Z^{\qeq\CV}_\text{1-loop}$ is divided into the bulk part~$Z'^{\qeq\CV }_\text{1-loop }$ 
and the boundary part~$Z^{\text{bdry}}$, which we now evaluate in turn.
First we turn to~$Z'^{\qeq\CV }_\text{1-loop }$ which is given by the formula~\eqref{n0plusIndfinal}. 
By changing the variable of integration  to the dimensionless parameter $\bar{t}:= t/\ell$, we obtain  an integral whose 
range of integration runs from $\epsilon/\ell$ to infinity, and one then extracts the  $\epsilon$-independent term,
which we now proceed to do.
The contribution of the bulk modes, i.e.~the third term in the formula~\eqref{n0plusIndfinal}, is captured by the  
index 
\be \label{defindtilde}
\text{ind} (D_{10})(t) \= \Tr_{\Phi}e^{tH}-\Tr_{\Psi}e^{tH}  \,.
\ee
This can be computed using the Atiyah-Bott fixed point formula applied to the field space with our prescribed boundary conditions.
Using these methods, the calculation of~$\text{ind} (D_{10})(t)$ 
reduces to the contribution from the fixed points of~$AdS_2\times S^2$ under the action of $H$.
Computing the~$t \to 0$ expansion of~$\text{ind} (D_{10})(t)$:
\be\label{a0index}
\frac14 \, \text{ind} (D_{10})(t) \= \cdots + \frac{a_{-2}}{t^2} + a^{\text{bulk}}_{0} + a_{2} \, t^{2} + \cdots \,,
\ee
the $\epsilon$-independent term  is given by the constant term in this expansion. 
In this manner we obtain that the third term of~\eqref{n0plusIndfinal} equals
\be\label{Zbulk}
2a^{\text{bulk}}_{0} \log \ell\,.
\ee
Using the definition of the boundary modes, the first two terms in~\eqref{n0plusIndfinal} give
\be\label{Zn0}
-\bigl(\nbd^{\text{bos}}  - \frac12 \nbd^{\text{fer}} \bigr) \log \ell\,.
\ee

The computation of the zero mode part in \eqref{factorizedZ} has been  performed in~\cite{Sen:2011ba} by 
associating these modes to asymptotic symmetries, and computing the Jacobian in transforming the variables 
to the parameters labelling the symmetries. This procedure yields the formula\footnote{Here we make the assumption 
that the same $\ell$ as for the bulk modes is also the relevant scale for the boundary modes. 
As we shall see the final answer is consistent with this assumption.}
\be\label{Zbeta}
\log  Z^{\text{bdry}}\=(\beta^\text{bos} \nbd^{\text{bos}} - \frac12 \beta^\text{fer} \nbd^{\text{fer}})\log \ell\,,
\ee
where $\beta$  and $\nbd$ are numbers associated with each type of field that has boundary modes, and is 
obtained on a case by case basis for each field, that we discuss in Appendix~\ref{App:bdrymodes}.

Summarising the equations \eqref{Zbulk}, \eqref{Zn0}, and  \eqref{Zbeta}, we get the 1-loop partition function 
\be
 Z^{\qeq\CV}_\text{1-loop}\=  \exp \bigl(-a_{0} \,\CK(\phi^{I}+\i p^I ) \bigr) \,,  
\ee
with
\be \label{a0bulkbdry}
a_{0} \= a_{0}^\text{bulk} + a_{0}^\text{bdry} \,,~~~~~~~
a_{0}^\text{bdry} \= \half (\beta^\text{bos} -1) \nbd^\text{bos} -\frac14 (\beta^\text{fer} -1) \nbd^\text{fer} \,.
\ee
In the rest of this section we calculate the contribution~$a_{0}^\text{bulk}$ to the one-loop 
determinant for a generic multiplet. In the next subsection we will assemble all the pieces to get the results for 
the full~$a_0$ for the various multiplets.

\subsubsection*{Contribution of bulk modes through the index}

Thus we focus on the fixed points of the $U(1)$ action~$H= (\partial_\tau -\partial_\phi) \equiv L_{0} - J_{0}$. 
The fixed points are given by~$w=0$, and~$z=0$ or~$1/z=0$ which are  the center of $AdS_2$, with the North Pole or South Pole of $S^2$ respectively.
The action of the operator~$e^{-\i Ht}$ on the spacetime coordinate is~$(w\,,z) \to (e^{ \i t/\ell} w\,,e^{ -\i t/\ell}z)$. Therefore the determinant factor in the denominator of~\eqref{ABFormula} is, 
with~$q = e^{\i t/\ell}$, 
\be
{\rm det}(1-\partial \wt x/\partial x) \= (1-q)^2(1-q^{-1})^2 \,.
\ee

Near the fixed points the space looks locally like~$\IR^{4}$, so we can assign the local coordinates
\be\ba{l}
w\= x_1+\i x_2\,, ~~~~~z\=x_3+\i x_4~~~~\mbox{at NP}\,,\\
w\= x_1+\i x_2\,, ~~~~1/z\=x_3+\i x_4~~~~\mbox{at SP}\,.\\
\ea\ee
For each local coordinates, we have an associated~$SO(4)=SU(2)_{+} \times SU(2)_{-}$ 
rotation symmetry.  A representation of the chiral and anti-chiral parts of the rotation generator, i.e. $\vec{J}^{(\pm)}$ of $SU(2)_\pm$,   can be given by 
\be
\frac{1}{8}\gamma_{ab}\left(1\pm\gamma_5\right)= \frac{1}{4}\left(\gamma_{ab}\mp \frac{1}{2}\e_{abcd}\gamma^{cd}\right)\,,
\ee
for our convention of chirality matrix $\gamma_5=\gamma_{1234}$. Therefore, since the representation of $L_0$ and $J_0$ is 
\be
L_0\= \frac{1}{4}\gamma_{12}\,,~~~~~J_0 \= \pm\frac{1}{4}\gamma_{34}  ~~~~\mbox{at NP/SP}\,,
\ee 
the action of~$H$ is identified with the Cartan generator of~$SU(2)_{+}$ at the North Pole, 
and with the Cartan of~$SU(2)_-$ at the South Pole:
\be
H\= L_0-J_0\= 2J^{(+)}_{3}\,~~~\mbox{at NP}\,,~~~~~~~H\= L_0-J_0=2J^{(-)}_3\,~~~\mbox{at SP}\,.
\ee
Furthermore, for a representation $(m,n)$ of the $SU(2)_{+} \times SU(2)_{-}$, at the north pole we have
\be\label{TrNP}
{\rm Tr}_{(m,n)}e^{ tH} \= n (q^{-|m-1|}+q^{-|m-1|+2}\cdots + q^{|m-1|-2}+ q^{|m-1|}) \,.
\ee
while at the south pole we have
\be\label{TrSP}
{\rm Tr}_{(m,n)}e^{ tH} \= m (q^{-|n-1|}+q^{-|n-1|+2}\cdots + q^{|n-1|-2}+ q^{|n-1|}) \,,
\ee

In the next subsection we compute the trace in numerator of the formula~\eqref{ABFormula} by
computing the charges of all the fields under these symmetry generators.
For all the supergravity multiplets the index takes the form 
\be
{\rm ind} (D_{10}) \= 2 \, \frac{c_2\,(q^2+q^{-2})+c_1\,(q+q^{-1})+c_0}{(1-q)^2 (1-q^{-1})^2} \,,
\ee
for some coefficients~$c_{2,1,0}$. In order to compute the coefficient~$a_{0}$, we see from Equation~\eqref{a0index} 
that we only need to compute the constant term in~$t\to 0$ expansion. We thus obtain 
\be \label{index2det}
a_{0}^\text{bulk} \= \frac{502 \, c_2 -38 \, c_1 +11 \, c_0}{1440}  \,.
\ee

\subsection{Results}

Near the fixed points i.e.~the north and the south pole, our local twisting construction of the 
previous section reduces to the twisting construction of~\cite{Nekrasov:2002qd} with respect to the usual 
global symmetries of~$\CN=2$ theories in flat space. As we present in detail in the appendix \ref{KillingS}, the Killing spinors play the role of locking the $SU(2)_R$ symmetry with one of the $SU(2)_+\times SU(2)_-$ local Lorentz rotation at the fixed points. In terms of the real coordinate system given by \eqref{metricreal},  at $\eta=0$ and $\psi=0$ (north pole), the chiral and anti-chiral part of the Killing spinor reduces to
\be
\ve^i_{+\alpha}=0\,,~~~~~~~~~\ve^{\,i}_{-\dot\alpha}=  \left(\sigma_3 \exp\left[{\i \frac{(\tau+\phi)}{2}\,  \sigma_3}\right]\right)^i{}_{\dot\alpha}\,,
\ee
and at $\eta=0$ and $\psi=\pi$ (south pole), 
\be
\ve^{\,i}_{+\alpha}=  \left(-\i \sigma_3 \exp\left[{\i \frac{(\tau+\phi)}{2}\,  \sigma_3}\right]\right)^i{}_{\alpha}\,,~~~~~~~~~~\ve^{\,i}_{-\dot\alpha}= 0\,.
\ee
Therefore, a representation of~$SU(2)_+ \times SU(2)_{-}\times SU(2)_R$ is twisted to the representation of $SU(2)_{+}\times SU(2)_{-R}$ and
$SU(2)_{+R}\times SU(2)_{-}$, at the north pole and the south poles, respectively. 
Here we denote the diagonal of~$SU(2)_{\pm} \times SU(2)_R$ as~$SU(2)_{\pm R}$. 
We can now compute the trace in the numerator of~\eqref{ABFormula}
for an arbitrary representation~$(m,n)$ at the north pole and south pole according to the \eqref{TrNP} and the \eqref{TrSP}.
\vspace{0.2cm}

\ndt {\bf Vector multiplet}
\vspace{0.1cm}

The twisted representation of the cohomological variables of the vector multiplet can be simply read off from 
the representation labels in Table~\ref{TwistedVectorcharges}. 
{\begin{table}[h] 
\begin{center}
\begin{tabular}{|c|c|}
\hline
\multirow{ 2}{*}{Elementary boson/fermion}&NP:~$SU(2)_+ \times SU(2)_{-R}$ rep\\
~&SP:~$SU(2)_{+ R} \times SU(2)_{-}$ rep \\
\hline\hline
$\widetilde{A}_\mu$&  $(2,2)$ \\
$\widetilde{X}_2$ &  $(1,1)$ \\
\hline
$\lambda^{ij}$ &  $(1,3)$\,\text{at NP}/ (3,1)\,\text{at SP} \\
$c$  & $(1,1)$\\
$b$  & $(1,1)$\\
\hline
\end{tabular}
\caption{The twisted representation labels of the elementary bosons and fermions of the vector multiplet.}
\label{TwistedVectorcharges}
\end{center}
\end{table}}
We have that the charges of the fields in $\Phi$ are~$(1,1)$, $(2,2)$, and those of~$\Psi$ are  
$(1,3)$, $(1,1)$, $(1,1)$ at the north pole and $(3,1)$, $(1,1)$, $(1,1)$ at south pole. Therefore the index is 
\be
{\rm ind}(D_{10}) \= 2 \, \frac{2q+2q^{-1}-4}{(1-q)^2(1-q^{-1})^2} \,.
\ee
From Equation~\eqref{index2det}, we obtain that the~$a_0^\text{vec, bulk}=-1/12$ for the vector multiplet. 
The only potential boundary contribution to the vector multiplet comes from the 1-form field, 
for which $\beta_\text{1-form}=1$ (see Appendix~\ref{App:bdrymodes}). This implies that~$a_0^\text{vec, bdry}=0$, and 
Equation~\eqref{a0bulkbdry} now yields
\be
a_0^{\text{vec}} \= -\frac{1}{12} \,.
\ee

\ndt {\bf Weyl multiplet}
\vspace{0.1cm}

Similarly the twisted representation of the cohomological variables of the Weyl multiplet can be read off from 
the representation labels in Table~\ref{TwistedWeylcharges}. 
{\begin{table}[h] 
\begin{center}
$\begin{tabular}{|c|c||c|c|}
\hline
\multirow{2}{*}{$\Phi$} & NP:~$SU(2)_+ \times SU(2)_{-R}$ rep & \multirow{2}{*}{$\Psi$} & NP:~$SU(2)_+ \times SU(2)_{-R}$ rep  \\
&SP:~$SU(2)_{+R} \times SU(2)_{-}$ rep&&SP:~$SU(2)_{+R} \times SU(2)_{-R}$ rep\\
\hline\hline
$\widetilde{e}_{\mu}^{\,a}$  & $(3,3)+(1,3)+(3,1)+(1,1)$ & $\psi_{\mu}$ & $(2,2)$\\
$\widetilde{A}_{\mu}^R$  & $(2,2)$ & $\psi_\mu^{ij}$ & $(2,4)$ at NP/$(4,2)$ at SP + $(2,2)$  \\
$\widetilde{A}_{\mu}^D$   &   $(2,2)$& $\chi$ & $(1,1)$ \\
 $\widetilde{T}^{+/-}_{ab}$at NP/SP& $(1,3)$ at NP/$(3,1)$ at SP & $c_\mu$ & $(2,2)$ \\
$c_Q$ & $(1,1)$ & $c_{M}^{ab}{}$ 
& $(1,3)+(3,1)$ \\ 
 $c_Q^{ij}$   &  $(1,3)$ at NP/$(3,1)$ at SP  & $c_D$ & $(1,1)$ \\
$b_{Q}$  & $(1,1)$ & $b_R$ &   $(1,1)$ \\
$b_{Q}{}_\mu$  & $(2,2)$ & $b_R{}^i_j$ &  $(1,3)$ at NP/$(3,1)$ at SP   \\
$b_{Q}{}^{ij}$ &  $(1,3)$ at NP/$(3,1)$ at SP  & $b_\mu$ &  $(2,2)$ \\
$b_{S}$  & $(1,1)$ & $b_D$ & $(1,1)$ \\
$b_{S}{}_\mu$  & $(2,2)$ & $b_K^a$  & $(2,2)$ \\
$b_{S}{}^{ij}$ & $(1,3)$ at NP/$(3,1)$ at SP  & $b_{M}^{ab}$ & $(1,3)$ + $(3,1)$
\\
\hline
\end{tabular}$
\caption{The twisted representation labels of the elementary bosons and fermions of the Weyl multiplet.}
\label{TwistedWeylcharges}
\end{center}
\end{table}}
Based on these charges, the index is
\be
{\rm ind}(D_{10}) \= \frac{2(q^2+q^{-2})-6 (q+q^{-1})+8}{(1-q^{-1})^2(1-q)^2}\times 2\,.
\ee
Using Equation~\eqref{index2det}, we see that the bulk contribution to the one-loop determinant \eqref{def1loop} is  governed by
\be
a_0^{\text{Weyl, bulk}} \= \frac{11}{12} \,.
\ee
The boundary contribution comes from the graviton for which~$\nbd^\text{grav}=-6$ and~$\beta_\text{grav}=2$, and the gravitini 
for which~$\nbd^\psi =-8$ and~$\beta_\psi =3$ (see Appendix~\ref{App:bdrymodes}). Putting all this together we obtain, from Equation~\eqref{a0bulkbdry},
$a_0^{\text{Weyl, bdry}} =1$, and therefore
\be
a_0^{\text{Weyl}} \= \frac{23}{12} \,, 
\ee
which is consistent with the on-shell computations~\cite{Sen:2011ba}.

\section{Outlook and speculations \label{sec:Outlook}}

We hope that this work brings some clarity to the idea of twisting and localization in supergravity, and that it 
may be useful in other directions. We briefly list some interesting directions that we think it may be related to.
\begin{enumerate}
\item \emph{Observables of quantum supergravity}. 
Our underlying assumption throughout this calculation is that there is a UV complete theory (like string theory) 
for which we can write an effective action
that commutes with a cutoff, with which we perform localization. This effective action is a formal object as it can contain an 
infinite number of terms with arbitrary derivatives. The results~\cite{deWit:2010za,Butter:2014iwa,Murthy:2013xpa} 
allows us to reduce the problem to 
a more controllable problem of an (infinite) series of F-terms. We can thus regard the right-hand-side 
of~\eqref{LocIntegral} and its non-perturbative completion 
as a \emph{definition} of the functional integral. 
With this viewpoint, we have a good definition for the class of observables in the $\qeq$-cohomology for 
any off-shell supergravity. The details of the functional integral measure remain to be worked out---in this regard 
our BRST procedure may be useful, as the measure should also be BRST-invariant.
\item \emph{Integers from supergravity}. Perhaps the most remarkable feature of the localization of quantum black hole 
entropy is the fact that one gets the integer degeneracies starting from a continuum calculation. The smooth localization 
configurations capture the summed-up perturbation series~~\cite{Dabholkar:2010uh, Dabholkar:2011ec}, and the orbifold 
configurations~\cite{Banerjee:2008ky, Murthy:2009dq, Dabholkar:2014ema} make up the remaining bit of the integer 
degeneracies. This suggests that 
our continuum results could be really some invariants of the $AdS_2 \times S^2$ manifold (with a dependence on the 
prepotential~$F$) that is computed by the twisted supergravity.
The results about the positivity of black hole degeneracies~\cite{Sen:2009vz, Dabholkar:2010rm, Bringmann:2012zr}, 
further suggests that this may actually be a counting problem. 
\item \emph{Quantum Black Hole entropy}. The OSV conjecture~\cite{Ooguri:2004zv} 
promoted the semi-classical observations of~\cite{LopesCardoso:1998wt} to a bold quantum statement
relating the microscopic black hole degeneracies and the topological string partition function~$Z_\text{BH}=|Z_\text{top}|^2$.
In the last ten years, we have begun to understand this equation as relating the microscopic and macroscopic 
computations of black hole entropy as a function of black hole charge (with \emph{a priori} different definitions):
\be \label{micmac}
Z_\text{BH}^\text{micro} (\vec{q}) \=Z_\text{BH}^\text{macro} (\vec{q}) \,.
\ee
The results of this paper suggest that both sides can be thought of as topological invariants (presumably the same!)
computed at different points in moduli space.
\item \emph{Relation to automorphic forms}. The left-hand side of Formula~\eqref{micmac} reduces to an (indexed) counting problem
in string theory. To see that the right-hand side is an integer is more difficult. In the cases where we do understand it, 
the integer
appears though an intricate relation to automorphic forms and analytical formulas for their Fourier 
coefficients~\cite{Dabholkar:2011ec,Murthy:2015zzy,Dabholkar:2014ema}, thus underlining their importance.
\item \emph{Twisted supergravity}. In this paper we construct the variables and transformation rules of 
twisted supergravity around a non-trivial supersymmetric background. 
The observables of the theory are in the cohomology of the operator~$\qeq$ that obeys the equivariant algebra. 
One could regard this theory as a generalization of the pure topological gravity studied in~\cite{Witten:1988xi}. 
One interesting difference with~\cite{Witten:1988xi} is that the action of our twisted theory contains an infinite 
number of higher-derivative terms, and can be thought of as capturing a protected sector of the full string theory.  
\item \emph{Exact~$AdS/CFT$}. 
The formula~\eqref{micmac} is of course the special case~$d=1$, using Sen's quantum 
entropy function~\cite{Sen:2008vm}, of the equality~$Z_{\text{CFT}_{d}}=Z_{\text{AdS}_{d+1}}$. 
It should be clear that our construction of~$\qeq$ applies equally well in any dimension. 
We hope that the ideas of this paper contribute to the understanding of 
an exact sector of $AdS_{d+1}/CFT_{d}$ holography, in which we can compute exact quantities using 
supersymmetry on both sides of the correspondence, and directly relate them. This idea has been 
recently discussed in the context of classical gravitational theories in~\cite{BenettiGenolini:2017zmu}, and in the 
context of topological worldsheet string theory in~\cite{Brennan:2017rbf}. Here we have a 
third angle on the story with a quantum bulk spacetime description, which may serve as another 
example of a ``missing corner'' of string theory in the sense of~\cite{Brennan:2017rbf}.  
\end{enumerate}


\section*{Acknowledgements}

We thank Atish Dabholkar, Bernard de Wit, Camillo Imbimbo, Jo\~ao Gomes, Rajesh Gopakumar, Rajesh Gupta, Sunil Mukhi,
Boris Pioline, Valentin Reys, Ashoke Sen for many interesting and useful discussions related to the topics discussed in this paper. 
This work was supported by the ERC Consolidator Grant N.~681908, ``Quantum black holes: A macroscopic window into the 
microstructure of gravity''.

\appendix

\section{Gamma matrices and spinors}\label{Gamma}
In Euclidean four dimensions we use the following gamma matrix conventions,
\be\ba{lll}
\gamma_{a}^{\dagger}=\gamma_{a}\,,~&~\\
\gamma_{a}^{*}=\gamma_{a}^{T}= C\gamma_{a}C^{-1}\,,~&~C^{\dagger}=C^{-1}\,,~&~C^{T}=-C\,~\Leftrightarrow~C^{*}C=-1\,,\\
\ea\label{Egamma}\ee 
with chirality operator
\be\label{g5}
\gamma_{5}:=\gamma_{1234}\,.
\ee

The Weyl condition of the spinors is compatible with the symplectic Majorana condition such that
\be
(\psi_\pm^{i})^{\dagger}=\bar\psi_{i \pm }\,,
\label{SMW}\ee
where the barred spinor with lower $SU(2)$ index $i$ is defined as the symplectic Majorana conjugate
\be
\bar\psi_{i \pm }:=\epsilon_{ij}\psi_\pm^{jT}C\,,
\label{SMconjugate}\ee 
and the subscript $\pm$ means chiral and anti-chiral projection of the spinors. \\

\ndt {\bf Useful relations}
\be
(C\gamma_{1\cdots n})^{T}= -(-)^{n(n-1)/2}C\gamma_{1\cdots n}\,.
\ee
\be
(C\gamma_{5})^{T}= -C\gamma_{5}\,.
\ee
For two symplectic Majorana spinors $\ve^i$ and $\eta^i$, \footnote{We use convention $(\theta_1 \theta_2)^\ast= \theta_1^\ast \theta_2^\ast$ for two grassmann numbers. If we want to use $(\theta_1 \theta_2)^\ast= \theta_2^\ast \theta_1^\ast$ and keep the reality \eqref{bispinorReality} and \eqref{bispinorReality2}, then we can use the symplectic Majorana condition $\i (\psi^{i})^{\dagger}= \bar\psi_{i  }$ instead of~\eqref{SMW}.}
\be\label{bispinorReality}
(\bar\eta_j \gamma_{a_1 a_2 \cdots a_n}\ve^i )^\ast = \epsilon_{ik}\epsilon^{jl}(\bar\eta_l \gamma_{a_1 a_2\cdots a_n}\ve^k)\,,
\ee
\be\label{bispinorReality2}
(\bar\eta_i \gamma_{a_1 a_2 \cdots a_n}\ve^i )^\ast = \bar\eta_i \gamma_{a_1 a_2\cdots a_n}\ve^i\,.
\ee
For the grassmann odd spinors,
\be
\bar\eta_j \gamma_{a_1 a_2 \cdots a_n}\ve^i= -(-1)^{n(n-1)/2}\epsilon_{jk}\epsilon^{il}\,\bar\ve_l \gamma_{a_1 a_2 \cdots a_n}\eta^k\,,
\ee
or for the grassmann even spinors,
\be\label{bispinorEven}
\bar\eta_j \gamma_{a_1 a_2 \cdots a_n}\ve^i= (-1)^{n(n-1)/2}\epsilon_{jk}\epsilon^{il}\,\bar\ve_l \gamma_{a_1 a_2 \cdots a_n}\eta^k\,.
\ee
The \eqref{bispinorEven} is followed by examples,
\be
\bar{\eta}_i \gamma_{a_1 \cdots a_n}\ve^i =(-1)^{n(n-1)/2} \bar\ve_i \gamma_{a_1 \cdots a_n}\eta^i\,,
\ee
\be
\bar\ve_i\gamma_{ab}\ve^i =0= \bar\ve_i \gamma_{abc}\ve^i \,,
\ee
\be
\bar\ve_i \ve^j = \frac{1}{2}\delta^j_i \bar\ve_k \ve^k\,,~~~~\bar\ve_i \gamma_a\ve^j = \frac{1}{2}\delta^j_i \bar\ve_k \gamma_a\ve^k\,.
\ee
For the choice of $\gamma_5$ in \eqref{g5}, 
\be
\gamma_a\gamma_5= \frac{1}{3!}\epsilon_{abcd}\gamma^{bcd}\,,~~~~\gamma_{ab}\gamma_5 = -\frac{1}{2}\epsilon_{abcd}\gamma^{cd}\,,~~~~\gamma_{abc}\gamma_5= -\epsilon_{abcd}\gamma^d\,,
\ee
which is followed by 
\be
T^{ab\pm}\gamma_{ab}\ve_{\pm}= 0\,,
\ee
where $T^{ab\pm}= \pm\half \epsilon^{abcd}T_{cd}^{\pm}$.

\section{Four-dimensional Euclidean $\cN=2$ supergravity \label{ConfGravity}}
In this appendix, we review the  off-shell Euclidean $d=4$ and $\cN =2$ supergravity. The $\cN=2$ supergravity 
was formulated as a gauge theory via the so-called superconformal calculus \cite{deRoo:1980mm,deWit:1980tn,deWit:1984px}.  
In particular, the Euclidean four dimensional supergravity was recently constructed in~\cite{deWit:2017cle} by 
performing time-like dimensional reduction of 5-dimensional supergravity. In the following subsections, 
we will present the superconformal algebra, and briefly review the superconformal construction with the 
Weyl multiplet and  the vector multiplets.
In the last subsection, we will present the relation to the Minkowskian 
4-dimensional supergravity.\footnote{Our presentation will follow the convention in appendix \ref{Gamma}.  
The difference from~\cite{deWit:2017cle} is that while we use the charge conjugation matrix 
satisfying~\eqref{Egamma} and the symplectic Majorana condition by~\eqref{SMW},  the~\cite{deWit:2017cle} 
uses the charge conjugation matrix satisfying $\gamma_a^{T}=-C\gamma_a C^{-1}$ and the symplectic 
Majorana condition by~$\bar{\psi}_{i}= -\e_{ij}(\psi^j)^T C$ where $\bar{\psi}_i := (\psi^i)^\dagger$. 
From what we present in this section, we can easily recover the results of~\cite{deWit:2017cle} by  changing 
$C\rightarrow -C\gamma_5$, i.e. by replacing $\bar{\psi}_i \rightarrow \bar{\psi}_i \gamma_5$ for any 
spinor $\psi^i$, and redefining the $S$-symmetry parameter~$\eta^i \rightarrow -\i \gamma_5 \eta^i$. 
These changes force us to use $(\bar{\eta}_j \gamma_{a_1 \cdots a_n}\ve^i)^\ast 
= -(-1)^n \e_{jk}\e^{il}\bar{\eta}_l \gamma_{a_1 \cdots a_n}\ve^k$ instead of~\eqref{bispinorReality} 
for two grassmann odd spinors.}

\subsection{Superconformal algebra}\label{superconfalg}
The superconformal algebra for  $d=4$ and $\cN=2$ is composed of the coformal symmetries, $P_a\,, M_{ab}\,, D\,,K_a$, supersymmetries, $Q^i\,,S^i\,,$ R-symmetries, $ SO(1,1)_R\,,SU(2)_R$, and possible central symmetry $Z$.
The symmetry transformations are
\be
\delta= \xi^a P_a + \ve^{ab}M_{ab}+ \Lambda_D D+  \Lambda_{K}^a K_a +{\bar\varepsilon}_{i}Q^i +{\bar\eta}_{i} S^i+ \Lambda_V{}^{i}{}_{j}V^j{}_{i} + \Lambda_A A +\i a Z\,.
\ee
The conformal algebra is
\begin{eqnarray}
&&[P_a\,, M_{bc}]= P_{[b} \eta_{c]a}\,,~~~~~~~~[M_{ab}\,,M_{cd}]= 2\eta_{[a[c}M_{b]d]}\,,  \\~~
 &&[K_a\,,M_{bc}]= K_{[b}\eta_{c]a}\,,~~~~~~~[P_a\,,K_b]= 2(\eta_{ab}D-2 M_{ab})\,, \\~
&&[D\,, P_a]= P_a\,,  ~~~~~~~~~~~~~~~~[D\,, K_a]= -K_a\,.
\end{eqnarray}
The commutators with supercharges are
\begin{eqnarray}
&&[M_{ab}\,, Q^i]= \frac{1}{4} \gamma_{ab}Q^i \,,  ~~~~~~~[M_{ab}\,, S^i]= \frac{1}{4}\gamma_{ab}S^i\,,\\
&&[V_\Lambda\,, Q]^i = \i (\sigma_\Lambda)^i{}_{j}Q^j\,,~~~~~~[V_\Lambda\,, S]^i = \i (\sigma_\Lambda)^i{}_{j}S^j\,,\\
&&[D\,,Q^i]=\half Q^i \,, ~~~~~~~~~~~~~~ [D\,,S^i]=-\half S^i \,,  \\
&&[A\,,Q^i]= \half \gamma_5Q^i \,, ~~~~~~~~~~~~[A\,,S^i]= -\half \gamma_5S^i\\
&&[K^a\,, Q^i]= \gamma_a \gamma_5S^i\,,~~   ~~~~~~~~ [P_a\,,S^i]= -\half \gamma_a \gamma_5Q^i\,,
\end{eqnarray}
Anticommutatiors are
\begin{eqnarray}
&\{Q^i\,, \overline{Q}_j\}= (\gamma^a P_a +\i Z) \delta^i _j\,,\\
& \{S^i\,, \overline{S}_j\}=\gamma^a K_a \delta^i_j\,,\\
& \{Q^i \,, \overline{S}_j\}=-\delta^i_j(\gamma^{ab}M_{ab}+D -\gamma_5 A) + 2V^{i}{}_{j}&\,.
\end{eqnarray}

\subsection{Weyl multiplet}\label{weylmultiplet}
The starting point is to construct superconformal gauge theory by promoting all the ${\cN=2}$ superconformal symmetries as local symmetries. The corresponding gauge fields and the symmetry parameters for each symmetry generators are listed in the table \ref{connection}.
The generic gauge field~$h^\a_\mu$ transforms under a generic gauge transformation with parameter~$\epsilon^{\a}$ as:
 \be
 \delta(\epsilon)h_{\mu}^\a \= \partial_\mu \epsilon^\a + \epsilon^\g\, h_{\mu}^{\,\b}\,\widetilde{f}_{\b\g}{}^{\a}\,,
 \label{GaugefieldTransf}
 \ee
 where $\widetilde{f}_{\b\g}{}^{\a}$ is the structure constant for the superconformal symmetries. 
 
{\scriptsize{\begin{table}[h]
\begin{center}
\begin{tabular}{c|cccccccc}
\hline
generator $T$~&~$P^{a}$&$M^{ab}$ &$D$&$K^{a}$&$Q^{i}$&$S^{i}$&$(V_{\Lambda})^{i}{}_{j}$&$A$ \\

Connection $h_{\mu}(T)$&~$e_{\mu}{}^{a}$&$\omega_{\mu}^{ab}$ &$A^D_{\mu}$&$f_{\mu}^{a}$&$\half\psi_{\mu}{}^{i}$&$\half \phi_{\mu}^{i}$&$-\half\cV_{\mu}{}^{i}{}_{j}$&$A^R_{\mu}$ \\
parameter &~$\xi^{a}$&$\varepsilon^{ab}$ &$\Lambda_{D}$&$\Lambda_{K}^{a}$&$\varepsilon^{i}$&$\eta^{i}$&$\Lambda_{V}{}^{i}{}_{j}$&$\Lambda_{A}$ \\
\hline

\end{tabular} 
\caption{ Table of superconformal gauge fields and transformation parameters  
}
\label{connection}
\end{center}
\end{table}}}

At this stage, the gauge fields are all independent fields. For the supergravity interpretation, the relation between them should be obtained  by imposing ``conventional constraint'', which we present in \eqref{ConvConst}. This determines $\omega_\mu^{ab}\,,\phi_\mu^i$ and $f_\mu^a$ in terms of the other fields, so that $e_\mu^a$ and~$\psi_\mu^i$ become the vielbein and the gravitini respectively. The constraints read to representing the translation $P_a$ as `covariant general coordinate transformation' \footnote{Note that $\xi^a$ is the symmetry parameter of the covariant general coordinate transformation and the $\xi^\mu$ is composite of the parameter and the inverse vielbein.  If we treat $\xi^\mu$ as a parameter, then $\delta_{\text{cgct}}(\xi)$ would not be covariant.}
\be
\xi^a P_a \= \delta_{\text{cgct}}(\xi) \= \delta_{\text{gct}}(\xi^\mu)- \sum_A\delta_A (\xi^\mu h_{\mu}^A)\,, ~~~~~~~~\xi^\mu = \xi^a e_a^{\mu}\,,
\label{CGCT}\ee
where the summation over $A$  denotes  all the gauge symmetries except the translation.
In fact the transformation (\ref{GaugefieldTransf}) of the vielbein $e_\mu{}^{a}$ is equivalent to the covariant general coordinate transformation (\ref{CGCT}) under the conventional constraint. For  non-gauge fields, $P_a$ acts as what we will call the covariant derivative
\be\label{covderiv}
P_a \phi \= D_a \phi \= e_a{}^{\mu}(\partial_\mu \phi- \delta_A (h^A_\mu)\phi ) \,.
\ee
This induces a change in the commutation relations of the original superconformal algebra.  
While translations in the original algebra did commute, now the covariant general coordinate transformation 
do not commute and instead give rise to the curvature:
 \be
 [D_a\,,D_b] \= -\delta_A(\widehat{R}^A_{ab})\,.
 \label{PPcommutation}
 \ee
Thus we see that the structure functions of the algebra are modified.
Using this, 
we can also check that the transformation (\ref{GaugefieldTransf}) with the translation parameter $\xi^a$ of the other gauge fields is 
equivalent to the covariant general coordinate transformation (\ref{CGCT}).

To match the fermionic and bosonic degrees of freedom we add the auxiliary tensor, fermions, and scalar field, $\bigl(T_{ab}^{\pm}\,, \chi^{i}\,, D \bigr)$.  
Thus we get  total $24+ 24$ physical degree of freedoms.  Now the independent fields are \footnote{The gauge field for the dilatation symmetry $D$ is usually denoted by $b_\mu$, but in this paper we use $A^D_\mu$ to avoid confusion with the anti-ghost field for the diffeomorphism. }
\be
\bigl( e_{\mu}^{a}\,, \psi_{\mu}^{i}\,,  A^D_{\mu}\,,  A^R_{\mu}\,, \CV_{\mu \, j}^{\, i}\,; \,T_{ab}^{\pm}\,, \chi^{i}\,, D \bigr) \,.
\ee
This is called the Weyl multiplet.  

{\scriptsize{\begin{table}[h]
\begin{center}
\begin{tabular}{c|cccccccc|ccc|cc}
\hline
~&~$e_{\mu}{}^{a}$&$\psi_{\mu \pm}^{i}$ &$A^D_{\mu}$&$A^R_{\mu}$&$\cV_{\mu}{}^{i}{}_{j}$&$T_{ab}^{\pm}$&$\chi^{i}_\pm $&$D$ &$\omega_{\mu}^{ab}$&$f_{\mu}^{a}$&$\phi_{\mu \pm}^{i}$&$\varepsilon^{i}_\pm$&$\eta^{i}_\pm $\\
\hline
$\omega$&  $-1$&$-\half$&$0$&$0$&$0$&$1$&$\textstyle{\frac{3}{2}}$&2&0&1&$\half$&$-\half$&$\half$\\
$c$  &$0$&$\mp\half$&0&0&0&$\pm 1$&$\mp \half$ &0&0&0&$\pm \half$&$\mp\half$&$\pm\half$ \\
\hline
\end{tabular}
\caption{ Weyl weight $\omega$ and $SO(1,1)_R$ weight $c$ for each the Weyl multiplet component field and supersymmetry parameters. } 
\label{Weyl}
\end{center}
\end{table}}}

The table \ref{Weyl} shows the charges of the Weyl multiplet fields as well as the composite fields and each supersymmetry parameters. 
The auxiliary tensor field satisfies self-dual and 
anti self-dual conditions
\be\label{SELFDUAL}
T_{ab}^{\pm}=\pm \half  \epsilon_{abcd}T^{cd \pm}\,,~~~~~~~~\epsilon^{1234}=1\,.
\ee
And the $SU(2)$ gauge fields $V_{\mu}{}^{i}{}_{j}$ satisfy the anti-hermitian and traceless condition
\be
\cV_{\mu}{}^{i}{}_{j}+ \cV_{\mu j }{}^{i}=0\,,  ~~~~~~~~\cV_{\mu}{}^{i}{}_{i}=0\,,~~~~~~~\mbox{where }~ \cV_{\mu j}{}^{i}:=(\cV_{\mu}{}^{j}{}_{i})^{*}=  -\epsilon_{jk}\cV_{\mu}{}^{k}{}_{l}\epsilon^{li}\,.
\ee
\\
{\bf Conventional constraints}\\
In order to relate $\omega_{\mu}^{ab}\,,\phi_{\mu}^{i}\,,f_{\mu}^{a}$ with other fields, we impose the following constraints,
\begin{eqnarray}
&&R_{\mu\nu}(P)^a\=0\,,\nn\\
&&\gamma^{\mu}(\widehat{R}_{\mu\nu}(Q)^{i} +\frac{1}{2}\gamma_{\mu\nu}\chi^{i})\=0\,,
\\
&&e_{b}{}^{\nu}\widehat{R}_{\mu\nu}(M)_{a}{}^{b}
-\frac{1}{2}\epsilon_{\mu a \lambda\rho}{\widehat{R}}^{\lambda \rho}(A^R)
+\frac{1}{16}T^+_{ab}T_{\mu b}^{-}-\frac{3}{2}D e_{\mu a}\=0\,.\nn
\label{ConvConst}
\end{eqnarray}
Here, the modified field strengths are
\begin{eqnarray} \label{curvatures}
&&\widehat{R}_{\mu\nu}(Q)^i \= 2\cD_{[\mu}\psi_{\nu]}^i + \gamma_{[\mu}\gamma_5 \phi_{\nu]}^i +\i \frac{1}{16}\gamma^{ab}(T^+_{ab}+ T^{-}_{ab})\gamma_{[\mu}\psi_{\nu]}^i \,,\nonumber\\
&&\widehat{R}_{\mu\nu}(A^R)\= 2\partial_{[\mu}A^R_{\nu]}- \frac{1}{2}\bar\psi_{[\mu i}\phi^i_{\nu]}-\frac{3}{4}\bar\psi_{[\mu i}\gamma_{\nu]}\gamma_5 \chi^i \,,\nonumber\\
&&\widehat{R}_{\mu\nu}(\cV)^i{}_{j}\= 2\partial_{[\mu}\cV_{\nu]}{}^i{}_{j}+ \cV_{[\mu}{}^i{}_{k}\cV_{\nu]}{}^k{}_{j} +2 \bar\psi_{[\mu j}\gamma_5 \phi_{\nu]}^i + 3 \bar\psi_{[\mu j}\gamma_{\nu]}\chi^i \nonumber\\
&&\qquad \qquad ~~~~~~~~~-\frac{1}{2}\delta^i_j \left(2 \bar\psi_{[\mu k}\gamma_5 \phi_{\nu]}^k + 3 \bar\psi_{[\mu k}\gamma_{\nu]}\chi^k\right)\,,\\
&&\widehat{R}_{\mu\nu}(M)^{ab}\=2\partial_{[\mu}\omega_{\nu]}^{ab}- 2\omega_{[\mu}^{ac}\omega_{\nu] c}{}^{b} - 4 f_{[\mu}{}^{[a}e_{\nu]}{}^{b} -\frac{1}{2}\bar\psi_{[\mu j}\gamma^{ab} \gamma_5\phi_{\nu]}^j\nonumber\\
&&~~~~~~~~~~ -\i \frac{1}{4} \bar\psi_{[\mu i +}\psi_{\nu] +}^{i} T^{ab + }-\i \frac{1}{4} \bar\psi_{[\mu i -}\psi_{\nu] -}^{i} T^{ab - } -\frac{3}{4} \bar\psi_{[\mu i }\gamma_{\nu]}\gamma^{ab}\chi^i - \bar\psi_{[\mu i}\gamma_{\nu]}\widehat{R}^{ab}(Q)^i\,,\nonumber
\end{eqnarray}
where the $\cD_{\mu}$ is defined as a covariant derivative with respect to $ M, D, A, V$.
Under the conventional constraints, (\ref{ConvConst}), the composite fields are expressed in terms of Weyl multiplet,
\be\ba{lll}
\omega_{\mu}^{ab}&=&-2 e^{\nu[a}\partial_{[\mu}e_{\nu]}{}^{b]}-e^{\nu[a}e^{b]\sigma}e_{\mu c}\partial_{\sigma} e_{\nu}{}^{c}-2 e_{\mu}{}^{[a} e^{b]\nu} A^D_{\nu}\\
&& -\frac{1}{4}(2 {\bar\psi}_{\mu i}\gamma^{[a}\psi^{b]i} + {\bar\psi}^{a}_{i}\gamma_{\mu}\psi^{bi})\,,
\\
\phi^{i}_{\mu}&=& \frac{1}{2}(\gamma^{\rho\sigma}\gamma_{\mu} -\frac{1}{3}\gamma_{\mu}\gamma^{\rho\sigma})\gamma_5(\cD_{\rho}\psi_{\sigma}^{i} +\i\frac{1}{32}\gamma^{ab}(T_{ab}^{+}+T_{ab}^{-})\gamma_{\rho}\psi_{\sigma }^i +\frac{1}{4}\gamma_{\rho\sigma}\chi^{i})\,,
\\
f_{\mu}{}^{a}&=& \half \widehat{R}_{\mu}{}^{a}-\frac{1}{4}(D+\frac{1}{3}\widehat{R})e_{\mu}{}^{a}
-\frac{1}{4}\epsilon_{\mu a \lambda \rho}\widehat{R}^{\lambda \rho}(A^R) 
+\frac{1}{32}T_{\mu b}^{-}T^{ab+}\,,
\label{composite}\ea\ee
where
\be
\widehat{R}_{\mu}{}^{a}=\widehat{R}(M)_{\mu\nu}{}^{ab}e_{b}{}^{\nu}|_{f=0}\,,~~~~~~~~~\widehat{R}=\widehat{R}_{\mu}{}^{a}e_{a}{}^{\mu}\,.
\ee
\\
{\bf The transformation laws and the superconformal algebra}
~\\~
The $Q-S-K-$ transformation rules for the elementary Weyl multiplet fields are
\begin{eqnarray}\label{QSKWeyl}
&&
\delta e_\mu{}^a = \bar\varepsilon_i \gamma^a \psi_{\mu}^i\,,\nonumber\\
&&\delta \psi_\mu^i = 2\cD_\mu \varepsilon^i +\i\frac{1}{16} \gamma_{ab}(T^{ab+}+ T^{ab-})\gamma_\mu  \varepsilon^i + \gamma_\mu \gamma_5 \eta^i\,,\nonumber\\
&& \delta A^D_\mu = -\frac{1}{2}\bar\varepsilon_i \gamma_5 \phi_\mu^i -\frac{3}{4}\bar\varepsilon_i \gamma_\mu \chi^i  -\frac{1}{2} \bar\eta_i \gamma_5 \psi_\mu^i +\Lambda_K^{~a}e_{\mu a}\,,\nonumber\\
&& \delta A^R_\mu = \frac{1}{2} \bar\varepsilon_i  \phi_\mu^i +\frac{3}{4} \bar\varepsilon_i \gamma_\mu \gamma_5\chi^i  +\frac{1}{2} \bar\eta_i  \psi_\mu^i\,, \nonumber\\
&& \delta \cV_{\mu}{}^{i}{}_{j}= -2 \bar\varepsilon_j \gamma_5 \phi_\mu^i  -3 \bar\varepsilon_j \gamma_\mu \chi^i + 2 \bar\eta_j \gamma_5 \psi_\mu^i  - \frac{1}{2}\delta^i_j \left( -2 \bar\varepsilon_k \gamma_5 \phi_\mu^k -3 \bar\varepsilon_k \gamma_\mu \chi^k  +2 \bar\eta_k \gamma_5 \psi_\mu^k \right)\,,\nonumber\\
&& \delta T^{\pm}_{ab}=-\i8 \bar\varepsilon_{i\mp}   \widehat{R}_{ab}(Q)^i_{\mp} \,,\\
&& \delta \chi^i = \i\frac{1}{24}\gamma_{ab}\slashed{D}(T^{ab+}+ T^{ab-})  \varepsilon^i +\frac{1}{6}\widehat{R}(\cV)^i{}_{j \mu\nu}\gamma^{\mu\nu}\varepsilon^j - \frac{1}{3} \widehat{R}(A^R)_{\mu\nu}\gamma^{\mu\nu}\gamma_5 \varepsilon^i \nonumber\\
&&~~~~~~~~~~~~~+ D\varepsilon^i +\i \frac{1}{24}(T^+_{ab}+ T^-_{ab})\gamma^{ab}\gamma_5\eta^i\,,\nonumber\\
&& \delta D= \bar\varepsilon_i \slashed{D}\chi^i\,. \nonumber
\end{eqnarray}
For the composite fields we have:
\be\nonumber
\ba{lll}
\delta \omega_{\mu}{}^{ab}&=&\frac{1}{2}{\bar\varepsilon}_{i}\gamma^{ab}\gamma_5\phi_{\mu}^i+ \i\frac{1}{4}T^{ab+}{\bar\varepsilon}_{i+}\psi_{\mu +}^{j}+ \i\frac{1}{4}T^{ab-}{\bar\varepsilon}_{i-}\psi_{\mu -}^{j}+\frac{3}{4}{\bar\varepsilon}_{i}\gamma_{\mu}\gamma^{ab}\chi^{i} \\
&&+{\bar\varepsilon}_{i}\gamma_{\mu}\widehat{R}^{ab}(Q)^{i}-\frac{1}{2}{\bar\eta}_{i}\gamma^{ab}\gamma_5\psi_{\mu }^i + 2 \Lambda_{K}^{[a}e_{\mu}{}^{b]} \,,
\\
\delta \phi_{\mu}{}^{i}&=&-2 f_{\mu}^{a}\gamma_{a}\gamma_5\varepsilon^{i}-\i\frac{1}{16}\slashed{D}(T_{cd}^{+}+T_{cd}^-)\gamma^{cd}\gamma_{\mu}\gamma_5\varepsilon^{i} \\
&&+\frac{3}{2}\left[  ({\bar\chi}_{j-}\gamma^{a}\varepsilon_+^{j})\gamma_{a}\psi_{\mu +}^{~i}-({\bar\chi}_{j-}\gamma^{a}\psi_{\mu +}^{~j})\gamma_{a}\epsilon_+^{i}  \right]-\frac{3}{2}\left[  ({\bar\chi}_{j+}\gamma^{a}\varepsilon_-^{j})\gamma_{a}\psi_{\mu -}^{~i}-({\bar\chi}_{j+}\gamma^{a}\psi_{\mu -}^{~j})\gamma_{a}\epsilon_-^{i}  \right]\\
&&+\frac{1}{4} \widehat{R}(\cV)_{cd}{}^{i}{}_{j}\gamma^{cd}\gamma_{\mu}\gamma_5\varepsilon^{j}+  \frac{1}{2}\widehat{R}(A^R)_{cd}\gamma^{cd}\gamma_{\mu}\gamma_5\varepsilon^{i}+2 \cD _{\mu}\eta^{i}+ \Lambda_{K}^{a}\gamma_{a}\gamma_5\psi_{\mu}^{i} \,, \\
\delta f_{\mu}^{a}&=& \i\frac{1}{4} {\bar\varepsilon}_{i+}\psi_{\mu +}^{~i}D_{b}T^{ba+}+\i\frac{1}{4} {\bar\varepsilon}_{i-}\psi_{\mu -}^{~i}D_{b}T^{ba-}-\frac{3}{4}e_{\mu}{}^{a}{\bar\varepsilon}_{i}\slashed{D}\chi^{i}-\frac{3}{4}{\bar\varepsilon}_{i}\gamma^{a}\psi_{\mu}^i D\\
&& +{\bar\varepsilon}_{i}\gamma_{\mu}D_{b}\widehat{R}^{ba}(Q)^{i}+\half {\bar\eta}_{i}\gamma^{a}\phi_{\mu }^i+\cD_{\mu}\Lambda_{K}^{a}\,.
\ea\ee
where  the covariant derivative $D_\mu$ is defined as \eqref{covderiv} and $\cD_\mu$ for the covariant derivative with respect to $M\,, D\,,A\,, V$. 
In particular
\begin{eqnarray}\label{Dve}
&&\cD_\mu \varepsilon^i=(\partial_{\mu}- \frac{1}{4}\omega_{\mu ab}\gamma^{ab} +\frac{1}{2}A^D_\mu  + \frac{1}{2}A^R_\mu \gamma_5)\varepsilon^i +\frac{1}{2}\cV_\mu{}^i{}_j \varepsilon^j\,.
\end{eqnarray}
\\
{\bf Supersymmetry algebra}
\be\label{algebra1}
[\delta_{Q}(\varepsilon_{1}), \delta_{Q}(\varepsilon_{2})]=\delta_{\text{cgct}}(\xi)+\delta_{M}(\varepsilon)+\delta_{K}(\Lambda_{K})+\delta_{S}(\eta)+\delta_{\text{gauge}}\,,
\ee
where 
$\delta_{\text{cgct}}(\xi)$ is defined in \eqref{CGCT}, 
 the composite parameters are
\be\ba{lll} \label{parameters1}
\xi^{\mu}&=& 2{\bar\varepsilon}_{2 i}\gamma^{\mu}\varepsilon_{1}^i\,, \\
\varepsilon^{ab}&=&\i \frac{1}{2}{\bar\varepsilon}_{2 i+}\varepsilon_{1+}^{i}T^{ab+ }
+\i \frac{1}{2}{\bar\varepsilon}_{2 i-}\varepsilon_{1-}^{i}T^{ab- }\,, \\
\Lambda_{K}^{a}&=&-\i\frac{1}{2}{\bar\varepsilon}_{2 i +}\varepsilon_{1+}^{i}D_{b}T^{ab+}
-\i\frac{1}{2}{\bar\varepsilon}_{2 i -}\varepsilon_{1-}^{i}D_{b}T^{ab-}
-\frac{3}{2}{\bar\varepsilon}_{2 i}\gamma^{a}\varepsilon_{1}^{i} D\,, \\
\eta^i&=& 3{\bar\varepsilon}_{[2+ j}\varepsilon^{j}_{1]+}\chi_-^{i}-  3{\bar\varepsilon}_{[2- j}\varepsilon^{j}_{1]-}\chi_+^{i} \,,
\ea\ee
and the $\delta_{\text{gauge}}$  in general includes additional abelian, non-abelian or central charge gauge transformations.  

\be\ba{lll} \label{algebra2}
[\delta_{S}(\eta), \delta_{Q}(\varepsilon)]&=&\delta_{M}\left({\bar\ve}_{i}\gamma^{ab}\gamma_5\eta^i\right)
+\delta_{D}\left(-{\bar\ve}_{i}\gamma_5\eta^{i}\right)+\delta_{A}\left({\bar\ve}_{i}\eta^{i}\right)\\
&&+\delta_{V}\left(2\bar\ve_j \gamma_5 \eta^i  - \delta^i_j \bar\ve_k \gamma_5 \eta^k\right)\,,
\ea\ee
\be \label{algebra3}
[\delta_{S}(\eta_{1}), \delta_{S}(\eta_{2})]=\delta_{K}\left(\Lambda_{K}^{a}\right)\,,~~~\mbox{with }
\Lambda_{K}^{a}={\bar\eta}_{2i}\gamma^{a}\eta^{i}_{1}\,,
\ee
\be
[\delta_K(\Lambda_K)\,, \delta_Q(\ve)]= \delta_S(\gamma_5 \gamma_a\ve^i \Lambda_K^a)\,.
\ee
\subsection{Vector multiplets} \label{App:vecmult}
Consider an abelian vector multiplet, which is consist of two  scalars $X$ and  $\bar{X}$,
$SU(2)_R$ doublet fermion $\lambda^{i}$, a vector gauge field $A_{\mu}$ and $SU(2)_R$ triplet auxiliary scalars $Y^{ij}$. 
Here the auxiliary fields  satisfy
\be
Y^{ij}=Y^{ji}\,, ~~~~ Y_{ij}=\epsilon_{ik}\epsilon_{jl}Y^{kl}\,,
\ee
where $Y_{ij }\equiv (Y^{ij})^*$.
The field contents and their charges are listed in the table \ref{vector}.  
{\scriptsize{\begin{table}[h]
\begin{center}
\begin{tabular}{c|ccccc}
\hline
~&~$X$&$\bar{X}$&$\lambda_\pm^{i}$ &$A_{\mu}$&$Y_{ij}$ \\
\hline
$\omega$&~$1$&$1$&$\textstyle{\frac{3}{2}}$ &$0$&$2$\\
$c$ &$-1$&$1$&$\mp\half$ &$0$&$0$\\
\hline
\end{tabular}
\caption{ Weyl weight $\omega$ and  $SO(1,1)_R$ weight $c$  for each vector multiplet component field } 
\label{vector}
\end{center}
\end{table}}}
The supersymmetry variations are:
\be\ba{ll}\label{susyVec}
\delta X&= \i\bar\varepsilon_{i+}\,\lambda_+^{i}\,,\\
\delta \bar{X}&=\i{\bar\varepsilon}_{i-}\,{\lambda}_-^{i}\,,\\
\delta A_{\mu}&= {\bar\varepsilon}_{i}\gamma_{\mu}\lambda^{i}  - 2\i \overline{\varepsilon}_{i+}\psi^i_{\mu +}\bar{X}- 2\i \overline{\varepsilon}_{i -}\psi^i_{\mu -}X\,,\\
\delta Y^{ij}&=-2 \varepsilon^{(i}C\gamma^aD_a{\lambda}^{j)}\,,\\
\ea\ee
\be\ba{ll}\nn
\delta\lambda_+^{ i}&=- 2 \i \gamma^{a}D_{a}X{\varepsilon}_-^{i}-\half \cF_{ab}\gamma^{ab}\varepsilon_+^{i}+Y^{ij}\epsilon_{jk}\varepsilon_+^{k}+2\i X\eta_+^{i}\,,\\
\delta{\lambda}_-^{ i}&= -2 \i \gamma^{a}D_{a}\bar{X}\varepsilon_+^{i}-\half \cF_{ab}\gamma^{ab}{\varepsilon}_-^{i} +Y^{ij}\epsilon_{jk}{\varepsilon}_-^{k}-2\i\bar{X}{\eta}_-^{i}\,,
\ea\ee
where the covariant derivatives are
\be\ba{ll}
D_{\mu}X&= (\partial_{\mu}-A^D_{\mu} +A^R_\mu)X-\i\frac{1}{2}\bar\psi_{\mu i+}\lambda_+^i \,,\\
D_{\mu}\bar{X}&= (\partial_{\mu}-A^D_\mu-A^R_{\mu})\bar{X}  -\i\frac{1}{2}{\bar\psi}_{\mu i-}{\lambda}_-^i\,,\\
D_{\mu}\lambda^{i}_+&=(\partial_{\mu}-\frac{1}{4} \omega_{\mu ab}\gamma^{ab}-\frac{3}{2}A_\mu^D+\frac{1}{2}  A^R_{\mu})\lambda_+^{i}+\half \cV_{\mu }{}^{i}{}_{j}\lambda_+^{j}\,\\
&~~~~ + \i \slashed{D}X{\psi}_{\mu -}^{\,i}  +\frac{1}{4}\cF^{ab}\gamma_{ab}\psi_{\mu+}^{\,i}  -\frac{1}{2}Y^{ij}\epsilon_{jk}\psi_{\mu+}^{\,k} - \i X \phi_{\mu+}^{\,i} \,,\\
D_{\mu}{\lambda}_-^{i}&=(\partial_{\mu}-\frac{1}{4} \omega_{\mu ab}\gamma^{ab}-\frac{3}{2}A_\mu^D-\frac{1}{2}  A^R_{\mu}){\lambda}_-^{i}+\half \cV_{\mu}{}^{i}{}_{j}{\lambda}_-^{j}\,,\\
&~~~~ + \i \slashed{D}\bar{X} {\psi}_{\mu+}^{\,i}  +\frac{1}{4}\cF^{ab}\gamma_{ab}{\psi}_{\mu-}^{\,i}  -\frac{1}{2}Y^{ij}\epsilon_{jk}{\psi}_{\mu-}^{\,k} + \i \bar{X} {\phi}_{\mu-}^i \,,
\ea\ee
and the covariant field strength $\cF_{\mu\nu}$ is defined as
\be\ba{l}
\cF_{\mu\nu}=F_{\mu\nu}-\left(\frac{1}{4} \bar{X}\,T^-_{\mu\nu}
+\frac{1}{4}  X\,{T}^+_{\mu\nu } + \bar\psi_{i[\mu}\gamma_{\nu]}{\lambda}^i- \i\bar{X}\,\bar\psi_{\mu i+}\psi_{\nu+}^{~i}- \i X \,{\bar\psi}_{\mu i-}{\psi}^{~i}_{\nu-}\right)\,,
\label{fieldstrength}\ea\ee
so that its variation is 
\be
\delta \cF_{ab}=- 2 \bar{\ve}_i \gamma_{[a}D_{b]}\lambda^i - \bar{\eta}_i \gamma_{ab}\gamma_5 \lambda^i\,.
\ee
The algebra (\ref{algebra1}) now includes the central charge gauge symmetry with its parameter,
\be
\delta_{\text{gauge}}(a)\,,~~~~
a = -4\i ({\bar\varepsilon}_{2i-}{\varepsilon}_{1-}^{~i} X + \bar\varepsilon_{2 i+}\varepsilon_{1+}^{~i}\bar{X})\,.
\label{centralgauge}\ee

\subsection{Relation to Minkowskian supergravity}\label{offsugra}

In this subsection we  present the relation to Minkowskian supergravity using the analytic continuation. We will relate our supergravity with the one presented in \cite{Mohaupt:2000mj}. Once we make the relation manifest, we can safely utilize a solution obtained in Minkowskian theory as the solution of Euclidean theory.

The complication comes from the fact that theories in different spacetime signatures have different reality properties for their field contents. Particularly for fermions, while 4-dimensional Minkowskian spacetime can allow Majorana and symplectic-Majorana representation yet not compatible with Weyl spinors,  the Euclidean space can allow symplectic Majorana representation which is compatible with Weyl spinors. The reality conditions for other bosonic fields can be set in order to be compatible with supersymmetries, and these are again in general different in Minkowkian and Euclidean theories. Therefore, to map the Euclidean to Minkowskian theory we first need to release the reality properties not imposing Majorana or symplectic Majorana conditions.
   
The Euclidean supergravity presented in this appendix does not contain  complex conjugation. Thus the action invariance and the algebra  are free from what reality condition we would impose.  They are also free under the change of spacetime signature by
\be\label{euclideanize}
x_0^{\scriptscriptstyle M}
= -\i \,x_4^{\scriptscriptstyle E}
\,.
\ee
Once we change the spacetime signature, we can impose a reality condition that is allowed in Minkowskian spacetime.  For generic spinors $\Psi^i$, we can impose the symplectic Majorana condition, which is not compatible with the chirality,  
\be\label{Majorana}
(\Psi_{\pm}^i)^\dagger \gamma_0 = \epsilon_{ij}(\Psi^j_{\mp})^T C\,,
\ee
using the same charge conjugation matrix $C$ that we have used in our Euclidean supergravity. The other bosonic fields will satisfy the reality condition in such a way that it is compatible with supersymmetries.

The resulting Minkowskian supergravity is equivalent to the one presented in~\cite{Mohaupt:2000mj} by following field redefinition. For fermions, starting from our Euclidean supergravity variables, we perform the redefinition as
\be\ba{ll}
\varepsilon^{{\scriptscriptstyle{M}}}_{i} \= +\i \epsilon_{ij} {\varepsilon}_-^{{\scriptscriptstyle E}j}\,,~~~~&\varepsilon^{{\scriptscriptstyle{M}}i}\= \varepsilon_+^{{\scriptscriptstyle E}i}\\
\psi^{\scriptscriptstyle{M}}_{\mu i}\= +\i \epsilon_{ij}{\psi}^{{\scriptscriptstyle E}j}_{\mu -}\,,~~~~~~~~~~&\psi^{{\scriptscriptstyle{M}}i}_{\mu}\= \psi^{{\scriptscriptstyle E}i}_{\mu +}\\
\chi^{\scriptscriptstyle{M}}_{i}\= +\i \epsilon_{ij}{\chi}_-^{{\scriptscriptstyle E}j}\,,~~~~&\chi^{{\scriptscriptstyle{M}}i}\= \chi_+^{{\scriptscriptstyle E}i}\\
\eta^{\scriptscriptstyle{M}}_{i}\= -\i \epsilon_{ij}\eta_+^{{\scriptscriptstyle E}j}~~~~&\eta^{{\scriptscriptstyle{M}}i}\= {\eta}_-^{{\scriptscriptstyle E}i}\\
\phi^{\scriptscriptstyle{M}}_{\mu i}\= -\i\epsilon_{ij}\phi^{{\scriptscriptstyle E}j}_{\mu +}~~~~&
\phi_{\mu}^{{\scriptscriptstyle{M}}i}\= {\phi}^{{\scriptscriptstyle E}i}_{\mu-}\,,\\
\Omega^{\scriptscriptstyle{M}}_{i}\= -\epsilon_{ij}\lambda_+^{{\scriptscriptstyle E}j}~~~~~& \Omega^{{\scriptscriptstyle{M}}i} \= \i \lambda_-^{{\scriptscriptstyle E}i}\,,
\label{Efermion}\ea\ee
together with the redefinition of the charge conjugation matrix as $\tilde{C}= \i C \gamma_5$ to satisfy  $\gamma^T= -\tilde{C}\gamma_a\tilde{C}^{-1}\,, ~~\tilde{C}^T= -\tilde{C}$. Then the symplectic Majorana condition \eqref{Majorana} is converted into the Majorana condition 
\be
\overline\Psi^{\scriptscriptstyle{M}}_i:=(\Psi^{{\scriptscriptstyle{M}}i})^\dagger \gamma_0=( \Psi^{{\scriptscriptstyle{M}}}_i )^T\tilde{C}\,,~~~~\overline\Psi^{{\scriptscriptstyle{M}}i}:=(\Psi^{{\scriptscriptstyle{M}} }_i)^\dagger \gamma_0=( \Psi^{{\scriptscriptstyle{M}}i})^T\tilde{C}\,.
\ee
 For  the bosonic fields, we redefine the abelian $R$-symmetry gauge field as
   \be\label{AE}
  A^{\scriptscriptstyle{M}}_\mu = -\i A^{\scriptscriptstyle{E}}_{\mu}\,.
  \ee
 to reflect that the Minlowskian theory has $U(1)_R$ symmetry while the Euclidean theory has  $SO(1,1)_R$ symmetry. 

For the Minkowskian spacetime, the self-duality condition should be re-expressed (See the appendix of \cite{Banerjee:2009af}).  Since the self-duality relation \eqref{SELFDUAL} is covariant, the same expression could be used after the coordinate change \eqref{euclideanize}. However we note that $1= \e^{1234}= \i\e^{1230}$ with the coordinate change \eqref{euclideanize}, and thus $\e^{0123}= \i$. Therefore, it is better to redefine 
\be
\e^{abcd}= \i \e_{\scriptscriptstyle{M}}^{~abcd}
\ee
such that we set $\e_{\scriptscriptstyle{M}}^{0123}=1$. It is followed by the self-duality condition in Minkowskian spacetime as
\be\label{MSELFDUAL}
T^{{\scriptscriptstyle{M}}\pm ab}=\pm \i \half  \epsilon_{\scriptscriptstyle{M}}^{~abcd}\,T^{{\scriptscriptstyle{M}}\pm}_{cd}\,,~~~~~~~~\epsilon_{\scriptscriptstyle{M}}^{~0123}=1\,.
\ee

 The superconformal algebra presented in appendix \ref{superconfalg} is also converted to the Minkowskian expression. 
 As we redefine  the supersymmetry parameters as (\ref{Efermion}), 
the supercharges are redefined as
\bea
&Q^{{\scriptscriptstyle{M}}i} \= Q^{{\scriptscriptstyle{E}}i}_-\,,~~~~~~&Q^{\scriptscriptstyle{M}}_i \= \i \e _{ij}Q^{{\scriptscriptstyle{E}}j}_+\,,\\
&S^{{\scriptscriptstyle{M}}i} \= S^{{\scriptscriptstyle{E}}i}_+\,,~~~~~~&S^{\scriptscriptstyle{M}}_i \= -\i \e_{ij} S_-^{{\scriptscriptstyle{E}}j}\,,
\eea
such 
that the Euclidean expression of the symmetries
$
\delta= \overline{\ve{\scriptscriptstyle^{E}}}_{i-}Q_-^{{\scriptscriptstyle{E}}i} +\overline{\ve^{\scriptscriptstyle{E}}}_{i+} Q_{+}^{{\scriptscriptstyle{E}}i} + \overline{\eta^{\scriptscriptstyle{E}}}_{i+} S_+^{{\scriptscriptstyle{E}}i} + \overline{\eta^{\scriptscriptstyle{E}}}_{i-} S_{-}^{{\scriptscriptstyle{E}}i}
$
becomes Minkiowskian expression 
$
\delta= \overline{\epsilon^{\scriptscriptstyle{M}}}_iQ^{{\scriptscriptstyle{M}}i} +\overline{\epsilon^{\scriptscriptstyle{M}}}^i Q^{\scriptscriptstyle{M}}_i + \overline{\eta^{\scriptscriptstyle{M}}}_i S^{{\scriptscriptstyle{M}}i} + \overline{\eta^{\scriptscriptstyle{M}}}^i S^{\scriptscriptstyle{M}}_i
$.
 For the abelian $R$-charge, following the redefinition \eqref{AE} the parameter is redefined as 
  $
  \Lambda_A^{\scriptscriptstyle{M}} = -\i \Lambda_A^{\scriptscriptstyle{E}}\,,$ and thus the generator is  redefine as 
\be
A^{\scriptscriptstyle{M}} \= \i A^{\scriptscriptstyle{E}}\,.
\ee
These also recover the algebra presented in~\cite{Mohaupt:2000mj}.


\section{Full transformation rules under~$\qeq$ \label{sec:transformations}}

In this appendix we present the full transformation rules of all the matter and ghost fields under the equivariant
supercharge~$\qeq$. Some of these equations are already present in Section~\ref{sec:Weyl}. 
 
The transformations of the $(b, B)$ ghost fields are:
\be \label{bBtransFull}
\begin{array}{ll}
\qeq b_\mu \= B_\mu \,, &\qquad \qeq B_\mu \=\cL_{\mathring{v}}b_\mu +\partial_\mu \mathring{\ve}^{ab}b_{ab}+\partial_\mu \mathring{\ve}_{D}b_{D}+\partial_\mu \mathring{\ve}_K^{a}b_{Ka}+\partial_\mu \mathring{\ve}_Rb_R+\partial_\mu \mathring{\ve}_R^{ij}b_{Rij} \,,\\
\qeq b_{ab}\= B_{ab}\,,&\qquad \qeq B_{ab}\=\cL_{\mathring{v}}b_{ab}+\mathring{\ve}_{a}{}^{c}b_{cb}+\mathring{\ve}_{b}{}^{c}b_{ac}+\mathring{\ve}_{K[b}b_{K a]}\,,\\
\qeq b_{D}\= B_{D}\,,&\qquad  \qeq B_{D}\= \cL_{\mathring{v}}b_{D}+\mathring{\ve}_{K}^{~a}b_{Ka}\,,\\
\qeq b_{Ka}\= B_{Ka}\,,&\qquad \qeq B_{Ka}\= \cL_{\mathring{v}}b_{Ka}+\mathring{\ve}_{a}{}^{b}b_{Kb}-\mathring{\ve}_D b_{Ka}\,,\\
\qeq b_R\= B_R\,,&\qquad \qeq B_R\= \cL_{\mathring{v}}b_R \,,\\
\qeq b_Q^{~i}\= B_Q^{~i}\,,&\qquad \qeq B_Q^{~i}\= \cL_{\mathring{v}}b_{Q}^{~i}+ \frac{1}{4}\mathring{\ve}^{ab}\gamma_{ab}b_{Q}^{~i} +\mathring{\ve}_R^{~i}{}_{j}b_{Q}^{~j}+\frac{1}{2}\mathring{\ve}_D b_{Q}^{~i}+\frac{1}{2}\mathring{\ve}_R\gamma_5 b_{Q}^{~i}+ \mathring{\ve}_K^{~a}\gamma_a\gamma_5 b_{S}^{~i}\,,\\
\qeq b_S^{~i}\= B_S^{~i}\,,&\qquad \qeq B_S^{~i}\= \cL_{\mathring{v}}b_{S}^{~i}+ \frac{1}{4}\mathring{\ve}^{ab}\gamma_{ab}b_{S}^{~i} +\mathring{\ve}_R^{~i}{}_{j}b_{S}^{~j}-\frac{1}{2}\mathring{\ve}_D b_{Q}^{~i}-\frac{1}{2}\mathring{\ve}_R\gamma_5 b_{S}^{~i}\,.
\end{array}\ee

The transformation rules of $c$ ghost fields are (with $\widetilde{e}_a^{~\mu}:= e_a^{~\mu}- \mathring{e}_{a}^{~\mu}$):
\be\label{cCtransFull}
\begin{array}{ll}
\qeq c^\mu &\= - 2\, \bar{\ve}_i \gamma^a c_Q^i \,\mathring{e}_{a}^{~\mu } - 2\, \bar{\ve}_i \gamma^a c_Q^i \,\widetilde{e}_{a}^{~\mu }  - \bar\ve_i \gamma^a \ve^i \widetilde{e}_a^{~\mu}+c^\nu\partial_\nu c^\mu- \bar{c}_{Qi}\gamma^a c_{Q}^ie_{a}^{\;\mu}\,,\\
 \qeq c^{ab}
&\= -\i \frac{1}{4}{\bar\ve}_{i+ }\ve^{i}_{+ }\widetilde{T}^{ab+}- \i \frac{1}{4}{\bar\ve}_{i- }\ve^{i}_{-}\widetilde{T}^{ab-}
- {\bar\ve}_{i }\gamma^{ab}\gamma_5 c_S^i-\overline{{c}_{Q}}_{ i }\gamma^{ab}\gamma_5 \eta^i- \overline{{c}_{Q}}_{ i }\gamma^{ab}\gamma_5 c_S^i
\\
&\qquad  -\i \frac{1}{2}{\bar\ve}_{i+ }c_{Q+}^{i}T^{ab+}- \i \frac{1}{2}{\bar\ve}_{i- }c_{Q-}^{i}T^{ab-}
   -\i \frac{1}{4}\overline{c_Q}_{i+ }c^{i}_{Q+ }{T}^{ab+}- \i \frac{1}{4}{\overline{c_Q}}_{i- }c^{i}_{Q-}{T}^{ab-}
\\
&\qquad +c^\mu \partial_\mu c^{ab} + ({\bar\ve+\overline{{c}_Q}})_i \gamma^\mu (\ve+c_Q)^i \omega_\mu^{ab}- \bar\ve_i \gamma^c \ve^i \mathring{e}_{c}^{~\mu}\mathring{\omega}_\mu^{ab}+c^{ac}c_{c}{}^{b} \,,
\\
 \qeq c_D
& \=\bar\ve_i \gamma_5 c_S^{\,i}+ \overline{c_Q}_i\gamma_5 \eta^{i} +\overline{c_Q}_i\gamma_5 c_{S}^{\,i} 
+ c^\mu \partial_\mu c_D +(\bar\ve+\overline{{c}_Q})_i \gamma^\mu (\ve+c_Q)^i A_\mu^{D} 
- \bar\ve_i \gamma^c \ve^i \mathring{e}_{c}^{~\mu}\mathring{A}_\mu^{D} \,, \\
\qeq c_Q^i 
& \=- \frac{1}{2}c_R\gamma_5 \ve^i+c_R^{~i}{}_{j}\,\ve^j -\frac{1}{2}c_D \,\ve^i  +\frac{1}{4}c^{ab}\gamma_{ab}(\ve+c_Q)^i  \\
&\qquad +c^\mu \partial_\mu(\ve+ c_Q)^i + \frac{1}{2}(\bar\ve+ \overline{{c}_Q})_j \gamma^\mu (\ve +c_Q)^j \psi_\mu^i- \frac{1}{2}c_R\gamma_5 c_Q^i  +c_R^{~i}{}_{j}c_{Q}^j -\frac{1}{2}c_D c_Q^i \,,
\\
  \qeq c_S^i
  &\=\half \bar{\ve}_j \gamma^a \ve^j \,\phi_a^i+c_R^{~i}{}_{j}\eta^j +\frac{1}{2}c_D \eta^i + \frac{1}{2}c_R\gamma_5 \eta^i+\frac{1}{4}c^{ab}\gamma_{ab}(\eta+c_S)^i
\\
&\qquad +c^\mu \partial_\mu (\eta+c_S)^i + \bar\ve_j \gamma^a c_Q^j\, \phi_a^i + \frac{1}{2}\overline{{c}_Q}_j \gamma^a c_Q^j \,\phi_a^i 
+c_R^{~i}{}_{j}c_{S}^j +\frac{1}{2}c_D c_S^i + \frac{1}{2}c_R\gamma_5 c_S^i\,
\\
&\qquad - \gamma_5\gamma_a (\ve+c_Q)^{i} c_K^a -\frac{3}{2}(\overline{\ve}+\overline{c_Q})_{j+}(\ve+ c_Q)^j_{+}\chi^i_{-}+\frac{3}{2}(\overline{\ve}+\overline{c_Q})_{j-}(\ve+ c_Q)^j_{-}\chi^i_{+}

\\
   \qeq c_R 
      &\=-\bar\ve_i c_{S}^{\,i} -\overline{c_Q}_i \eta^i  + c^\mu \partial_\mu c_R +(\bar\ve +\bar{c}_Q)_i\gamma^\mu (\ve+c_Q)^i A_\mu^R - \bar\ve_i \gamma^a \ve^i \mathring{e}_{a}^{~\mu}\mathring{A}_\mu^R -\overline{{c}_{Q}}_i c_S^i \,,
   \\
\qeq c^i{}_j 
&\= -2\bar\ve_j \gamma_5 c_S^{\,i}- 2\overline{c_Q}_j \gamma_5 \eta^{i } +\delta^i_j ( \bar{\ve}_k\gamma_5 c_S^{\,k}+ \overline{c_Q}_k\gamma_5 \eta^{k}) -2\overline{{c}_Q}_j \gamma_5 c_S^{\,i} +\delta^i_j (\overline{{c}_Q}_k \gamma_5 c_S^k)\\
&\qquad +c^\mu \partial_\mu c^i{}_j -\frac{1}{2}(\bar\ve +\overline{{c}_Q})_k\gamma^\mu (\ve+c_Q)^k \cV_\mu{}^i{}_{j} +\frac{1}{2} \bar\ve_k \gamma^a \ve^k \mathring{e}_{a}^{~\mu}\mathring{\cV}_\mu{}^i{}_{j}  + c^i{}_{k}c^k{}_{j}
\\
\qeq c_K^a 
&\= -\bar\eta_i \gamma^a  c_S^i  -\frac{1}{2}\overline{{c}_S}_i \gamma^a c_S^i \\
&\qquad
+ c^\mu \partial_\mu c_K^a + (\bar\ve+\overline{{c}_Q})_i\gamma^\mu (\ve+c_Q)^i f_\mu^a- \bar\ve_i\gamma^c \ve^i \mathring{e}_{c}^{~\mu} \mathring{f}_\mu^a +c^{a}{}_{b}c_{K}^b + c_D c_K^a\\
&\qquad +\i \frac{1}{4}(\bar\ve +\overline{{c}_Q})_{i+ }(\ve +c_Q)^{i}_{+ }D_bT^{ab+}+ \i \frac{1}{4}(\bar\ve +\overline{{c}_Q})_{i- }(\ve +c_Q)^{i}_{-}D_bT^{ab-}\\
&\qquad-\i \frac{1}{4}\bar\ve_{i+ }\ve^{i}_{+ }\mathring{D}_b\mathring{T}^{ab+}- \i \frac{1}{4}\bar\ve_{i- }\ve^{i}_{-}\mathring{D}_b\mathring{T}^{ab-} +\frac{3}{4}(\bar\ve +\overline{{c}_Q})_i \gamma^a(\ve +c_Q)^i D-\frac{3}{4}\bar\ve_i \gamma^a \ve^i \mathring{D}\,.
\end{array}
\ee

The transformation rules of the Weyl multiplet fields are (with $\wt \cD_\mu\equiv\cD_\mu  - \mathring{\cD}_\mu $ 
and~$T^{ab} \equiv T^{ab+}+T^{ab-}$, $\wt \gamma_{\mu} \equiv \gamma_{a} \, \wt e_{\mu}^{~a}$, $\mathring{\gamma_{\mu}} \equiv \gamma_{a} \,  \mathring{e}_{\mu}^{~a}$): 
\be\nonumber
\begin{array}{ll}
\qeq  \widetilde{e}_{\mu}{}^{a}
	&\= \bar\ve_i \gamma^a \psi_\mu^{\;i}+ c^\nu \partial_\nu e_{\mu}^a+\partial_\mu c^\nu e_{\nu}^a+ c^{ab}e_{\mu b} 
	-c_D e_\mu^a +\overline{{c}_Q}_{i} \gamma^a \psi_\mu^{~i}\,, \\
\qeq \psi_\mu^{~i}&\= 2\cD_{\mu}(\ve+c_{Q})^i + c^\nu \partial_\nu \psi_\mu^{~i} +\partial_\mu c^\nu \psi_\nu^{~i} + \frac{1}{4}c^{ab}
	\gamma_{ab}\psi_{\mu}^{~i}-\frac{1}{2}c_D \psi_{\mu}^{~i}-\frac{1}{2}c_R \gamma_5 \psi_{\mu}^{~i}\\
& \qquad +\;c^i{}_{j}\psi_{\mu}^{~i}+\i \frac{1}{16}  T^{ab} \gamma_{ab}\gamma_\mu (\ve+c_Q)^i 
	+\gamma_\mu \gamma_5 (\eta+c_S)^i\,,\\
&\= 2\wt \cD_{\mu}\ve^i+\mathring{\gamma_\mu} \gamma_5 c_S^i+\i \frac{1}{16}  \gamma_{ab}(T^{ab}\gamma_\mu - \mathring{T^{ab}} \mathring{\gamma_\mu}) \ve^i+
2\cD_{\mu}c_{Q}^i + c^\nu \partial_\nu \psi_\mu^{~i} +\partial_\mu c^\nu \psi_\nu^{~i} \\
& \qquad + \frac{1}{4}c^{ab}
	\gamma_{ab}\psi_{\mu}^{~i} -\frac{1}{2}c_D \psi_{\mu}^{~i}-\frac{1}{2}c_R \gamma_5 \psi_{\mu}^{~i}+\;c^i{}_{j}\psi_{\mu}^{~i} +
 \i \frac{1}{16}  \gamma_{ab} T^{ab}  \gamma_\mu c_Q^i 
	+\wt\gamma_\mu \gamma_5 c_S^i + \wt \gamma_\mu \gamma_5 \eta^{i} \,, \\
\qeq \widetilde{A}_{\mu}^R
	&\=\frac{1}{2}\bar\ve_i\phi_\mu^{~i} +\frac{3}{4}\bar\ve_i \gamma_\mu\gamma_5 \chi^i 
	+\frac{1}{2}\bar\eta_{i} \psi_\mu^{\,i}
	\\
	&\qquad  + c^\nu \partial_\nu A_\mu^{R} +\partial_\mu c^\nu A_\nu^{R} +\partial_\mu c_R
	+\frac{1}{2}\overline{{c}_Q}_i\phi_\mu^{~i} +\frac{3}{4}\overline{{c}_Q}_i \gamma_\mu\gamma_5 \chi^i 
	+\frac{1}{2}\overline{{c}_{S}}_{i} \psi_\mu^{\,i}\,,\\
\end{array}
\ee
\be\label{WeyltransFull}
\begin{array}{ll}
\qeq \widetilde{A}_\mu^D &\=-\frac{1}{2}\bar\ve_i \gamma_5\phi_\mu^{~i} -\frac{3}{4}\bar\ve_i \gamma_\mu \chi^i 
	-\frac{1}{2}\bar\eta_{i}\gamma_5 \psi_\mu^{\,i} + c_{K a} e_{\mu }{}^a
	\\
	&\qquad  + c^\nu \partial_\nu A_\mu^{D} +\partial_\mu c^\nu A_\nu^{D} +\partial_\mu c_D
	-\frac{1}{2}\overline{{c}_Q}_i \gamma_5 \phi_\mu^{~i} -\frac{3}{4}\overline{{c}_Q}_i \gamma_\mu \chi^i 
	-\frac{1}{2}\overline{{c}_{S}}_{i}\gamma_5 \psi_\mu^{\,i}\,,\\
\qeq \chi^i 
	&\= D\varepsilon^i + c^\mu \partial_\mu \chi^i +\frac{1}{4}c^{ab}\gamma_{ab}\chi^i +\frac{3}{2}c_D \chi^i 
	-\frac{1}{2} c_R \gamma_5 \chi^i +c^i{}_{j}\chi^j\\
&\qquad  +\i\frac{1}{24}\gamma_{ab}\slashed{D}(T^{ab+}+ T^{ab-})  (\ve+ c_Q)^i 
	+\frac{1}{6}\widehat{R}(\cV)^i{}_{j \mu\nu}\gamma^{\mu\nu}(\ve+ c_Q)^j  \\
&\qquad - \frac{1}{3} \widehat{R}(A^R)_{\mu\nu}\gamma^{\mu\nu}\gamma_5 (\ve+c_Q)^i
	+ D \,c_Q^{\,i} +\i \frac{1}{24}(T^+_{ab}+ T^-_{ab})\gamma^{ab}\gamma_5 (\eta+c_S)^i\,,\\
\qeq \widetilde{T}^{\pm}_{ab}
	&\= - 8\i \bar\ve_{i\mp}\widehat{R}_{ab}(Q)^i_{\mp}\\
	&\qquad+ c^\mu \partial_\mu T^{\pm}_{ab}+ c_a{}^c T^{\pm}_{cb}+ c_{b}{}^{c}T^{\pm}_{ac}
	+c_D T^\pm_{ab}\pm c_R T^{\pm}_{ab} -8\i \overline{{c}_Q}_{i\mp} \widehat{R}_{ab}(Q)^i_{\mp}\,,\\
	&\= - 8\i \bar\ve_{i\mp}\gamma_{[a}\gamma_5 \phi_{b]}^i - 8\i \bar\ve_{i\mp}\left( 2e_{[a}^{\,\mu}e_{b]}^{\,\nu}\cD_{\mu}\psi_{\nu}^{\;i}  +\i \frac{1}{16}\gamma^{cd}(T_{cd}^+ +T_{cd}^-)\gamma_{[a}\psi_{b]}^{\;i}\right)\\
	&\qquad+ c^\mu \partial_\mu T^{\pm}_{ab}+ c_a{}^c T^{\pm}_{cb}+ c_{b}{}^{c}T^{\pm}_{ac}
	+c_D T^\pm_{ab}\pm c_R T^{\pm}_{ab} - 8\i \overline{{c}_Q}_{i\mp} \widehat{R}_{ab}(Q)^i_{\mp}\,,
	\\
\qeq \widetilde{\cV}_{\mu}{}^i{}_{j}
	&\= -2 \bar\ve_j \gamma_5 \phi_\mu^{~i} -3 \bar\ve_j \gamma_\mu \chi^i + 2 \bar\eta_j\gamma_5 \psi_\mu^{\,i} -\frac{1}{2}\delta^i_j \bigl[ -2 \bar\ve_k \gamma_5 \phi_\mu^{~k} -3 \bar\ve_k \gamma_\mu \chi^k 
	+ 2 \bar\eta_k\gamma_5 \psi_\mu^{\,k} \bigr] \\
	&\qquad -2 \overline{c_Q}_j \gamma_5 \phi_\mu^{~i} -3 \overline{c_Q}_j \gamma_\mu \chi^i + 2 \overline{c_S}_j\gamma_5 \psi_\mu^{\,i} -\frac{1}{2}\delta^i_j \bigl[ -2 \overline{c_Q}_k \gamma_5 \phi_\mu^{~k} -3 \overline{c_Q}_k \gamma_\mu \chi^k 
	+ 2 \overline{c_S}_k\gamma_5 \psi_\mu^{\,k} \bigr] \\
	& \qquad +c^\nu \partial_\nu \cV_\mu{}^{i}{}_{j} +\partial_\mu c^\nu \cV_\nu{}^{i}{}_{j}
	-2 \partial_{\mu}c^{i}{}_{j} -2 c^{i}{}_{k}\cV_{\mu}{}^{k}{}_{j}+2\cV_{\mu}{}^{i}{}_{k}c^{k}{}_{j}\,,\\
\qeq \widetilde{D}&\=  (\bar\varepsilon+\overline{{c}_Q})_i \slashed{D}\chi^i+c^\mu \partial_\mu D \,,
\end{array}
\ee
Here the covariant derivative~$\cD_\mu\ve^i$ is given in~\eqref{Dve}, the curvatures $\widehat{R}_{\mu\nu}(Q)^i$, $\widehat{R}_{\mu\nu}(A^R)$ $\widehat{R}^i{}_{j \mu\nu}(\cV)$ are in\eqref{curvatures}
and the composite field $\phi_\mu^{\,i}$ is in \eqref{composite}.

\section{$AdS_2\times S^2$ and the Killing spinor}{\label{KillingS}}
The Euclidean  $AdS_{2} \times S^{2}$ configuration in unit radius  is  
\bea \label{metric}
&& ds^2 \= \bigl(d\eta^2 + \sinh^2\eta \, d\tau^2 \bigr) + \bigl(d\psi^2 + \sin^2\psi \, d\phi^2 \bigr)\, ,\\
&& F^{I}_{ \eta\tau} = -\i \,e^I_* \sinh \eta \, , \quad {F}^I_{\psi\phi} = p^I\sin\psi \, , \quad X^I = X^I_* \, , 
\quad T_{\eta\tau}^- =- \i \, 4 \sinh \eta \, , \nonumber
\eea
with all other fields not related by symmetries set to zero.
Here~$F^I_{\mu\nu}$ is the field strength of the~$U(1)$ vector field in the vector multiplet~$I$ with electric field
and magnetic charge given by~$(e^I_*,p^I)$, respectively. The constant values~$X^I_*$ of the scalar fields are 
given by the \emph{attractor equations}:
\be \label{attraceq}
X^I_* + \bar{X}^I_* \= e^I_* \,, \qquad X^I_* - \bar{X}^I_* \= \i \, p^I \, ,  \qquad   F_I - \bar{F}_I   \= \i \, q_I  \, .
\ee
The killing spinor equations are obtained from the variation of the gravitino. On the $AdS_2 \times S^2$ configuration given by \eqref{metric}, the equation becomes
\begin{eqnarray}\label{KE}
&&D_\mu\ve^{i}= -\frac{1}{2}\gamma_{12}\gamma_{\mu}{\ve}^{i} \,.
\end{eqnarray}
With the choice of gamma matrices,
\be\label{gamma}
\gamma^{1}=\tau_1 \otimes 1\,,~~~\gamma^{2}=\tau_2 \otimes 1\,,~~~\gamma^{3}=\tau_3 \otimes \sigma_1\,,~~~\gamma^{4}=\tau_3\otimes\sigma_2\,,~~~\gamma_5=\gamma_{1234}=-\tau_3\otimes\sigma_3\,,
\ee
the equation \eqref{KE} is  solved by $8$ sets of symplectic Majorana spinors. We choose a set of Killing spinors,
\be
\ve^1 =\frac{1}{\sqrt{2}}e^{\i(\tau+\phi)/2}\begin{pmatrix}  \cosh\frac{\eta}{2}\cos\frac{\psi}{2} \\ -\i \cosh\frac{\eta}{2}\sin\frac{\psi}{2} \\ \i \sinh \frac{\eta}{2}\cos\frac{\psi}{2}\\ \sinh\frac{\eta}{2}\sin\frac{\psi}{2}\end{pmatrix}\,,~~~~~\ve^2=\frac{1}{\sqrt{2}}e^{-\i(\tau+\phi)/2}\begin{pmatrix}  \sinh\frac{\eta}{2}\sin\frac{\psi}{2}\\ \i \sinh\frac{\eta}{2}\cos\frac{\psi}{2} \\ \i \cosh \frac{\eta}{2}\sin\frac{\psi}{2}\\ -\cosh\frac{\eta}{2}\cos\frac{\eta}{2} \end{pmatrix}\,,
\ee
satisfying the symplectic Majorana condition,
\be
(\ve^i)^\ast = -\i \e_{ij}(\tau_1 \otimes \sigma_2) \ve^j\,.
\ee
Then, the corresponding the Killing vector  is
\be
v^\mu\partial_{\mu}= \bar\ve_i\gamma^\mu \ve^i \partial_{\mu}=\partial_{\tau}-\partial_{\phi}\,,
\ee
and  fermionic bilinears are
\be
\bar\ve_i\ve^i= \cosh\eta\,,~~~~~~~~ \bar{\ve}_i\gamma_5\ve^i= -\cos\psi\,. 
\ee

At $\eta=0$ and $\psi=0$ (North Pole), the chiral and anti-chiral part of the Killing spinor reduces to
\be
\ve^i_{+\alpha}=0\,,~~~~~~~~~\ve^{\,i}_{-\dot\alpha}=  \left(\sigma_3 \exp\left[{\i \frac{(\tau+\phi)}{2}\,  \sigma_3}\right]\right)^i{}_{\dot\alpha}\,,
\ee
and at $\eta=0$ and $\psi=\pi$ (South Pole), 
\be
\ve^{\,i}_{+\alpha}=  \left(-\i \sigma_3 \exp\left[{\i \frac{(\tau+\phi)}{2}\,  \sigma_3}\right]\right)^i{}_{\alpha}\,,~~~~~~~~~~\ve^{\,i}_{-\dot\alpha}= 0\,.
\ee
Therefore, the  $SU(2)_R$ symmetry is identified with the inverse of $SU(2)_{-}$ of the rotation symmetry $SO(4)$ at North Pole, and with the inverse of $SU(2)_{+}$ at the South Pole.

\section{Counting the number of boundary modes \label{App:bdrymodes}}
In this appendix, 
we shall count the number of boundary modes for the 1-form field $\wt{A}_\mu$, the graviton $\wt{g}_{\mu\nu}$, 
and the gravitino $\psi_\mu^i$ on $AdS_2 \times S^2$. Denoting a generic field as $\phi$ and its boundary modes as $\phi^{\text{bdry}}$, 
the number of the boundary modes  for each field can be counted using the definition
\be
\nbd^\phi \; := \;  \Tr_{\phi^{\text{bdry}}}e^{tH} \Big{|}_{t^0} \,,
\ee
which we justified in the main text. As we see below, the trace for each field turns out to be regular at~$t=0$ and therefore we can use
\be\label{Nzero}
\nbd^\phi\=\lim_{t\rightarrow 0} \Tr_{ \phi^{\text{bdry}}}e^{tH} \,,
\ee
using the bosonic generator $H$ coming from the equivariant algebra. 
The value of~$\beta$ has been calculated~\cite{Sen:2012cj} for a 1-form field, the graviton, and the gravitino~$\psi$  in four dimensions
to be, respectively, 
\be
\b_\text{1-form} \= 1 \,, \qquad \b_\text{grav} \= 2 \,, \qquad \b_\psi \= 3 \,.
\ee 
Once we calculate~$\nbd$ for each field, we can evaluate the formula~\eqref{Zbeta}.

The number of boundary modes for each field on  $AdS_2\times S^2$ can be counted by decomposing 
the field into various two dimensional fields on $AdS_2$, i.e. $1$-forms, scalars, graviton, and spinors on $AdS_2$, 
and looking at the massless fields among them.  
For example, the 1-form $\wt{A}_\mu$, is decomposed to a vector $v_m$ and two scalars $\phi_p$, where~$\mu$ is the 
four-dimensional index and~$m$ is the two-dimensional~$AdS_2$ index.
As explained in~\cite{Banerjee:2009af,Sen:2012cj}, a more conceptual manner of understanding these boundary modes is to associate them 
with asymptotic symmetries on~$AdS_2$ which can be summarized as the modes of currents. 
In the 1-form example, we should associate a~$U(1)$ current with modes~$j_n$, $n\in \IZ$.
Here~$n$ is the eigenvalue of~$L_0$ which is equal to~$H$ in our formalism. 
The zero mode~$j_0$ is a global symmetry and so we should not count it as a zero mode. Thus we obtain:\\
{\bf $1$-form: Spin-$1$ current ($j_n$)}
\begin{eqnarray}\label{N01-form}
\nbd^\text{1-form}&\=&\lim_{t\rightarrow 0} \sum_{n \in \IZ, \atop n\neq 0}^{ } q^{n} \\
&\=& \lim_{t\rightarrow 0} \left[\frac{q}{1-q} + \frac{q^{-1}}{1-q^{-1}}\right] \=-1 \,,\nn
\end{eqnarray}
where we used the geometric summation over $q$ and $q^{-1}$.

The symmetries associated with the Weyl multiplet are present in every theory and 
are generated by~$L_n$, $n \in \IZ$, $G^\mu_r$, $\mu=1,\cdots,4$ $r\in \IZ + \half$, and~$J^a_n$, $a=1,2,3$, $n \in \IZ$,
which obey the chiral~$\CN=4$ algebra of a two-dimensional SCFT in the NS sector.
The global part of this algebra is generated by~$L_{0,\pm1}$, $G^\mu_{\pm \half}$, $J^a_0$ and these should be not 
be counted as boundary modes. This we obtain:\\
{\bf Graviton: Spin-$2$ current ($L_n$)}
\begin{eqnarray}\label{N0grav}
\nbd^{\text{grav},2}&\=&\lim_{t\rightarrow 0} \sum_{n \in \IZ \atop n\neq 0\,,\pm 1}^{ } q^{n}
\\
&\=&\lim_{t\rightarrow 0}\left[ \sum_{n\neq 0}q^{n} - q -q^{-1}\right]=-3\nn
\end{eqnarray}
{\bf Graviton: Spin-$1$ current ($J^a_n$, $a=1,2,3$)}
\be
\nbd^{\text{grav},1} \= 3 \times (-1) \,,
\ee
where we have used the above calculation of a generic spin-1 (1-form) field.
Note that this spin-1 current is really a part of the graviton and therefore should have the same~$\beta$ as the graviton. 
In the spacetime picture, this can be thought of as the graviton ${g}_{\mu\nu}$,  
decomposed into a graviton $h_{mn}$ and 3 massless vectors $v_m k_a$, where $k_a$ are the three Killing vectors of~$S^2$. 

\ndt The boundary modes of the gravitino are associated with the fermionic currents:\\
{\bf Spin-$3/2$ current ($G^\mu_r$, $\mu = 1,\cdots, 4$)}
\begin{eqnarray}\label{N03/2}
\nbd^{\psi}&\=&4 \times \lim_{t\rightarrow 0} \sum_{r \in \IZ + \half \atop r \neq \pm \half}^{ } q^{r} \\
&\=&4 \times \lim_{t\rightarrow 0}\, q^{-1/2} \left(\frac{q^{2}}{1- q} + \frac{q^{-1}}{1-q^{-1}}\right)\nn\\
&\=&4 \times \lim_{t\rightarrow 0} \,q^{-1/2} (-1- q)\=-8\,.\nn
\end{eqnarray}

To summarize, the final result for the number of zero modes for 1-form, graviton, and gravitino are
\be
\nbd^\text{1-form}=-1 \,,~~~~~\nbd^\text{grav}= -6\,,~~~~~\nbd^\psi= -8 \,.
\ee
These results agree with the results obtained in~\cite{Sen:2012cj} which used a different regularisation scheme suitable to the on-shell analysis.

\bibliographystyle{JHEP}

\begin{thebibliography}{10}

\bibitem{dWMR}
{Bernard de Wit, Sameer Murthy, and Valentin Reys}, ``{BRST quantization and equivariant cohomology: 
localization with asymptotic boundaries}.'' 
[\href{https://arxiv.org/abs/1806.03690}{{\ttfamily 1806.03690}}].

\bibitem{Dabholkar:2010uh}
A.~Dabholkar, J.~Gomes and S.~Murthy, \emph{{Quantum black holes, localization
  and the topological string}},
  \href{https://doi.org/10.1007/JHEP06(2011)019}{\emph{JHEP} {\bfseries 1106}
  (2011) 019}, [\href{https://arxiv.org/abs/1012.0265}{{\ttfamily 1012.0265}}].

\bibitem{Dabholkar:2011ec}
A.~Dabholkar, J.~Gomes and S.~Murthy, \emph{{Localization $\&$ Exact
  Holography}}, \href{https://doi.org/10.1007/JHEP04(2013)062}{\emph{JHEP}
  {\bfseries 1304} (2013) 062},
  [\href{https://arxiv.org/abs/1111.1161}{{\ttfamily 1111.1161}}].

\bibitem{Sen:1995in}
A.~Sen, \emph{Extremal black holes and elementary string states}, {\emph{Mod.
  Phys. Lett.} {\bfseries A10} (1995) 2081--2094},
  [\href{https://arxiv.org/abs/hep-th/9504147}{{\ttfamily hep-th/9504147}}].

\bibitem{Strominger:1996sh}
A.~Strominger and C.~Vafa, \emph{Microscopic origin of the bekenstein-hawking
  entropy}, {\emph{Phys. Lett.} {\bfseries B379} (1996) 99--104},
  [\href{https://arxiv.org/abs/hep-th/9601029}{{\ttfamily hep-th/9601029}}].

\bibitem{Duistermaat:1982vw}
J.~J. Duistermaat and G.~J. Heckman, \emph{{On the Variation in the cohomology
  of the symplectic form of the reduced phase space}},
  \href{https://doi.org/10.1007/BF01399506}{\emph{Invent. Math.} {\bfseries 69}
  (1982) 259--268}.

\bibitem{Berline:1982}
N.~Berline and M.~Vergne, \emph{{Classes caract\'eristiques \'equivariantes.
  {F}ormule de localisation en cohomologie \'equivariante}}, {\emph{C. R. Acad.
  Sci. Paris S\'er. I Math.} {\bfseries 295} (1982) no.~9 539--541}.

\bibitem{Atiyah:1984px}
M.~F. Atiyah and R.~Bott, \emph{{The Moment map and equivariant cohomology}},
  \href{https://doi.org/10.1016/0040-9383(84)90021-1}{\emph{Topology}
  {\bfseries 23} (1984) 1--28}.

\bibitem{Witten:1988ze}
E.~Witten, \emph{{Topological Quantum Field Theory}},
  \href{https://doi.org/10.1007/BF01223371}{\emph{Commun. Math. Phys.}
  {\bfseries 117} (1988) 353}.

\bibitem{Witten:1988xj}
E.~Witten, \emph{{Topological Sigma Models}},
  \href{https://doi.org/10.1007/BF01466725}{\emph{Commun. Math. Phys.}
  {\bfseries 118} (1988) 411}.

\bibitem{Nekrasov:2002qd}
N.~A. Nekrasov, \emph{{Seiberg-Witten prepotential from instanton counting}},
  \href{https://doi.org/10.4310/ATMP.2003.v7.n5.a4}{\emph{Adv. Theor. Math.
  Phys.} {\bfseries 7} (2003) 831--864},
  [\href{https://arxiv.org/abs/hep-th/0206161}{{\ttfamily hep-th/0206161}}].

\bibitem{Baulieu:1988xs}
L.~Baulieu and I.~M. Singer, \emph{{Topological Yang-Mills symmetry}},
  \href{https://doi.org/10.1016/0920-5632(88)90366-0}{\emph{Nucl. Phys. Proc.
  Suppl.} {\bfseries 5B} (1988) 12--19}.

\bibitem{Grisaru:1981xm}
M.~T. Grisaru and W.~Siegel, \emph{{Supergraphity Part 1. Background field
  formalism}}, \href{https://doi.org/10.1016/0550-3213(81)90121-8}{\emph{Nucl.
  Phys.} {\bfseries B187} (1981) 149--183}.

\bibitem{Grisaru:1983rg}
M.~T. Grisaru and D.~Zanon, \emph{{Quantum Superfield Supergravity With
  Off-shell Background Fields}},
  \href{https://doi.org/10.1016/0550-3213(84)90014-2}{\emph{Nucl. Phys.}
  {\bfseries B237} (1984) 32--58}.

\bibitem{deRoo:1980mm}
M.~de~Roo, J.~W. van Holten, B.~de~Wit and A.~{Van Proeyen}, \emph{{Chiral
  superfields in $\mathcal{N}=2$  supergravity}},
  \href{https://doi.org/10.1016/0550-3213(80)90449-6}{\emph{Nucl. Phys.}
  {\bfseries B173} (1980) 175}.

\bibitem{deWit:1980tn}
B.~de~Wit, J.~W. van Holten and A.~{Van Proeyen}, \emph{{Structure of
  $\mathcal{N}=2$ Supergravity}},
  \href{https://doi.org/10.1016/0550-3213(81)90211-X}{\emph{Nucl. Phys.}
  {\bfseries B184} (1981) 77}.

\bibitem{deWit:1984px}
B.~de~Wit, P.~G. Lauwers and A.~{Van Proeyen}, \emph{{Lagrangians of
  $\mathcal{N}=2$ Supergravity - Matter Systems}},
  \href{https://doi.org/10.1016/0550-3213(85)90154-3}{\emph{Nucl. Phys.}
  {\bfseries B255} (1985) 569}.

\bibitem{Berkovits:1993hx}
N.~Berkovits, \emph{{A Ten-dimensional superYang-Mills action with off-shell
  supersymmetry}},
  \href{https://doi.org/10.1016/0370-2693(93)91791-K}{\emph{Phys.Lett.}
  {\bfseries B318} (1993) 104--106},
  [\href{https://arxiv.org/abs/hep-th/9308128}{{\ttfamily hep-th/9308128}}].

\bibitem{Baulieu:2007ew}
L.~Baulieu, N.~J. Berkovits, G.~Bossard and A.~Martin, \emph{{Ten-dimensional
  super-Yang-Mills with nine off-shell supersymmetries}},
  \href{https://doi.org/10.1016/j.physletb.2007.05.027}{\emph{Phys. Lett.}
  {\bfseries B658} (2008) 249--254},
  [\href{https://arxiv.org/abs/0705.2002}{{\ttfamily 0705.2002}}].

\bibitem{Pestun:2007rz}
V.~Pestun, \emph{{Localization of gauge theory on a four-sphere and
  supersymmetric Wilson loops}},
  \href{https://doi.org/10.1007/s00220-012-1485-0}{\emph{Commun. Math. Phys.}
  {\bfseries 313} (2012) 71--129},
  [\href{https://arxiv.org/abs/0712.2824}{{\ttfamily 0712.2824}}].

\bibitem{Baulieu:2012jj}
L.~Baulieu, M.~Bellon and V.~Reys, \emph{{Twisted N=1, d=4 supergravity and its
  symmetries}},
  \href{https://doi.org/10.1016/j.nuclphysb.2012.10.007}{\emph{Nucl. Phys.}
  {\bfseries B867} (2013) 330--353},
  [\href{https://arxiv.org/abs/1207.4399}{{\ttfamily 1207.4399}}].

\bibitem{Bae:2015eoa}
J.~Bae, C.~Imbimbo, S.-J. Rey and D.~Rosa, \emph{{New Supersymmetric
  Localizations from Topological Gravity}},
  \href{https://doi.org/10.1007/JHEP03(2016)169}{\emph{JHEP} {\bfseries 03}
  (2016) 169}, [\href{https://arxiv.org/abs/1510.00006}{{\ttfamily
  1510.00006}}].

\bibitem{Costello:2016mgj}
K.~Costello and S.~Li, \emph{{Twisted supergravity and its quantization}},
  [\href{https://arxiv.org/abs/1606.00365}{{\ttfamily 1606.00365}}].

\bibitem{Imbimbo:2018duh}
C.~Imbimbo and D.~Rosa, \emph{{The topological structure of supergravity: an
  application to supersymmetric localization}},
  \href{https://doi.org/10.1007/JHEP05(2018)112}{\emph{JHEP} {\bfseries 05}
  (2018) 112}, [\href{https://arxiv.org/abs/1801.04940}{{\ttfamily
  1801.04940}}].

\bibitem{Hama:2012bg}
N.~Hama and K.~Hosomichi, \emph{{Seiberg-Witten Theories on Ellipsoids}},
  \href{https://doi.org/10.1007/JHEP09(2012)033,
  10.1007/JHEP10(2012)051}{\emph{JHEP} {\bfseries 1209} (2012) 033},
  [\href{https://arxiv.org/abs/1206.6359}{{\ttfamily 1206.6359}}].

\bibitem{Banerjee:2009af}
N.~Banerjee, S.~Banerjee, R.~K. Gupta, I.~Mandal and A.~Sen,
  \emph{{Supersymmetry, Localization and Quantum Entropy Function}},
  \href{https://doi.org/10.1007/JHEP02(2010)091}{\emph{JHEP} {\bfseries 1002}
  (2010) 091}, [\href{https://arxiv.org/abs/0905.2686}{{\ttfamily 0905.2686}}].

\bibitem{Gupta:2015gga}
R.~K. Gupta, Y.~Ito and I.~Jeon, \emph{{Supersymmetric Localization for BPS
  Black Hole Entropy: 1-loop Partition Function from Vector Multiplets}},
  \href{https://doi.org/10.1007/JHEP11(2015)197}{\emph{JHEP} {\bfseries 11}
  (2015) 197}, [\href{https://arxiv.org/abs/1504.01700}{{\ttfamily
  1504.01700}}].

\bibitem{Murthy:2015yfa}
S.~Murthy and V.~Reys, \emph{{Functional determinants, index theorems, and
  exact quantum black hole entropy}},
  \href{https://doi.org/10.1007/JHEP12(2015)028}{\emph{JHEP} {\bfseries 12}
  (2015) 028}, [\href{https://arxiv.org/abs/1504.01400}{{\ttfamily
  1504.01400}}].

\bibitem{Gomis:1994he}
J.~Gomis, J.~Paris and S.~Samuel, \emph{{Antibracket, antifields and gauge
  theory quantization}},
  \href{https://doi.org/10.1016/0370-1573(94)00112-G}{\emph{Phys.Rept.}
  {\bfseries 259} (1995) 1--145},
  [\href{https://arxiv.org/abs/hep-th/9412228}{{\ttfamily hep-th/9412228}}].

\bibitem{Seiberg:1993vc}
N.~Seiberg, \emph{{Naturalness versus supersymmetric nonrenormalization
  theorems}}, \href{https://doi.org/10.1016/0370-2693(93)91541-T}{\emph{Phys.
  Lett.} {\bfseries B318} (1993) 469--475},
  [\href{https://arxiv.org/abs/hep-ph/9309335}{{\ttfamily hep-ph/9309335}}].

\bibitem{Freedman:2012zz}
D.~Z. Freedman and A.~Van~Proeyen, \emph{{Supergravity}}.
\newblock Cambridge Univ. Press, Cambridge, UK, 2012.

\bibitem{Kibble:1961ba}
T.~W.~B. Kibble, \emph{{Lorentz invariance and the gravitational field}},
  \href{https://doi.org/10.1063/1.1703702}{\emph{J. Math. Phys.} {\bfseries 2}
  (1961) 212--221}.

\bibitem{Pestun:2016zxk}
V.~Pestun et~al., \emph{{Localization techniques in quantum field theories}},
  \href{https://doi.org/10.1088/1751-8121/aa63c1}{\emph{J. Phys.} {\bfseries
  A50} (2017) 440301}, [\href{https://arxiv.org/abs/1608.02952}{{\ttfamily
  1608.02952}}].

\bibitem{David:2016onq}
J.~R. David, E.~Gava, R.~K. Gupta and K.~Narain, \emph{{Localization on
  AdS$_{2} \times$ S$^{1}$}},
  \href{https://doi.org/10.1007/JHEP03(2017)050}{\emph{JHEP} {\bfseries 03}
  (2017) 050}, [\href{https://arxiv.org/abs/1609.07443}{{\ttfamily
  1609.07443}}].

\bibitem{Sen:2011ba}
A.~Sen, \emph{{Logarithmic Corrections to N=2 Black Hole Entropy: An Infrared
  Window into the Microstates}},
  \href{https://doi.org/10.1007/s10714-012-1336-5}{\emph{Gen. Rel. Grav.}
  {\bfseries 44} (2012) 1207--1266},
  [\href{https://arxiv.org/abs/1108.3842}{{\ttfamily 1108.3842}}].

\bibitem{Sen:2008vm}
A.~Sen, \emph{{Quantum Entropy Function from AdS(2)/CFT(1) Correspondence}},
  \href{https://doi.org/10.1142/S0217751X09045893}{\emph{Int.J.Mod.Phys.}
  {\bfseries A24} (2009) 4225--4244},
  [\href{https://arxiv.org/abs/0809.3304}{{\ttfamily 0809.3304}}].

\bibitem{Gupta:2012cy}
R.~K. Gupta and S.~Murthy, \emph{{All solutions of the localization equations
  for N=2 quantum black hole entropy}},
  \href{https://doi.org/10.1007/JHEP02(2013)141}{\emph{JHEP} {\bfseries 1302}
  (2013) 141}, [\href{https://arxiv.org/abs/1208.6221}{{\ttfamily 1208.6221}}].

\bibitem{Banerjee:2008ky}
N.~Banerjee, D.~P. Jatkar and A.~Sen, \emph{{Asymptotic Expansion of the N=4
  Dyon Degeneracy}},
  \href{https://doi.org/10.1088/1126-6708/2009/05/121}{\emph{JHEP} {\bfseries
  05} (2009) 121}, [\href{https://arxiv.org/abs/0810.3472}{{\ttfamily
  0810.3472}}].

\bibitem{Murthy:2009dq}
S.~Murthy and B.~Pioline, \emph{{A Farey tale for $\mathcal{N}=4$ dyons}},
  \href{https://doi.org/10.1088/1126-6708/2009/09/022}{\emph{JHEP} {\bfseries
  09} (2009) 022}, [\href{https://arxiv.org/abs/0904.4253}{{\ttfamily
  0904.4253}}].

\bibitem{Dabholkar:2014ema}
A.~Dabholkar, J.~Gomes and S.~Murthy, \emph{{Nonperturbative black hole entropy
  and Kloosterman sums}},
  \href{https://doi.org/10.1007/JHEP03(2015)074}{\emph{JHEP} {\bfseries 03}
  (2015) 074}, [\href{https://arxiv.org/abs/1404.0033}{{\ttfamily 1404.0033}}].

\bibitem{LeeSJ}
S.~Lee, ``Index, supersymmetry, and localization, {Lectures at the Pyeong-Chang
  Summer School}.'' \url{http://psi.kias.re.kr/2013/sub02/sub02_01.php}, 2013.

\bibitem{Hosomichi:2015jta}
K.~Hosomichi, \emph{{The localization principle in SUSY gauge theories}},
  \href{https://doi.org/10.1093/ptep/ptv033}{\emph{PTEP} {\bfseries 2015}
  (2015) 11B101}, [\href{https://arxiv.org/abs/1502.04543}{{\ttfamily
  1502.04543}}].

\bibitem{Atiyah:1974}
M.~F. Atiyah, \emph{{Elliptic operators and compact groups}}.
\newblock Lecture Notes in Mathematics, Springer Verlag, Vol 401, 1974.

\bibitem{Assel:2016pgi}
B.~Assel, D.~Martelli, S.~Murthy and D.~Yokoyama, \emph{{Localization of
  supersymmetric field theories on non-compact hyperbolic three-manifolds}},
  \href{https://doi.org/10.1007/JHEP03(2017)095}{\emph{JHEP} {\bfseries 03}
  (2017) 095}, [\href{https://arxiv.org/abs/1609.08071}{{\ttfamily
  1609.08071}}].

\bibitem{David:2018pex}
J.~R. David, E.~Gava, R.~K. Gupta and K.~Narain, \emph{{Boundary Conditions and
  Localization on AdS: Part 1}},
  [\href{https://arxiv.org/abs/1802.00427}{{\ttfamily 1802.00427}}].

\bibitem{Banerjee:2010qc}
S.~Banerjee, R.~K. Gupta and A.~Sen, \emph{{Logarithmic Corrections to Extremal
  Black Hole Entropy from Quantum Entropy Function}},
  \href{https://doi.org/10.1007/JHEP03(2011)147}{\emph{JHEP} {\bfseries 1103}
  (2011) 147}, [\href{https://arxiv.org/abs/1005.3044}{{\ttfamily 1005.3044}}].

\bibitem{deWit:2010za}
B.~de~Wit, S.~Katmadas and M.~van Zalk, \emph{{New supersymmetric
  higher-derivative couplings: Full N=2 superspace does not count!}},
  \href{https://doi.org/10.1007/JHEP01(2011)007}{\emph{JHEP} {\bfseries 01}
  (2011) 007}, [\href{https://arxiv.org/abs/1010.2150}{{\ttfamily 1010.2150}}].

\bibitem{Butter:2014iwa}
D.~Butter, B.~de~Wit and I.~Lodato, \emph{{Non-renormalization theorems and N=2
  supersymmetric backgrounds}},
  \href{https://doi.org/10.1007/JHEP03(2014)131}{\emph{JHEP} {\bfseries 03}
  (2014) 131}, [\href{https://arxiv.org/abs/1401.6591}{{\ttfamily 1401.6591}}].

\bibitem{Murthy:2013xpa}
S.~Murthy and V.~Reys, \emph{{Quantum black hole entropy and the holomorphic
  prepotential of N=2 supergravity}},
  \href{https://doi.org/10.1007/JHEP10(2013)099}{\emph{JHEP} {\bfseries 10}
  (2013) 099}, [\href{https://arxiv.org/abs/1306.3796}{{\ttfamily 1306.3796}}].

\bibitem{Sen:2009vz}
A.~Sen, \emph{{Arithmetic of Quantum Entropy Function}},
  \href{https://doi.org/10.1088/1126-6708/2009/08/068}{\emph{JHEP} {\bfseries
  08} (2009) 068}, [\href{https://arxiv.org/abs/0903.1477}{{\ttfamily
  0903.1477}}].

\bibitem{Dabholkar:2010rm}
A.~Dabholkar, J.~Gomes, S.~Murthy and A.~Sen, \emph{{Supersymmetric Index from
  Black Hole Entropy}},
  \href{https://doi.org/10.1007/JHEP04(2011)034}{\emph{JHEP} {\bfseries 1104}
  (2011) 034}, [\href{https://arxiv.org/abs/1009.3226}{{\ttfamily 1009.3226}}].

\bibitem{Bringmann:2012zr}
K.~Bringmann and S.~Murthy, \emph{{On the positivity of black hole degeneracies
  in string theory}},
  \href{https://doi.org/10.4310/CNTP.2013.v7.n1.a2}{\emph{Commun. Num. Theor
  Phys.} {\bfseries 07} (2013) 15--56},
  [\href{https://arxiv.org/abs/1208.3476}{{\ttfamily 1208.3476}}].

\bibitem{Ooguri:2004zv}
H.~Ooguri, A.~Strominger and C.~Vafa, \emph{{Black hole attractors and the
  topological string}},
  \href{https://doi.org/10.1103/PhysRevD.70.106007}{\emph{Phys.Rev.} {\bfseries
  D70} (2004) 106007}, [\href{https://arxiv.org/abs/hep-th/0405146}{{\ttfamily
  hep-th/0405146}}].

\bibitem{LopesCardoso:1998wt}
G.~{Lopes Cardoso}, B.~de~Wit and T.~Mohaupt, \emph{Corrections to macroscopic
  supersymmetric black-hole entropy}, {\emph{Phys. Lett.} {\bfseries B451}
  (1999) 309--316}, [\href{https://arxiv.org/abs/hep-th/9812082}{{\ttfamily
  hep-th/9812082}}].

\bibitem{Murthy:2015zzy}
S.~Murthy and V.~Reys, \emph{{Single-centered black hole microstate
  degeneracies from instantons in supergravity}},
  \href{https://doi.org/10.1007/JHEP04(2016)052}{\emph{JHEP} {\bfseries 04}
  (2016) 052}, [\href{https://arxiv.org/abs/1512.01553}{{\ttfamily
  1512.01553}}].

\bibitem{Witten:1988xi}
E.~Witten, \emph{{Topological Gravity}},
  \href{https://doi.org/10.1016/0370-2693(88)90704-6}{\emph{Phys. Lett.}
  {\bfseries B206} (1988) 601--606}.

\bibitem{BenettiGenolini:2017zmu}
P.~Benetti~Genolini, P.~Richmond and J.~Sparks, \emph{{Topological AdS/CFT}},
  \href{https://doi.org/10.1007/JHEP12(2017)039}{\emph{JHEP} {\bfseries 12}
  (2017) 039}, [\href{https://arxiv.org/abs/1707.08575}{{\ttfamily
  1707.08575}}].

\bibitem{Brennan:2017rbf}
T.~D. Brennan, F.~Carta and C.~Vafa, \emph{{The String Landscape, the
  Swampland, and the Missing Corner}},
  [\href{https://arxiv.org/abs/1711.00864}{{\ttfamily 1711.00864}}].

\bibitem{deWit:2017cle}
B.~de~Wit and V.~Reys, \emph{{Euclidean supergravity}},
  \href{https://doi.org/10.1007/JHEP12(2017)011}{\emph{JHEP} {\bfseries 12}
  (2017) 011}, [\href{https://arxiv.org/abs/1706.04973}{{\ttfamily1706.04973}}].

\bibitem{Mohaupt:2000mj}
T.~Mohaupt, \emph{{Black hole entropy, special geometry and strings}},
  {\emph{Fortsch. Phys.} {\bfseries 49} (2001) 3--161},
  [\href{https://arxiv.org/abs/hep-th/0007195}{{\ttfamily hep-th/0007195}}].

\bibitem{Sen:2012cj}
  A.~Sen,
  \emph{{Logarithmic Corrections to Rotating Extremal Black Hole Entropy in Four and Five Dimensions}},
      \href{https://doi.org/10.1007/s10714-012-1373-0}{\emph{Gen. Rel. Grav.} {\bfseries 44} (2012) 1947},
  [\href{https://arxiv.org/abs/hep-th/1109.3706}{{\ttfamily 1109.3706}}].

\end{thebibliography}

\providecommand{\href}[2]{#2}\begingroup\raggedright\endgroup

\end{document}